\documentclass[pdflatex,sn-nature]{sn-jnl}

\usepackage{graphicx}%
\usepackage{multirow}%
\usepackage{amsmath,amssymb,amsfonts}%
\usepackage{amsthm}%
\usepackage{mathrsfs}%
\usepackage[title]{appendix}%
\usepackage{xcolor}%
\usepackage{textcomp}%
\usepackage{manyfoot}%
\usepackage{booktabs}%
\usepackage{algorithm}%
\usepackage{algorithmicx}%
\usepackage{algpseudocode}%
\usepackage{listings}%

\theoremstyle{thmstyleone}%
\theoremstyle{thmstyletwo}%

\theoremstyle{thmstylethree}%

\newcounter{extfig}

\begin{document}

\def \nuprocess{$\nu$-process}
\def \nodata{. . .}
\def \degree{$^{\circ}$}
\def \Msolar{M$_{\odot}$}
\def \alphafe{[$\alpha$/Fe]}
\def \HI{H\ion{I}}
\def \sion{\ion{II}}
\def \vninety{v$_{90}$}
\def \Lbol{L$_{\rm bol}$}
\def \Mstar{M_{\star}}
\def \logMstar{\log(M_{\star}/\mathrm{M}_{\odot})}
\def \logRimp{log(R$_{\rm imp}$/kpc)}
\def \logLbol{log(L$_{\rm AGN}$/erg s$^{-1}$)}
\def \logsSFR{log(\mathrm{sSFR / yr}^{-1})}
\def \kms{km s$^{-1}$}
\def \zabs{z$_{\rm abs}$}
\def \zem{z$_{\rm em}$}
\def \Rimp{$\rho_{\rm imp}$}
\def \Rvir{$\rho_{\rm vir}$}
\def \deltaEW{${\rm \Delta log(EW/m\AA)}$}
\def \RadRat{$f_{\rm AGN}/f_{\rm HM01}$}
\def \mnfe{[Mn/Fe]$_{\rm DC}$}
\def \Mgeqw{W$_{0}^{2796}$}
\def \Feeqw{W$_{0}^{2600}$}
\def \fracMgFe{ ${\rm{W}_{0}^{2796}}$/${\rm{W}_{0}^{2600}}$}
\def \omegaDLA{$\Omega_{\rm H \textsc{i}}$}
\def \ndla{30}
\def \npdla{46}
\def \nlpdla{41}
\def \nxpdla{5}
\def \nmdla{27}
\def \nlmdla{21}
\def \nxmdla{6}
\def \CosmoZ{$\langle Z/Z_{\odot} \rangle$}
\def \fNX{$f(N,X)$}
\def \deltasfms{\Delta \langle \log(\mathrm{SFR})\rangle_{\mathrm{MS}}}
\def \dlogsfms{\Delta \langle \log(\mathrm{SSFR})\rangle_{\mathrm{MS}}}
\def \fgas{f_{\mathrm{gas}}}
\def \mstar{M_{\star}}
\def \logsfr{\log({\mathrm{SFR}})}
\def \dlogsfr{\Delta\log({\mathrm{SFR}})}
\def \logssfr{\log({\mathrm{SSFR}})}


\newcommand{\actaa}{Acta Astron.} 
\newcommand{\araa}{Annu. Rev. Astron. Astrophys.} 
\newcommand{\aar}{Astron. Astrophys. Rev.} 
\newcommand{\ab}{Astrobiol.} 
\newcommand{\aj}{Astron. J.} 
\newcommand{\apj}{Astrophys. J.} 
\newcommand{\apjl}{Astrophys. J. Lett.} 
\newcommand{\apjs}{Astrophys. J. Suppl. Ser.} 
\newcommand{\ao}{Appl. Opt.} 
\newcommand{\apss}{Astrophys. Space Sci.} 
\newcommand{\aap}{Astron. Astrophys.} 
\newcommand{\aapr}{Astron. Astrophys. Rev.} 
\newcommand{\aaps}{Astron. Astrophys. Suppl.} 
\newcommand{\baas}{Bull. Am. Astron. Soc.} 
\newcommand{\caa}{Chinese Astron. Astrophys.} 
\newcommand{\cjaa}{Chinese J. Astron. Astrophys.} 
\newcommand{\cqg}{Class. Quantum Gravity} 
\newcommand{\gal}{Galaxies} 
\newcommand{\gca}{Geochim. Cosmochim. Acta} 
\newcommand{\icarus}{Icarus} 
\newcommand{\jcap}{J. Cosmol. Astropart. Phys.} 
\newcommand{\compastro}{Comput. Astrophys.} 
\newcommand{\jgr}{J. Geophys. Res.} 
\newcommand{\jgrp}{J. Geophys. Res.: Planets} 
\newcommand{\jqsrt}{J. Quant. Spectrosc. Radiat. Transf.} 
\newcommand{\memsai}{Mem. Soc. Astron. Italiana} 
\newcommand{\mnras}{Mon. Not. R. Astron. Soc.} 
\newcommand{\nat}{Nature} 
\newcommand{\nastro}{Nat. Astron.} 
\newcommand{\ncomms}{Nat. Commun.} 
\newcommand{\nphys}{Nat. Phys.} 
\newcommand{\nrevphys}{Nat. Rev. Phys.}
\newcommand{\na}{New Astron.} 
\newcommand{\nar}{New Astron. Rev.} 
\newcommand{\physrep}{Phys. Rep.} 
\newcommand{\pra}{Phys. Rev. A} 
\newcommand{\prb}{Phys. Rev. B} 
\newcommand{\prc}{Phys. Rev. C} 
\newcommand{\prd}{Phys. Rev. D} 
\newcommand{\pre}{Phys. Rev. E} 
\newcommand{\prl}{Phys. Rev. Lett.} 
\newcommand{\psj}{Planet. Sci. J.} 
\newcommand{\planss}{Planet. Space Sci.} 
\newcommand{\pnas}{Proc. Natl Acad. Sci. USA} 
\newcommand{\procspie}{Proc. SPIE} 
\newcommand{\pasa}{Publ. Astron. Soc. Aust.} 
\newcommand{\pasj}{Publ. Astron. Soc. Jpn} 
\newcommand{\pasp}{Publ. Astron. Soc. Pac.} 
\newcommand{\rmxaa}{Rev. Mexicana Astron. Astrofis.} 
\newcommand{\sci}{Science} 
\newcommand{\sciadv}{Sci. Adv.} 
\newcommand{\solphys}{Sol. Phys.} 
\newcommand{\sovast}{Soviet Ast.} 
\newcommand{\ssr}{Space Sci. Rev.} 
\newcommand{\uni}{Universe} 

\title{Widespread Inflows Reveal Baryonic Cycling in Star-forming and Quiescent Galaxies}

\author*[1]{\fnm{Hassen M.} \sur{Yesuf}}\email{yesufh@shao.ac.cn}
\author[2]{\fnm{Ravi}\sur{Joshi}}
\author[1,3]{\fnm{Yuxuan}\sur{Zou}}
\author[4]{\fnm{Feng}\sur{Yuan}}
\author[5,6]{\fnm{Luis C.}\sur{Ho}}
\author[1]{\fnm{Lin}\sur{Lin}}
\author[1]{\fnm{Lei}\sur{Hao}}
\author[1]{\fnm{Shiyin}\sur{Shen}}
\author[7]{\fnm{Connor} \sur{Bottrell}}
\author[1]{\fnm{Fulai} \sur{Guo}}
\author[8]{\fnm{John D.} \sur{Silverman}}

\affil[1]{\orgdiv{Astrophysics Division}, \orgname{Shanghai Astronomical Observatory, Chinese Academy of Sciences}, \orgaddress{\street{80 Nandan Road}, \city{Shanghai}, \postcode{200030}, \country{China}}}

\affil[2]{\orgname{Indian Institute of Astrophysics (IIA)}, \orgaddress{\city{Koramangala}, \state{Bangalore}, \postcode{560034}, \country{India}}}

\affil[3]{\orgname{University of Chinese Academy of Sciences}, \orgaddress{No. 19A Yuquan Road, \city{Beijing} \postcode{100049}, \country{China}}}

\affil[4]{\orgdiv{Center for Astronomy and Astrophysics and Department of Physics}, \orgname{Fudan University}, \orgaddress{\city{Shanghai}, \postcode{200438}, \country{China}}}

\affil[5]{\orgdiv{Kavli Institute for Astronomy and Astrophysics}, \orgname{Peking University}, \orgaddress{\city{Beijing}, \postcode{100871}, \country{China}}}

\affil[6]{\orgdiv{Department of Astronomy, School of Physics}, \orgname{Peking University}, \orgaddress{\city{Beijing}, \postcode{100871}, \country{China}}}

\affil[7]{\orgdiv{International Centre for Radio Astronomy Research}, \orgname{University of Western Australia}, \orgaddress{\street{35 Stirling Hwy}, \city{Crawley}, \postcode{6009}, \state{WA}, \country{Australia}}}

\affil[8]{\orgdiv{Kavli Institute for the Physics and Mathematics of the Universe}, \orgname{University of Tokyo}, \orgaddress{\street{5-1-5 Kashiwanoha Campus}, \city{Kashiwa}, \postcode{277-8583}, \state{Chiba}, \country{Japan}}}

\date{\today}

\newcommand{\arcsec}{^{\prime\prime}}

\maketitle

\textbf{Cool-gas inflows, required to sustain star formation and black-hole growth, have been fundamental in simulations yet remained observationally elusive. Using DESI spectroscopy of $\sim30{,}000$ galaxies, we identify coherent inflowing gas ($\sim100~\mathrm{km\,s^{-1}}$) in $\sim20$--$50\%$ of the sample, yielding the first population-level census of galactic gas flows. We uncover a striking inversion: inflows are most frequently detected in quiescent galaxies, whereas star-forming systems are dominated by gravitationally bound outflows. At fixed stellar age, galaxies with inflows, outflows, or no/weak flows share similar masses, environments, and structures, indicating that these longer-lived host-galaxy properties do not strongly differentiate the observed flow states. Instead, gas-flow state is more closely linked to stellar population age and recent evolutionary history, consistent with age-dependent gas flows operating in two regimes.  In some star-forming galaxies, elevated star formation surface densities drive outflows that likely recycle on short timescales ($\sim0.5$\,Gyr), consistent with a galactic fountain. In quiescent systems, low-level ``drizzling'' inflows persist, consistent with supply by slowly cooling enriched halo gas and often accompanied by weak, radio-mode nuclear activity. The broad gas-phase metallicity distributions---and the absence of a pristine dilution signature---indicate that the detected inflows are predominantly recycled or previously enriched. Detectability is strongly modulated by dust shielding, ionization, and anisotropic geometry: in star-forming disks, inflowing gas lies near the disk plane, where it is often obscured or ionized, while outflow hosts exhibit higher dust and metal content. As star formation declines, cold-outflow signatures weaken, and gas that is likely recycled or slowly cooling is more readily detected as inflow. Post-starburst galaxies provide direct snapshots of this rapid transition. Together, our results resolve the long-standing scarcity of observed inflows, provide evidence for widespread gas accretion and recycling in present-day galaxies, and establish an observational framework linking gas flows to star formation, chemical evolution, and galaxy structure.}

\section*{Main}

\begin{figure}
\includegraphics[width=\textwidth]{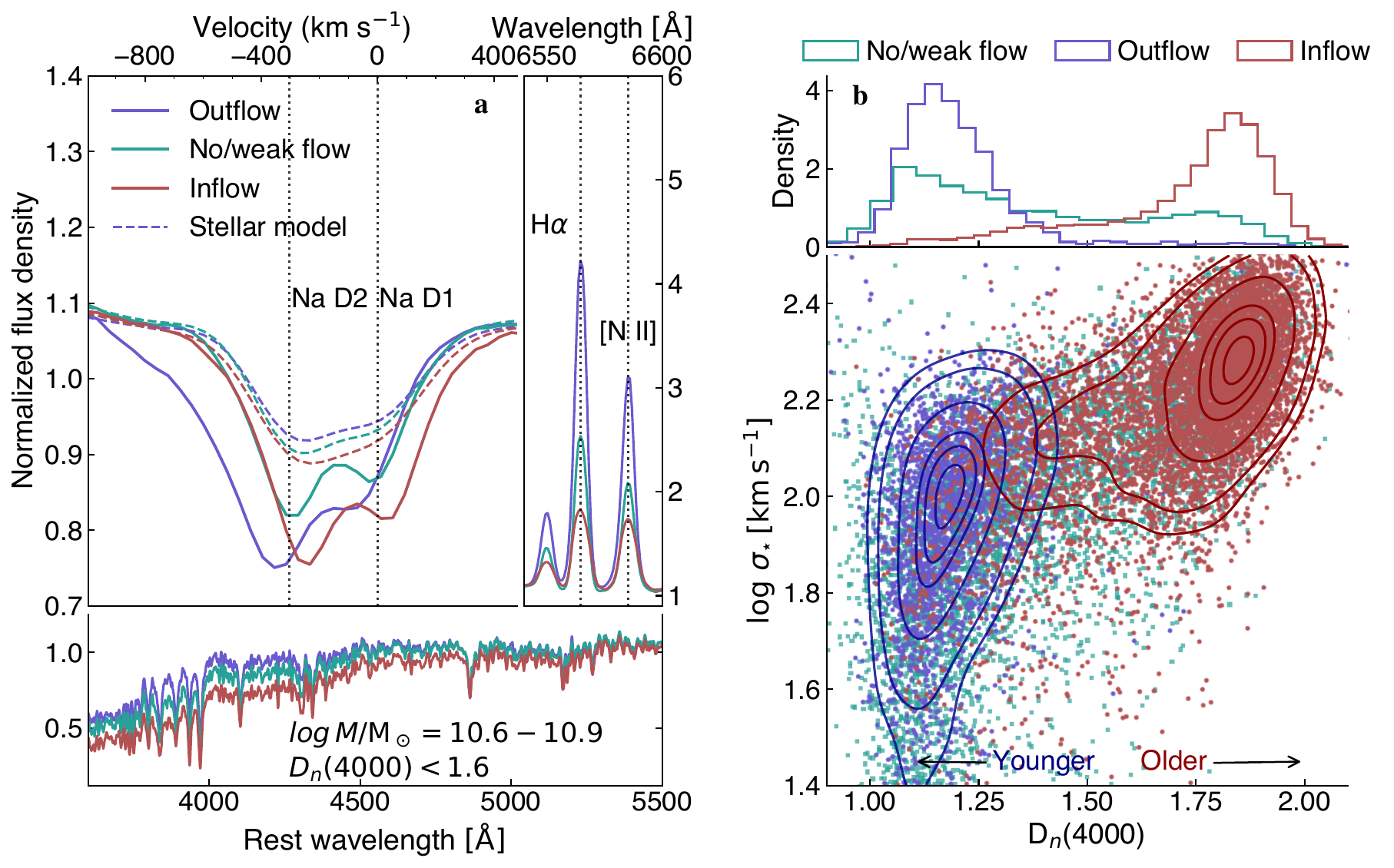}
\caption{\textbf{Spectroscopic signatures of gas flows and their stellar populations.} \textbf{a,} Zoomed-in example stacked spectra of young galaxies ($D_n(4000) < 1.6$) with stellar masses $\log (M_\star/M_\odot)=10.6$--10.9, shown around the Na\,\textsc{i}\,D absorption doublet (top), H$\alpha$ and [N\,II] emission (right), and the emission-subtracted blue stellar continuum (bottom). Young galaxies are classified by gas-flow state prior to stacking in this narrow-mass range: outflow (blue, $N=441$), no/weak flow (teal, $N=855$), and inflow (red, $N=733$ galaxies). Dashed curves indicate the modeled stellar photospheric Na\,\textsc{i}\,D absorption; residual absorption traces interstellar neutral gas. Doppler shifts in Na\,\textsc{i}\,D relative to the systemic rest frame identify cool gas inflows and outflows, while the ionized-gas emission lines remain centered at systemic velocity. Differences in spectral shape and emission-line strengths are modest, as the stacks are restricted to young galaxies within a narrow stellar-mass range, highlighting the connection between flow state and galaxy properties. \textbf{b,} Distribution of the 4000\,\AA\ break index, $D_n(4000)$, for the full sample, shown as histograms for each flow category and as $D_n(4000)$ versus stellar velocity dispersion. Contours indicate the 5, 15, 25, 50, 75, and 85\% number-density levels of the inflow or outflow hosts. Unlike panel~(a), no restriction in mass or age is applied. This panel places individual galaxies with inflows, outflows, and no/weak flows in the broader stellar-population and kinematic/structural space of the full sample, showing how the inflow and outflow states occupy distinct regions of this diagnostic diagram.}\label{fig1}
\end{figure}

\begin{figure}
\includegraphics[width=1.15\textwidth]{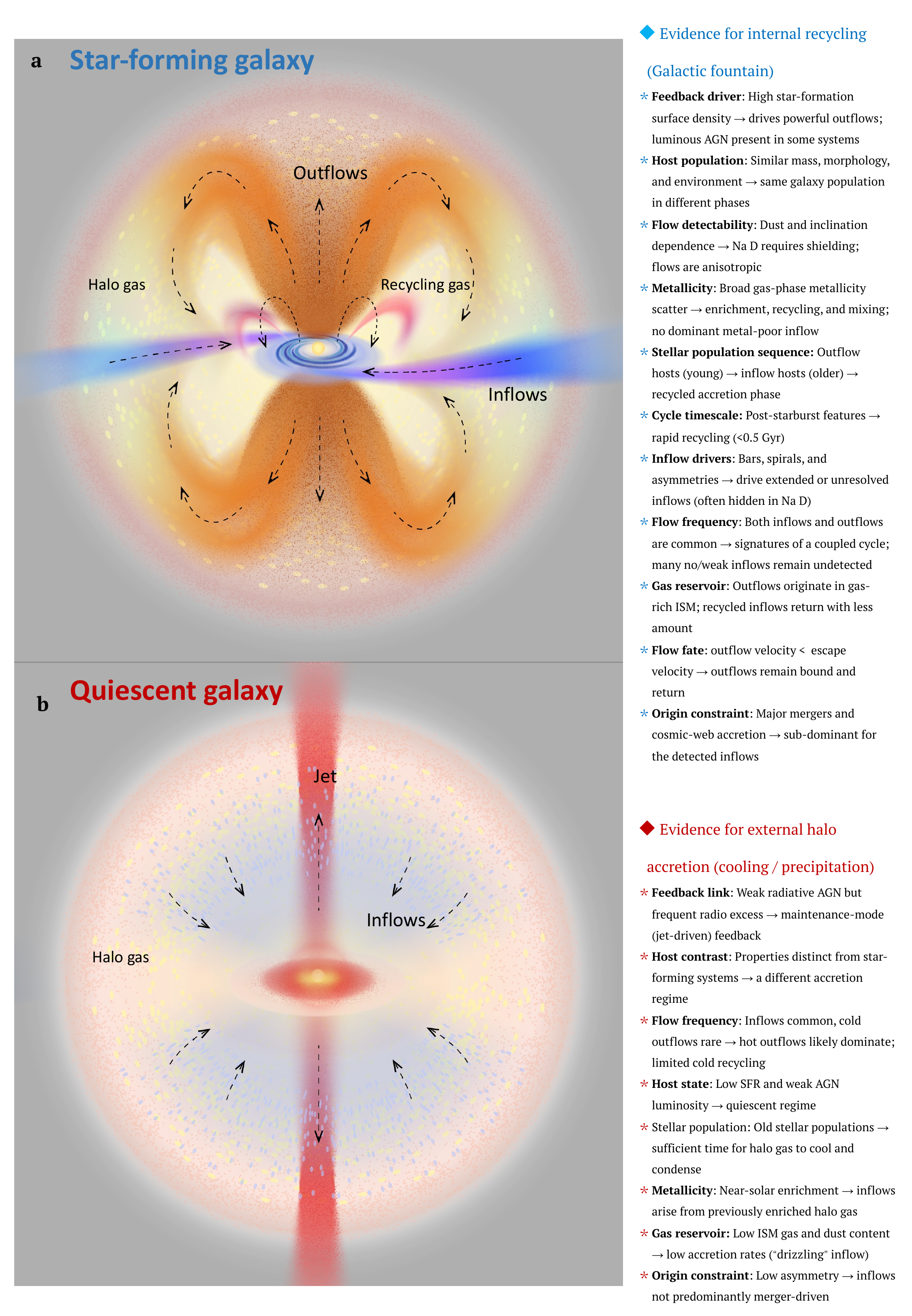}
\caption{\textbf{Schematic of the unified framework for cool gas flows}. The framework comprises two regimes: \textbf{a}, In star-forming galaxies, stellar feedback drives outflows that recycle on short timescales, forming a galactic fountain. \textbf{b}, In quiescent systems, the cycle shifts to a ``drizzling'' mode governed by cooling from hot halos. Observational support for the framework is listed on the right. The illustration is not to scale.}
\end{figure}

\begin{figure}
 \includegraphics[width=\textwidth]{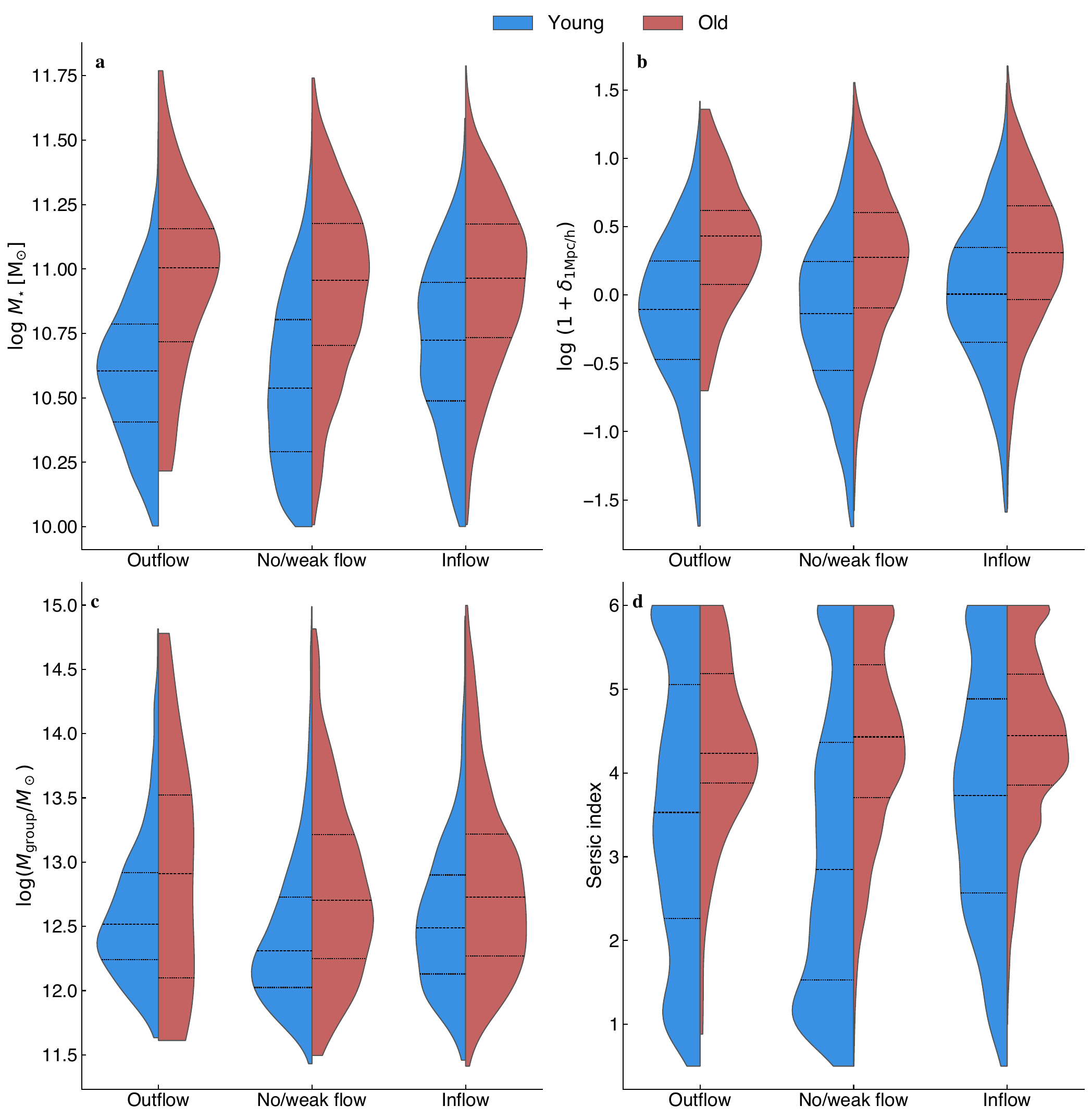}
\caption{\textbf{Similarity of host-galaxy demographics at fixed stellar population age.}
\textbf{a},  the stellar mass distributions of young ($D_n(4000)<1.6$) and old ($D_n(4000) \ge 1.6$) galaxies. 
\textbf{b}, the large-scale environmental overdensity (defined as the stellar-mass excess within $1\,h^{-1}\,\mathrm{Mpc}$); \textbf{c}, dark-matter halo mass; \textbf{d}, S\'{e}rsic index, a proxy for bulge prominence. 
In each panel, dashed vertical lines indicate the 25th, 50th, and 75th percentiles of the corresponding distributions. As expected, old galaxies preferentially inhabit more massive,
bulge-dominated systems, more massive haloes, and denser environments than young galaxies.
However, at fixed stellar age, galaxies classified as outflow, no flow, and inflow occupy
similar host galaxies and environments. These global properties therefore do not uniquely
determine the instantaneous gas-flow state. Taken together with Fig.~1, this suggests that
flow states in young galaxies are regulated by recent variability in gas accretion, feedback, and recycling, superimposed on long-term galaxy evolution. The sample is restricted to $M_\star > 10^{10}\,\mathrm{M_\odot}$ here. Stellar mass and structural measurements are available for all galaxies; environmental-density and group mass estimates are available for 93\% and 48\% of the sample, respectively.}
\end{figure}

Galactic outflows of cool, neutral gas, often traced by the Na\,\textsc{i}\,D doublet, are commonly observed in the local Universe \citep{Veilleux+20}. They have been detected in star-forming and starburst galaxies \citep{MartinCL05,Rupke+05,ChenY+10,Concas+19,Roberts-Borsani+20}, as well as in galaxies hosting active galactic nuclei (AGN) \citep{Sarzi+16,Nedelchev+19}. These outflows typically reach velocities of a few hundred km\,s$^{-1}$ and are primarily driven by stellar feedback in star-forming galaxies, with additional contributions from AGN in some systems \citep{Sarzi+16,Nedelchev+19}. The incidence and observational strength of cool-gas outflows generally increase with star formation activity and dust attenuation, as reflected by higher detection rates and stronger absorption-line equivalent widths, although these trends exhibit substantial scatter \citep{MartinCL05,ChenY+10,Concas+19}. The observed prevalence of cool-gas outflows decreases with increasing disk inclination because galactic winds are preferentially launched along the minor axis. Although the Na\,\textsc{i}\,D absorption strength, which is largely produced by the interstellar medium, generally increases toward edge-on galaxies, projection effects and disk absorption make outflow signatures more difficult to identify \citep{ChenY+10,Concas+19}.

On the other hand, despite routinely predicted in cosmological simulations, direct detections of cool gas inflows remain rare \cite{Krug+10,Martin+12,Rubin+12,Roy+21,Rupke+21,Fernandez-Figueroa+25}, likely reflecting observational challenges rather than true absence. Short gas depletion times ($\sim 1$--2~Gyr) \cite{SaintongeCatinella22} imply that star-forming galaxies must continuously replenish their gas supply. This gas may arise from internal mass loss from evolved stars, accretion of cooling halo gas, or gas-rich mergers, with stellar mass loss expected to dominate in many nearby systems \cite{LeitnerKravtsov+11}. 

To address the conundrum on how galaxies fuel star formation for longer than their depletion time essentially requires a population-level analysis that jointly characterizes inflows and outflows across the full range of galaxy evolutionary states. Leveraging the statistical power of the DESI survey and its improved spectral resolution, we present the first comprehensive census of cool gas inflows and outflows traced by Na\,\textsc{i}\,D absorption, and interpret the results within a unified framework that connects Na\,\textsc{i}\,D absorption properties and kinematics to the underlying baryon cycle while accounting for selection effects.

In Milky Way–like spiral galaxies, simulations predict a dynamic cycle of gas flows in which fresh and recycled (chemically enriched) material accretes preferentially near the stellar disk, while feedback-driven outflows emerge primarily perpendicular to the disk \citep{Keres+05, FraternaliBinney+08, Oppenheimer+10, AnglesAlcazar2017, Grand+19}. A key challenge is that inflows are expected to arise from multiple physical channels operating across different environments and scales. Disk-plane inflows are typically slow ($\lesssim$10\,km\,s$^{-1}$; \citep{Schmidt+16, DiTeodoroPeek21, Trapp+22}), much slower than random gas motions or rotation, and generally insufficient to sustain star formation alone. Additional supply may come from extraplanar cold streams, high-velocity clouds in inner halos, or condensation near the disk--halo boundary \citep{Fraternali17, Armillotta+16}. In massive ellipticals and galaxy groups, precipitation from cooling hot gas drives low-level inflows that maintain residual star formation and long-term black hole accretion \citep{Gaspari+18}. Bars and spiral arms further channel gas inward \citep{YuHo22}, producing intermittent central inflows that concentrate gas and enhance star formation. Together, these processes create a multi-phase, time-variable, and strongly geometry-dependent gas supply, explaining the difficulty of directly detecting inflows.

As an example, Fig.~1a shows stacked spectra of young, Milky Way–mass galaxies, revealing clear and coherent Doppler shifts in Na\,\textsc{i}\,D absorption ($|v| \sim 100$--200~km~s$^{-1}$) for both inflow and outflow populations. The presence of these shifts in the stacked spectra demonstrates that the flows are directly observed and common signatures, rather than relying on detailed modeling. Nebular emission lines such as H$\alpha$ and [N\,\textsc{II}] remain centered near the systemic redshift, whereas Na\,\textsc{i}\,D absorption exhibits significant Doppler shifts, indicating that the cool, Na-bearing gas is kinematically decoupled from the bulk stellar and ionized interstellar medium (ISM). The stacked spectra also reveal broadly similar stellar ages and metallicities, with only minor differences associated with recent star formation activity. Additional stacked examples spanning a wider range of stellar masses and ages are shown in Extended Data Fig.~1, while Extended Data Figs.~2 and 3 present individual spectra of recently quenched (post-starburst) galaxies exhibiting inflows and outflows.

After detailed spectral modeling of individual galaxies using stellar population synthesis combined with a two-component absorption model (interstellar medium plus flowing gas; see Methods), we find that $\sim$20\% of our $\sim$30{,}000 galaxies host high-confidence cool-gas inflows ($P>0.85$, $v_{\rm flow} > 30~\mathrm{km\,s^{-1}}$), with the intrinsic incidence likely exceeding 50\% when intermediate-probability inflows are included.

Fig.~1b shows the full sample as a function of the 4000\,\AA\ break index, $D_n(4000)$, a proxy for light-weighted stellar age, and stellar velocity dispersion, $\sigma_\star$, colored by flow state. Outflows are preferentially associated with younger stellar populations, whereas inflows are dominated by older galaxies, with a tail extending to intermediate ages. Galaxies with no or weak flows ($-30 < v < 30$~km~s$^{-1}$) are distributed across the full age range with comparable frequency. Older inflow hosts are more kinematically dispersion-dominated than young outflow systems, indicating that gas-flow mode depends on both evolutionary state and galaxy structure (Fig.~1b).

We use the empirical trends in Fig.~1 to motivate our framework (Fig.~2) and test it using multiple independent diagnostics. Stellar population age and star formation properties trace temporal sequencing; feedback indicators (star formation surface density and AGN activity) identify driving mechanisms; metallicity constrains gas origin; and dust content and inclination probe geometric and shielding effects. Bars and spirals drive internal gas flows, while asymmetries and environment probe external influences. If inflows and outflows were unrelated phenomena tied to fundamentally different populations, they would exhibit distinct host and environmental properties at fixed stellar mass. Indeed, such separation is observed between star-forming and quiescent populations. However, within the young population, matched samples of inflow, outflow, and weak/no-flow galaxies share similar global properties, differing primarily in star formation intensity, stellar age, dust content, and geometry. This is consistent with a plausible picture in which the observed association between gas-flow state and stellar population age reflects the interplay between feedback, accretion, and recent galaxy evolution: feedback-associated outflows and inflow signatures in star-forming galaxies with different stellar population ages represent different states of gas supply and cycling in at least some galaxies, whereas flows in quiescent galaxies likely reflect a different accretion regime.

This bimodality in Fig.~1b is reflected in star formation rate and star formation surface density, mass-weighted stellar age, and stellar assembly history (Extended Data Figs.~4 and 5). Outflows preferentially occur in galaxies with high star formation surface density and young stellar populations, consistent with feedback-driven launching, whereas inflows are predominantly found in older, more dynamically evolved systems with higher velocity dispersions. Because structural evolution proceeds on Gyr timescales, comparable to the $\sim7\,\mathrm{Gyr}$ difference in median light-weighted stellar age between the populations, inflow and outflow hosts occupy largely distinct evolutionary regimes. Consistently, rapid transitions between these states are rare in present-day galaxies (Extended Data Fig.~4; \citep{Yesuf+14,Tanaka+24}).

Where, then, are the initial inflows that fueled star formation and the readily observed outflows that follow from it in the gas cycle? A small number of recently quenched, young galaxies exhibit all three flow states (Extended Data Figs.~2--4 and Supplementary Information), providing direct snapshots of gas cycling and showing that some gas is newly accreted or recycled rather than residual from pre-quenching. 

We next examine the tail of the inflow distribution at young stellar ages (Fig.~1b), highlighting its connection to the broader outflow population. To isolate the role of assembly history, we divide galaxies into young and old populations at $D_n(4000)=1.6$ (corresponding to characteristic ages of $\sim1$–2\,Gyr). Fig.~3 shows that, at fixed $D_n(4000)$, the different flow classes have comparable distributions in stellar mass, environment (quantified by stellar-mass excess within $1\,h^{-1}\,\mathrm{Mpc}$ \citep{Yesuf2022}), group halo mass, and S\'{e}rsic index ($n$). These results are robust to more stringent matching tests (Supplementary Information), and environmental measures spanning 0.25--$10\,h^{-1},\mathrm{Mpc}$ yield similar results. Within the young sample, inflow, outflow, and no/weak-flow galaxies therefore occupy similar ranges of stellar mass, environment, and internal structure, consistent with a shared underlying population observed at different stages of the gas cycle.

\begin{figure}
\includegraphics[width=\textwidth]{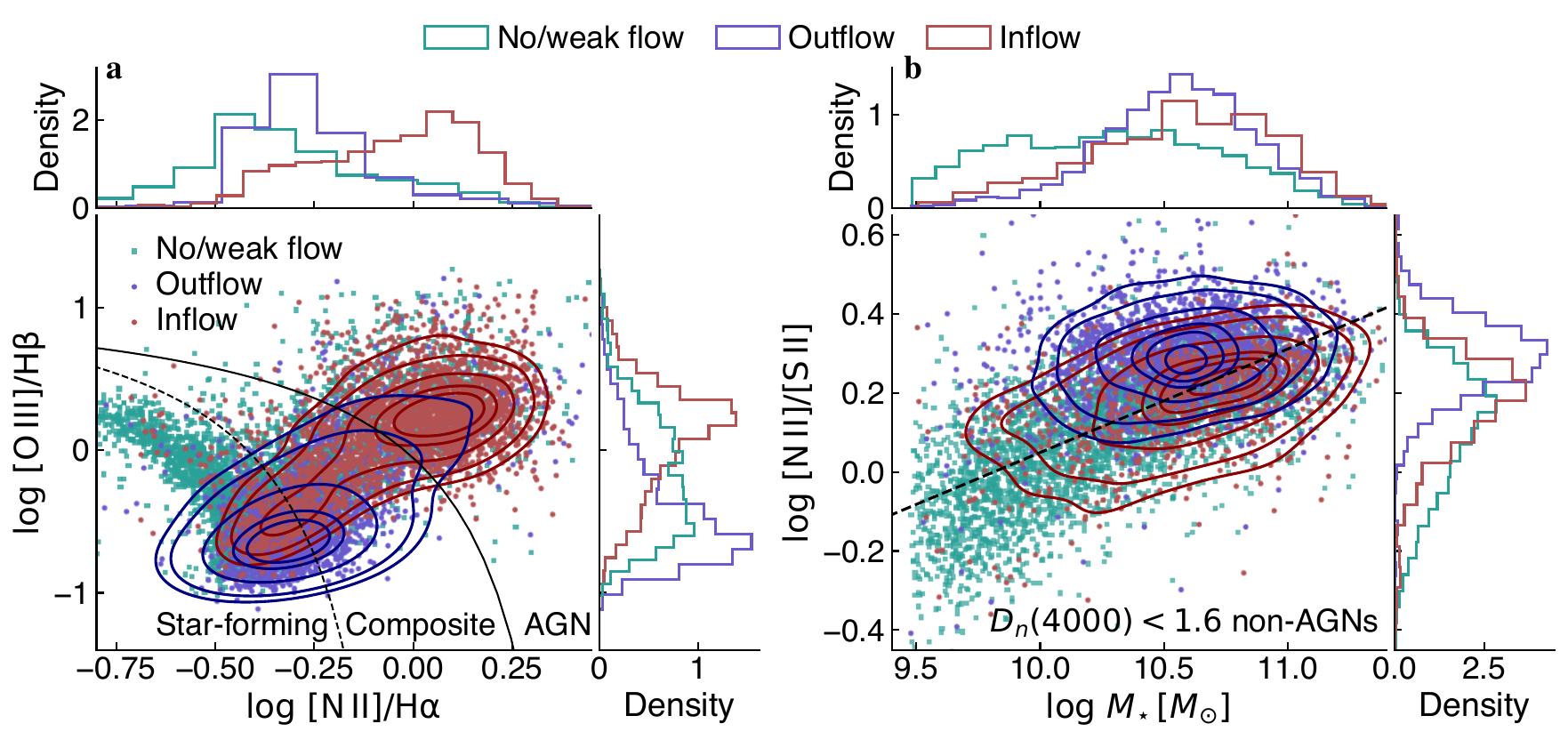}
\caption{\textbf{Gas excitation and chemical enrichment across different gas-flow states.} Inflow, outflow, and no/weak-flow galaxies are shown in red, blue, and teal, respectively. Contours indicate the 15th, 25th, 50th, 75th, and 85th percentiles of the data distribution. \textbf{a}, Emission-line diagnostic diagram separating ionization by star formation and black hole accretion. The solid black curve marks the theoretical maximum starburst boundary, above which line ratios require dominant AGN photoionization \citep{Kewley+01}. The dashed curve indicates the empirical demarcation below which galaxies are predominantly star-formation dominated \citep{Kauffmann+03}. \textbf{b}, Stellar mass versus the [N\,\textsc{ii}]/[S\,\textsc{ii}] line ratio, used as a proxy for chemical enrichment, for young non-AGN galaxies. The dashed line shows the median regression fit to the full young non-AGN galaxy sample.}\label{fig:BPT_MZR}
\end{figure}

Structural differences reflect differences in stellar population age. Old inflow hosts are dominated by dense, pressure-supported spheroids, with higher velocity dispersions and S\'{e}rsic indices ($n\!\approx\!3$--5.5; $\sigma_\star \!\approx\!180~\mathrm{km\,s^{-1}}$), whereas the young population spans a broader range of morphologies and lower velocity dispersions, with medians of $\sigma_\star \!\approx\!120~\mathrm{km\,s^{-1}}$ for inflows and $\sigma_\star \!\approx\!100~\mathrm{km\,s^{-1}}$ for outflows. Systems with no or weak flows are the least centrally concentrated. The outflow velocities in young galaxies are only a few times the stellar velocity dispersion, implying that much of the ejected material remains gravitationally bound and likely recycles. Galaxy mass, potential depth, and environment therefore do not uniquely determine whether a galaxy hosts an inflow or outflow. The key question is what regulates the gas-flow state and its observability.

We address this in Fig.~4, linking gas-flow state to nuclear activity and chemical enrichment using a standard optical line ratio diagnostic (panel a) and the mass--metallicity relation (MZR) traced by [N\,\textsc{ii}]/[S\,\textsc{ii}] in young non-AGN galaxies (panel b). All flow states appear in both star-forming and AGN-host galaxies, indicating that AGN activity is not the dominant driver of large-scale cool outflows. Nevertheless, weak AGN are more common among inflow hosts ($\sim$60\%) than among outflow ($\sim$13\%) or no/weak-flow systems ($\sim$20\%), a trend that persists in the subset of young galaxies ($D_n(4000)<1.6$).

Outflows show a stronger association with star formation. They preferentially occur in galaxies exceeding a star formation surface density of $\Sigma_{\rm SFR} \sim 0.1~{\rm M_\odot~yr^{-1}~kpc^{-2}}$, whereas inflow and no/weak-flow systems lie below this threshold (Supplementary Information). This scale is consistent with efficient wind launching in star-forming disks \citep{Heckman+15}, indicating that gas-flow state in the young population is closely linked to central star formation intensity rather than galaxy mass or environment.

In older galaxies, a distinct regime emerges. Inflow hosts show modestly enhanced nuclear radio emission ($\sim3\times10^{22}$~W~Hz$^{-1}$; Extended Data Fig.~6c,d), consistent with low-level AGN activity. Because low-Eddington-ratio black holes are preferentially radio-loud \citep[e.g.,][]{Ho2002}, these inflows may trace the fuel supply sustaining maintenance-mode feedback, in which halo cooling is regulated rather than quenched \citep{Gaspari+18,guo18}. Additional inflow channels may arise from condensation in hot buoyant outflows \citep{guo18,duan24}, while AGN-driven hot winds are unlikely to be detectable in Na\,\textsc{i}\,D absorption \citep{guo12,guo18,zhang20}.

Gas-phase metallicity provides an additional constraint. Many no/weak-flow galaxies occupy the high-[O\,\textsc{iii}]/H$\beta$ wing of the star-forming sequence (Fig.~4a), reflecting low gas-phase metallicities. We quantify enrichment via offsets from the mass--metallicity relation (Fig.~4b and Extended Data Fig.~7), using \([\mathrm{N\,II}]/[\mathrm{S\,II}]\) as a proxy (Methods). At fixed stellar mass, outflow galaxies show systematically elevated \([\mathrm{N\,II}]/[\mathrm{S\,II}]\), indicating more enriched gas, whereas inflow galaxies span a broad range; the most enriched systems also show stronger Na\,\textsc{i}\,D, consistent with recycled accretion.

Offsets and scatter about the MZR imply that enrichment and accretion vary on short timescales, with much inflowing gas evading direct detection. Inflow systems lie roughly symmetrically above and below the relation, contrary to the expectation of predominantly metal-poor accretion. Instead, $\sim$30\% of galaxies below the relation show no/weak flow signatures, consistent with dilution by low-metallicity accretion or minor mergers. Only $\sim$7\% of galaxies below MZR show high-confidence inflows, rising to $\sim$30\% including intermediate cases, with outflow fractions remaining comparable.

\begin{figure}
 \includegraphics[width=\textwidth]{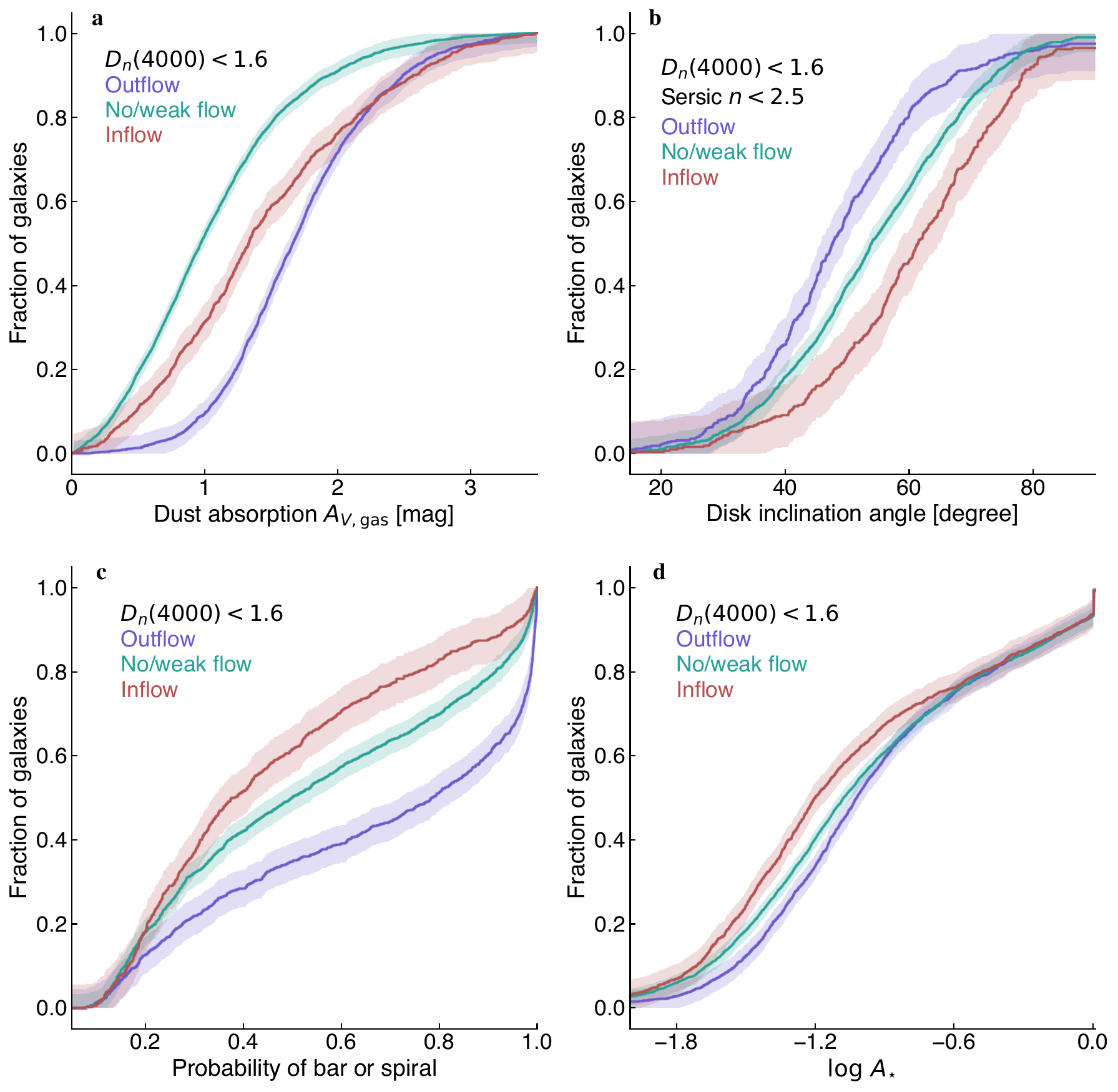}
\caption{Cumulative distributions of galaxy properties associated with gas flows in young systems.} Inflow, outflow, and no/weak-flow galaxies are shown in red, blue, and teal, respectively. Shaded regions indicate Dvoretzky--Kiefer--Wolfowitz confidence bands at $\alpha = 0.05$.
\textbf{a}, Nebular dust attenuation inferred from the H$\alpha$/H$\beta$ Balmer decrement. \textbf{b}, Disk inclination angle derived from $r$-band axis ratios. \textbf{c}, Probability of hosting spiral arms or a bar from machine-learning morphological classifications. \textbf{d}, $z$-band asymmetry of the stellar light distribution.
\label{fig:visibility}

\end{figure}

Observational biases further limit inflow detectability. Na\,\textsc{i}\,D absorption is easily suppressed by photoionization unless dust-shielded, and weakened by resonant scattering in low-dust environments \citep{Prochaska+11}. Consistently, in young galaxies the strongest nebular dust attenuation occurs in outflow hosts, while no/weak-flow systems show the lowest $A_V$, even when strongly star forming (Fig.~5a).

Detection also depends strongly on geometry. In disk-dominated galaxies (S\'{e}rsic $n<2.5$), inclination cleanly separates flow states ($p<10^{-6}$): inflows are preferentially detected in edge-on systems, whereas outflows dominate face-on disks (Fig.~5b). This trend holds across stellar ages, indicating that cool gas flows are highly anisotropic and that inflows are most visible when sightlines probe near the disk plane. The observed inflow fraction in star-forming galaxies therefore likely underestimates the true incidence of accretion. 

A further key empirical result is the morphological inversion (Fig.~5c,d): inflow hosts are systematically smoother and less asymmetric than outflow hosts (median $A \approx 0.1$, $p<10^{-6}$), while strongly disturbed systems ($A>0.3$) remain rare (8--20\%; Supplementary Information). This contradicts the expectation that inflows should induce asymmetric or lopsided structures via disruptive accretion \citep{Bournaud+05, Yesuf+21, Bottrell+24}. Instead, the low asymmetry indicates that Na\,\textsc{i}\,D traces settled, recycled gas---such as galactic fountains or cooled circumgalactic material---that has aligned with the disk’s angular momentum \citep{Grand+19, ZouYX+26}. This explains how inflows produce coherent kinematics without leaving strong morphological disturbance.

By contrast, enhanced asymmetry in outflow systems reflects their association with intense star formation (powered by hidden gas accretion) and external dynamical perturbations. The low asymmetry of observed inflow hosts instead supports a temporal sequence in which earlier disturbances have already relaxed, so recycled inflows are observed after morphological settling on $\lesssim$Gyr timescales \citep{Lotz+10}.

Consistent with this, among young galaxies, bars and spiral arms---often linked to gas inflow---are more frequent in outflow hosts (bar fraction $44\%$, spiral fraction $54\%$) than in inflow (30\%, 31\%) or no/weak-flow systems (31\%, 40\%). These inverted structural trends suggest that the initial, dynamically disruptive accretion phase is either short-lived, deposited at large radii, migrating inward slowly, or not traced in Na\,\textsc{i}\,D, while detected inflows correspond to later, recycled or dynamically settled gas. This reconciles the low observed inflow fraction in young star-forming systems with the continuous gas supply required to sustain star formation (Extended Data Fig.~5).

\section*{\textbf{Discussion}}

We present a population-level census of cool gas inflows and outflows traced by Na\,\textsc{i}\,D absorption across a large and diverse sample of nearby galaxies. Our key result is the widespread detection of inflows in aged star-forming systems and quiescent galaxies. This indicates that cold gas accretion is not confined to actively star-forming disks, but extends into more evolved populations. More broadly, inflow and outflow signatures extend across a wide range of stellar ages and structural properties (Fig.~1), supporting the view that cool gas flows are a common component of galaxy evolution.
Recent analyses of DESI DR2 similarly report widespread gas inflows and connections between gas-flow states and offsets in the stellar mass--SFR and stellar mass--gas-phase metallicity relations \citep{Weng+16a,Weng+16b}. Beyond these related efforts, our study presents a unified population-level comparison of gas-flow states across diverse galaxy populations, and investigates galaxy substructure, star formation histories, stellar ages and metallicities, AGN activity, multiscale environments, dust, and gas content. 

We interpret these results within a self-consistent framework (Fig.~2), in which Na\,\textsc{i}\,D kinematic states provide observational signatures of different phases of the galactic gas cycle. Although inflow, outflow, and no/weak-flow galaxies are broadly matched in stellar mass, gravitational potential, and large-scale environment, they differ systematically in stellar age, gas-phase metallicity, dust content, certain structural characteristics, including axis ratio, bar and spiral features, and viewing geometry (Figs.~1–5). This demonstrates that global properties (mass and environment) alone do not determine the observed flow state (Fig.~3). Instead, internal processes---such as bar-driven torques, spiral structure \citep{YuHo22}, and minor mergers \citep{Yesuf+21, Bottrell+24}---regulate gas transport and star formation, while dust content and viewing angle govern detectability (Fig.~5).

In star-forming galaxies, internal processes likely sustain a baryon cycle regulated by feedback-driven recycling. The similarity of host properties across flow classes (Fig.~3), together with the coexistence of multiple flow signatures in post-starburst systems (Extended Data Figs.~2–3), supports a scenario in which inflows and outflows represent different stages of a connected gas-cycling process. This is naturally explained by a galactic fountain in which feedback-driven outflows remain gravitationally bound and recycle on short timescales \citep{Bregman80, FraternaliBinney+08}. In this picture, inflows are consistent with recycled material returning on $\sim$ 0.1--1\,Gyr timescales \citep{Fraternali17, Marasco+22}. The observed older stellar ages and near-solar metallicities are consistent with recycling and/or mixing of accreted gas with the enriched interstellar medium, rather than a dominant contribution from pristine gas inflow.

Another key empirical result is the morphological inversion shown in Fig.~5: young inflow hosts are smoother and less asymmetric than outflow hosts. Previous studies have linked gas inflows to mergers, tidal interactions, lopsidedness, and metallicity dilution \citep[][]{Bournaud+05, Kewley+06merger, Reichard+09, Yesuf+21, Bottrell+24}. However, young inflow hosts in our sample do not exhibit enhanced asymmetry or strong signatures of disturbed morphology relative to outflow and no/weak-flow samples. This suggests that the inflows traced by Na\,\textsc{i}\,D are not predominantly associated with strongly disruptive accretion events. The combination of weak morphological disturbances, the absence of a metal-poor ISM signature, and strong inclination dependence indicates that the detected inflows are preferentially associated with gas near the disk plane rather than with highly disturbed, isotropically distributed accretion. These trends are consistent with models in which recycled gas contributes substantially to the detected inflow population in younger galaxies, while slowly cooling halo gas may become increasingly important in older and more passive systems. Strongly disruptive accretion may therefore be less common, shorter lived, or less efficiently traced by the neutral gas phase probed here. Together, these results are consistent with a picture in which feedback-associated outflows and later inflow signatures are linked through a common gas-cycling process in at least some star-forming galaxies, while viewing geometry modulates their observability.

In quiescent galaxies, the decline in cold gas reservoirs is expected to reduce the efficiency of stellar-feedback-driven recycling and quasar-mode AGN feedback, consistent with a shift toward a complementary gas supply channel through AGN-regulated cooling and precipitation from hot halos \citep{Gaspari+18, DonahueVoit22}.
This process yields ``drizzling'' inflows that fuel low-luminosity AGN, which help maintain the cycle. Outflows in this regime are predominantly hot and diffuse \citep{guo12,guo18,zhang20}, making them largely undetectable in Na\,\textsc{i}\,D absorption. The observed predominance of inflows over outflows in quiescent galaxies, together with their $\sim 100\,\mathrm{km\, s^{-1}}$ inflow velocities, lack of strong internal gas-transport features such as bars or spiral structures, absence of recent major merger signatures, and radio excess relative to the star formation–radio correlation, is broadly consistent with expectations from halo cooling and precipitation models.

The flow statistics presented here do not directly measure duty cycles or provide a complete census of the baryon cycle; instead, they represent observationally selected snapshots whose detectability and classification are shaped by data quality, gas phase, dust content, and viewing geometry. Na\,\textsc{i}\,D absorption, measured here from central galaxy spectra, primarily probes neutral, dust-shielded gas already present in the interstellar medium, including gas associated with inflows, outflows, and recycling, while being less sensitive to ionized, diffuse, extended, or geometrically unfavorable phases. The kinematic classifications therefore reflect both physical state and observational selection, as reinforced by the inclination dependence: inflows are preferentially detected in edge-on systems, whereas outflows dominate face-on orientations.

The emerging picture of gas recycling from the observations presented here is supported by high-resolution simulations \citep{ZouYX+26} (Supplementary Information). These simulations reproduce the geometric segregation of inflows and outflows and find that recycled material dominates over pristine accretion. In the simulations, inflows and outflows coexist within individual galaxies and are primarily governed by internal properties---angular momentum transport, gravitational potential, and feedback---rather than halo mass alone.

Importantly, our results provide empirical benchmarks for theory rather than validation of a single predictive model. While current simulations reproduce individual components of the baryon cycle, they do not yet provide a first-principles mapping between multi-dimensional galaxy properties and observable gas-flow properties. In particular, inflow and outflow incidence, host-galaxy trends, inclination dependence, morphological inversion, and phase-dependent detectability have not yet been jointly reproduced in existing frameworks under like-to-like mock observational conditions. These limitations reflect a broader gap between theory and observation. A fully predictive framework will require simulations that self-consistently couple multiphase gas physics with realistic radiative transfer, explicitly resolving how feedback regulates the ionization state, phase structure, and dust and gas content of the interstellar and circumgalactic media.

\section*{Summary}

We present a comprehensive census of cool gas flows in $\sim$\,30,000 galaxies from the DESI survey, providing the first statistically robust comparison of inflows and outflows at the population level. Using Na\,\textsc{i}\,D absorption profiles, we classify galaxies into three flow states---inflows, outflows, and no/weak flows---across diverse galaxy populations. Our analysis addresses a central question: do observed inflows and outflows trace distinct galaxy populations, or are they different manifestations of a common gas cycle?

We find systematic trends across multiple observables. Outflows are associated with high star formation surface density and enhanced chemical enrichment. In contrast, inflow and no/weak-flow systems exhibit broad metallicity distributions without clear dilution signatures, indicating a mixture of recycled gas and unresolved inflow rather than dominant pristine accretion. Across carefully matched subsamples, inflow, outflow, and no/weak flow systems exhibit similar stellar masses, large-scale environments, and structural properties---including axis ratios---while differing markedly in recent star formation activity and stellar age. These results are consistent with the interpretation that observed flow states represent transient stages of gas cycling rather than distinct galaxy populations characterized by different long-lived properties.

To interpret these trends in a broader evolutionary context, we combine two existing gas-cycling pathways within a common framework. In star-forming and recently quenched galaxies, recycling driven by stellar feedback likely operates through a short-timescale fountain cycle. In massive, passive systems, the cycle shifts to low-level accretion through cooling and precipitation from the hot halo, yielding a sustained ``drizzling'' inflow mode that can maintain weak nuclear activity. Together, these regimes offer a plausible explanation for the observed inflow and outflow states across galaxy populations and their dependence on host properties. At the same time, the observed diversity of flow signatures reflects not only underlying baryon cycling processes but also observational effects—including dust attenuation, ionization conditions, and viewing geometry—that affect their detectability.

Our gas-flow framework motivates several follow-up tests. Spatially resolved spectroscopy and future multi-phase observations can probe the kinematic and spatial relationships between outflowing and inflowing gas in galaxies, and assess how the relative importance of recycling and halo cooling varies with galaxy properties and evolutionary state. These new measurements will help refine our understanding of how galaxies acquire gas, regulate star formation, and fuel supermassive black hole growth, thereby providing a more complete picture of the baryon cycle shaped by gravitational accretion, cooling, and stellar and AGN feedback processes.

\renewcommand{\thefigure}{\arabic{figure}}  
\renewcommand{\figurename}{Extended Data Fig.}  
\setcounter{figure}{0}

\begin{figure}
 \includegraphics[width=0.48\textwidth]{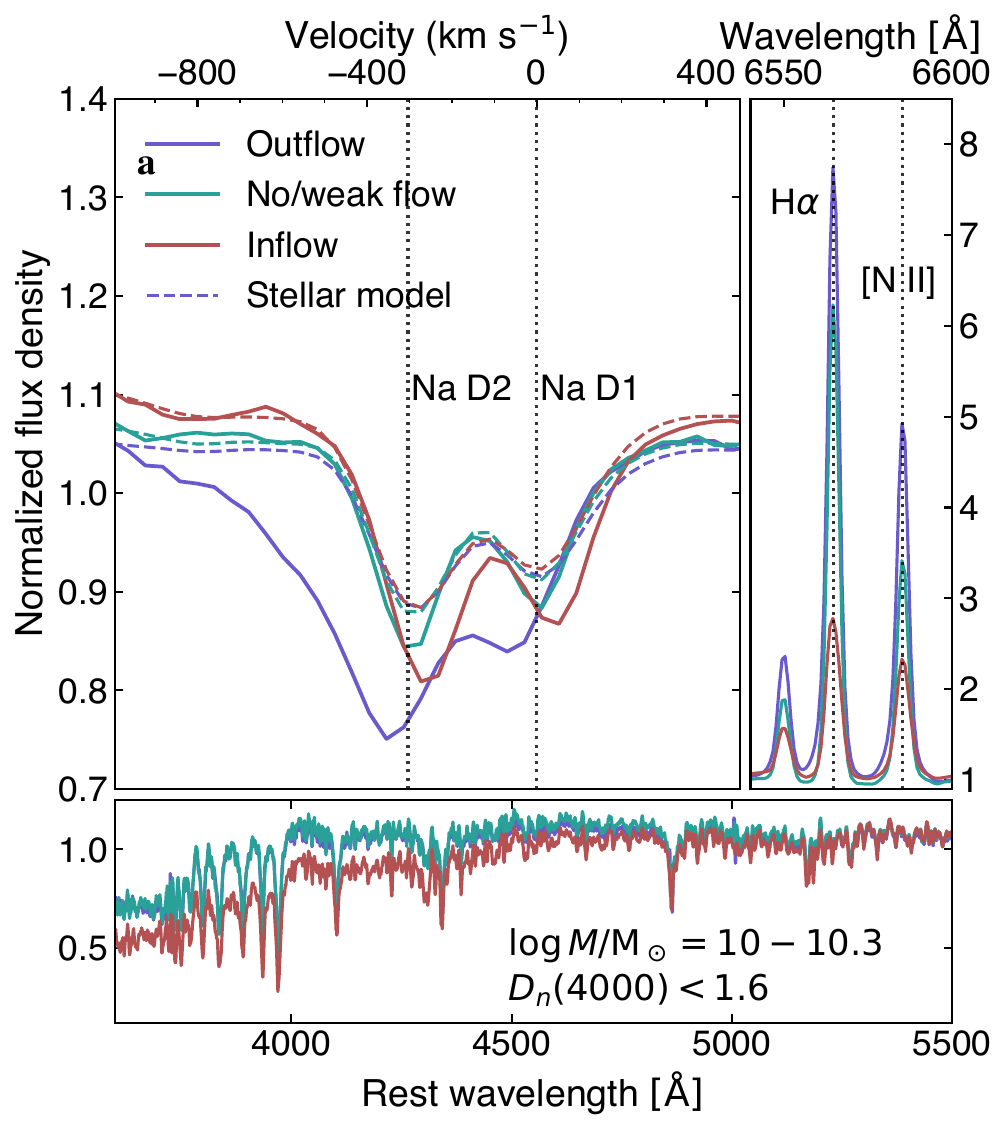}
 \includegraphics[width=0.49\textwidth]{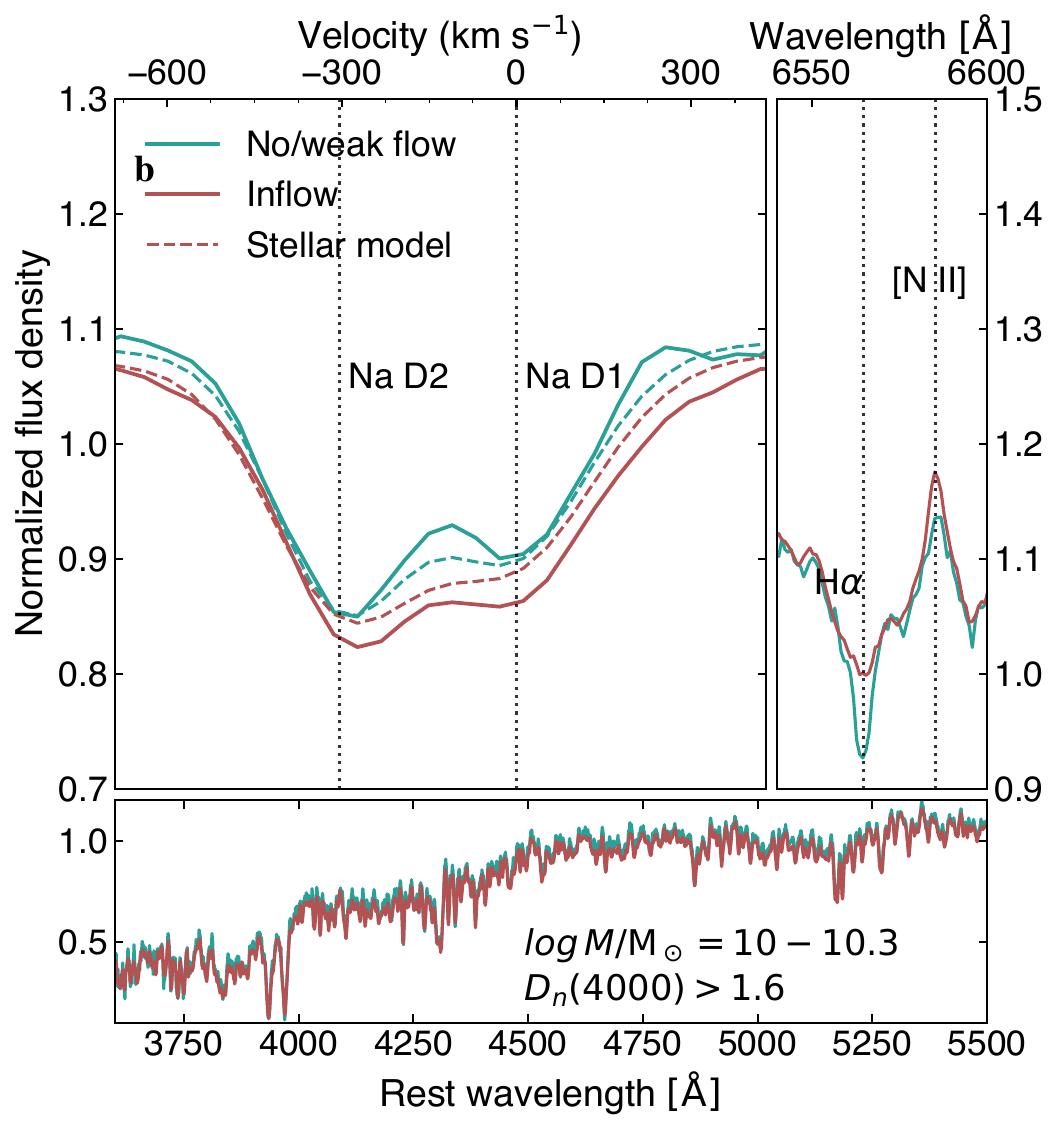}
 \includegraphics[width=0.48\textwidth]{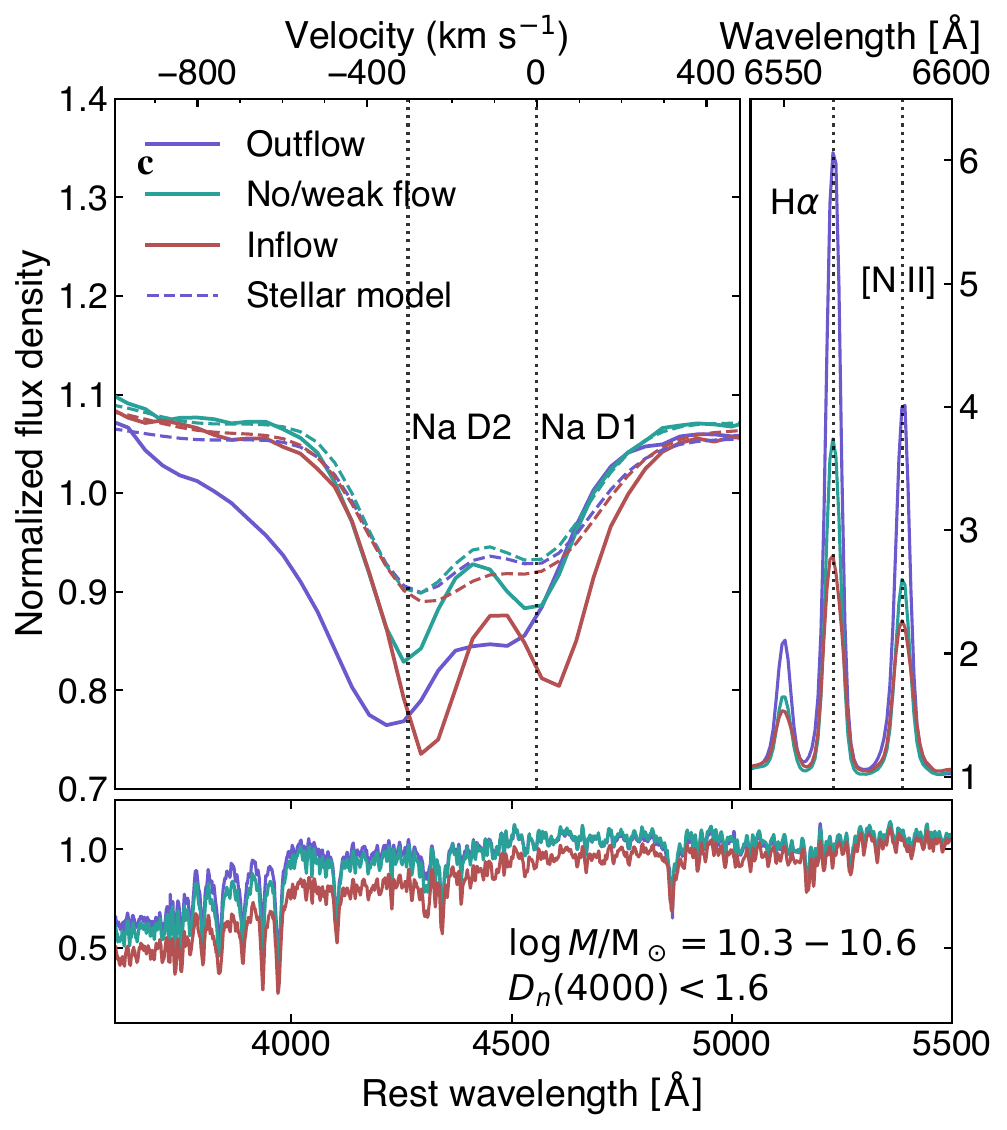}
 \includegraphics[width=0.49\textwidth]{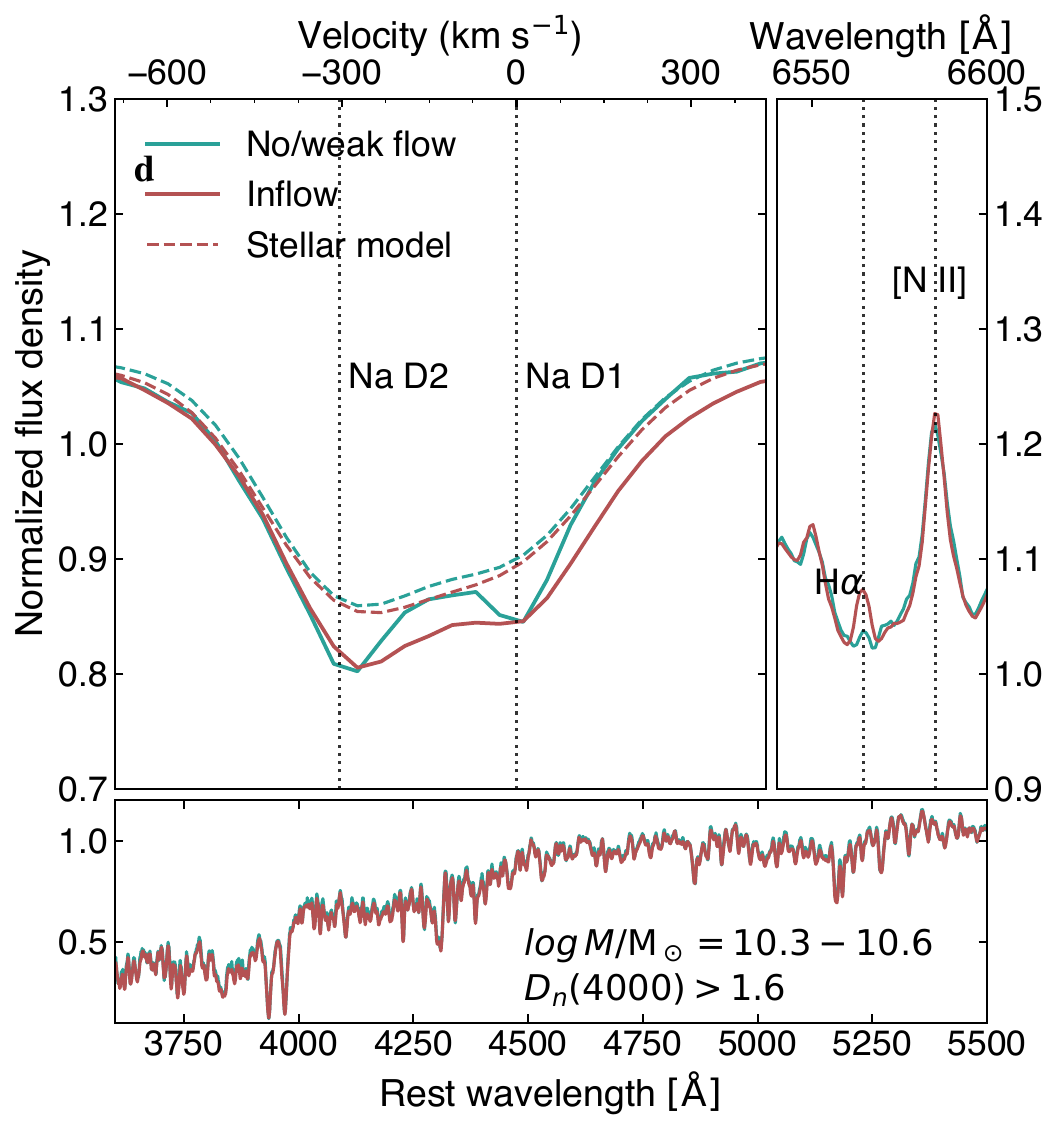}
 \caption{\textbf{Widespread spectroscopic signatures of galactic gas flows across stellar mass and age.}
Rows correspond to stellar-mass bins $\log\,(M_\star/\mathrm{M_\odot}) = 10.0$--$10.3$ (top) and $10.3$--$10.6$ (bottom), while columns separate younger ($D_n(4000) < 1.6$; left) and older ($D_n(4000) > 1.6$; right) stellar populations. Main panels show stacked spectra centred on the Na\,\textsc{i}\,D absorption doublet. Marginal panels display the corresponding H$\alpha$ emission and the emission-subtracted blue stellar continuum. Prior to stacking, galaxies are classified by their gas-flow state: outflow (blue), no flow (teal), and inflow (red). In the main panels, dashed curves represent the stellar photospheric Na\,\textsc{i}\,D component; the remaining absorption is attributed to interstellar gas. The stacked inflow samples contain 169 galaxies in \textbf{a} and 344 galaxies in \textbf{b}, while the corresponding outflow and no/weak-flow samples are approximately $2$--$5\times$ larger. Likewise, the inflow samples in older galaxies contain 147 galaxies in \textbf{c} and 611 galaxies in \textbf{d}, whereas the corresponding no/weak-flow samples are approximately $2$--$3\times$ smaller. Outflow signatures are rare among older galaxies in both mass bins ($N < 10$); their low signal-to-noise stacked spectra are therefore omitted from the left panels for clarity.}
 \label{ExFig:stackedspec}
\end{figure}

\begin{figure}
\includegraphics[width=\textwidth]{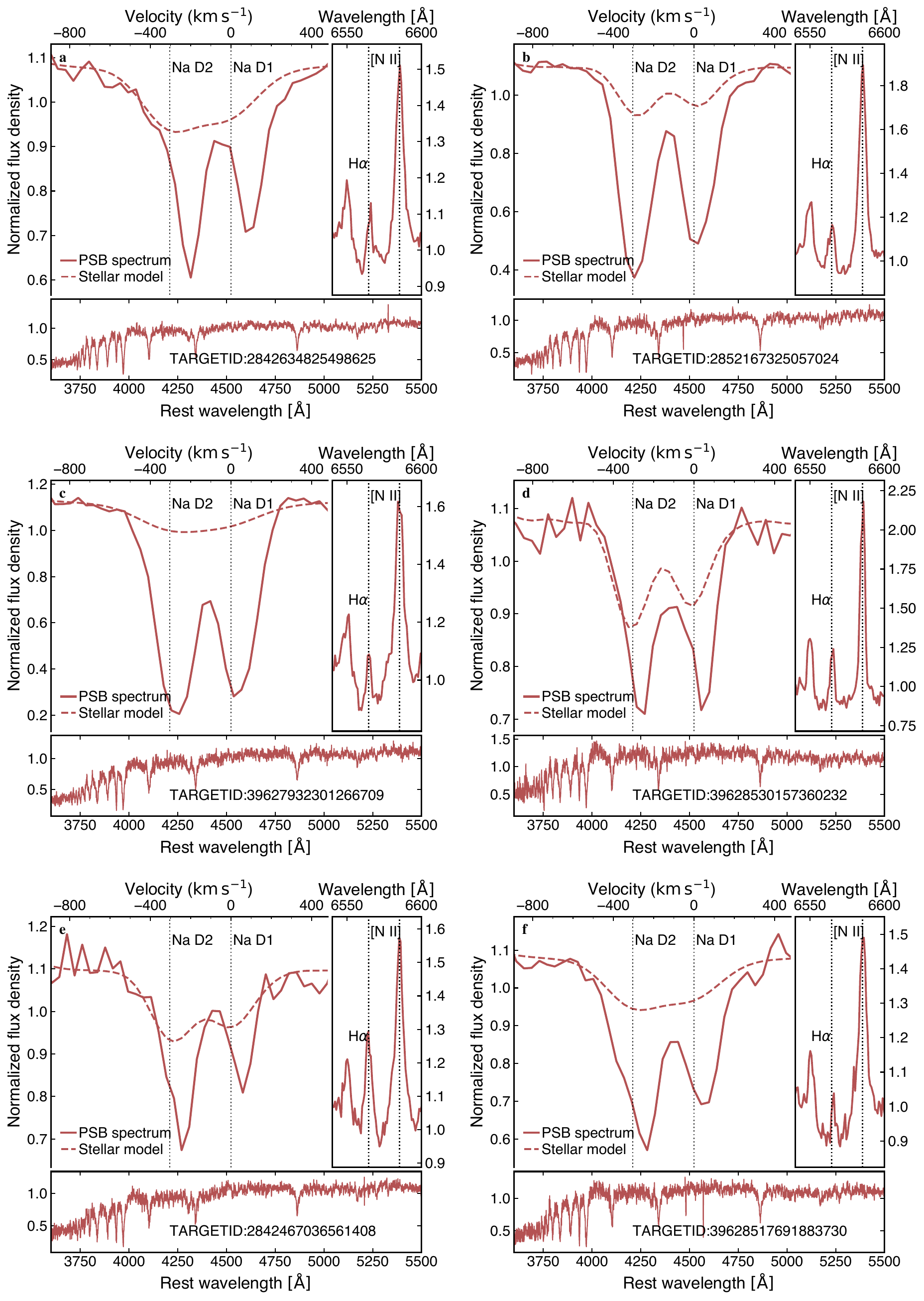}
\caption{\textbf{Example post-starburst galaxies exhibiting inflow signatures.}
Main panels show spectra centered on the Na\,\textsc{i}\,D absorption doublet.
Dashed curves indicate the stellar photospheric Na\,\textsc{i}\,D component, while the excess absorption is attributed to interstellar gas. Marginal panels display the corresponding spectral region around H$\alpha$ emission and the emission-subtracted blue stellar continuum.}
\label{ExFig:PSB_inflow}
\end{figure}

\begin{figure}
\includegraphics[width=\textwidth]{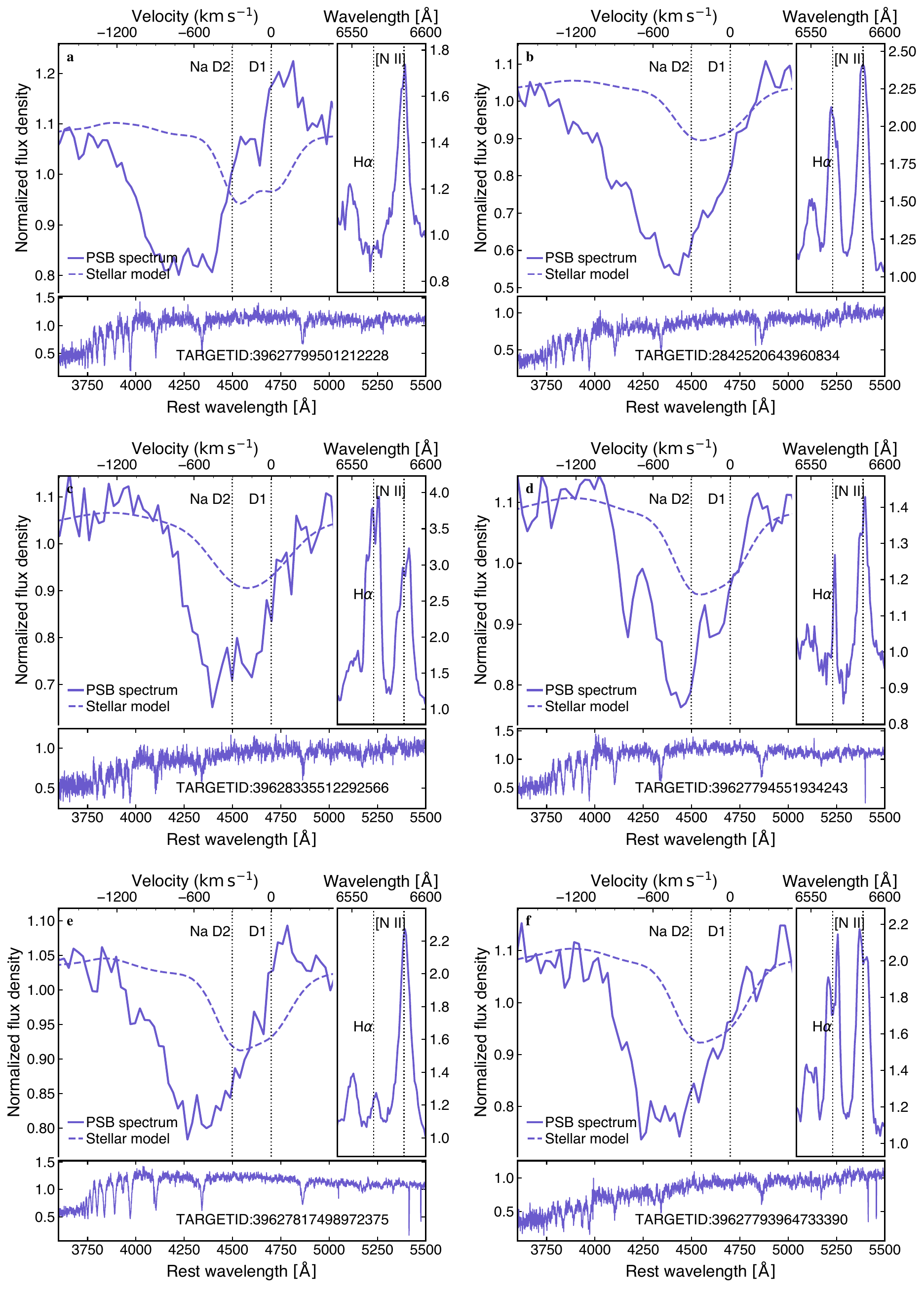}
 \caption{
\textbf{Example post-starburst galaxies exhibiting outflow signatures.} The examples shown were selected to illustrate systems with particularly large outflow velocities. Main panels show spectra centered on the Na\,\textsc{i}\,D absorption doublet. Dashed curves indicate the stellar photospheric Na\,\textsc{i}\,D component, while blueshifted excess absorption is attributed to outflowing interstellar gas.
Marginal panels display the corresponding spectral region around H$\alpha$ emission and the emission-subtracted blue stellar continuum. 
}\label{ExFig:PSB_outflow}
\end{figure}

\begin{figure}
\includegraphics[width=\textwidth]{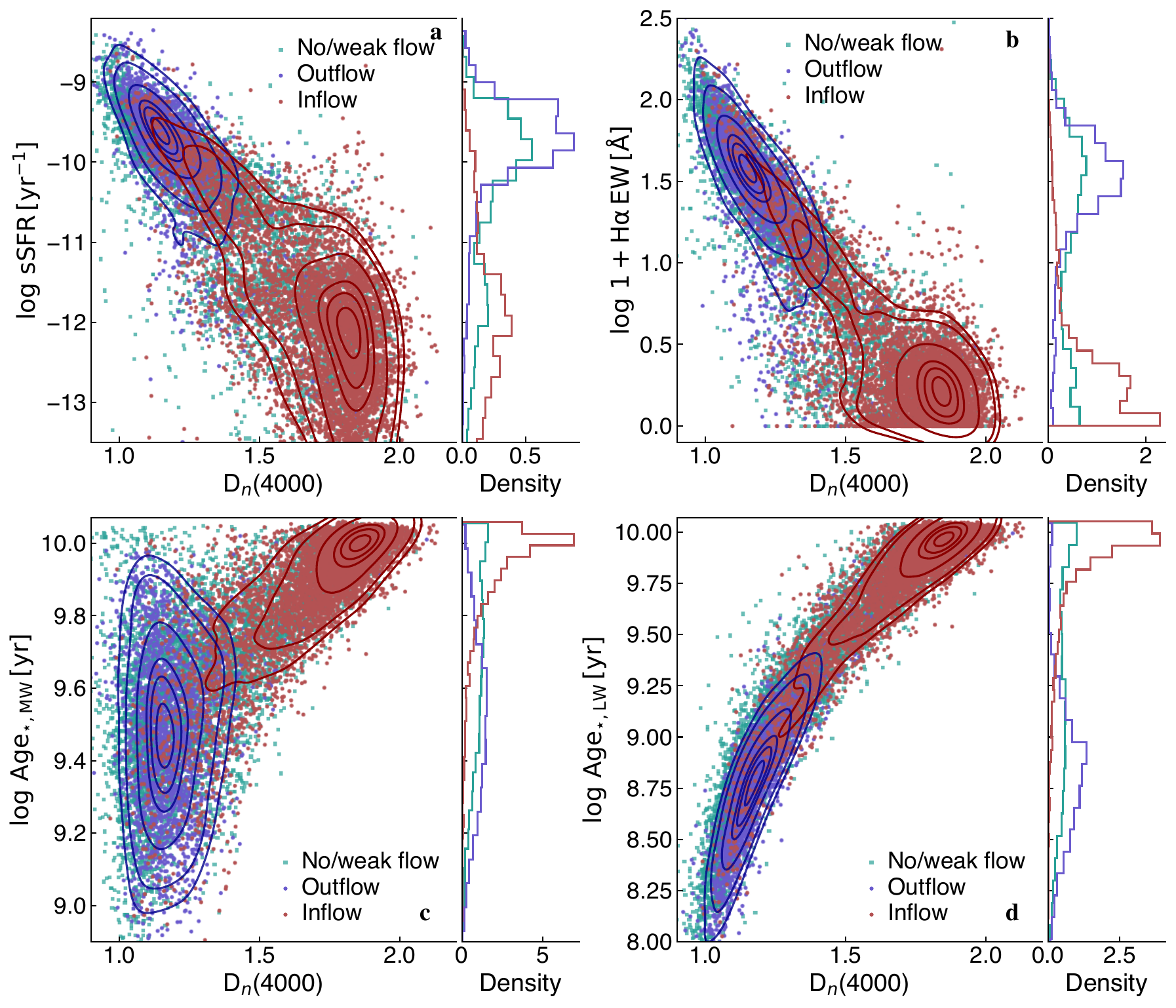}
\caption{\textbf{Star formation and stellar age diagnostics across different gas-flow states.} Inflow, outflow, and no/weak-flow galaxies are shown in red, blue, and teal, respectively. Contours indicate the 5th, 15th, 25th, 50th, 75th, and 85th percentiles of the data distribution.
\textbf{a}, Central $D_n(4000)$ index versus global specific star formation rate derived from spectral energy distribution fitting.
\textbf{b}, Central $D_n(4000)$ index versus central H$\alpha$ equivalent width.
\textbf{c}, $D_n(4000)$ index versus mass-weighted stellar age derived from spectral stellar population analysis.
\textbf{d}, $D_n(4000)$ index versus light-weighted stellar age.}
\label{ExFig:SFRAge}
\end{figure}

\begin{figure}
\includegraphics[width=\textwidth]{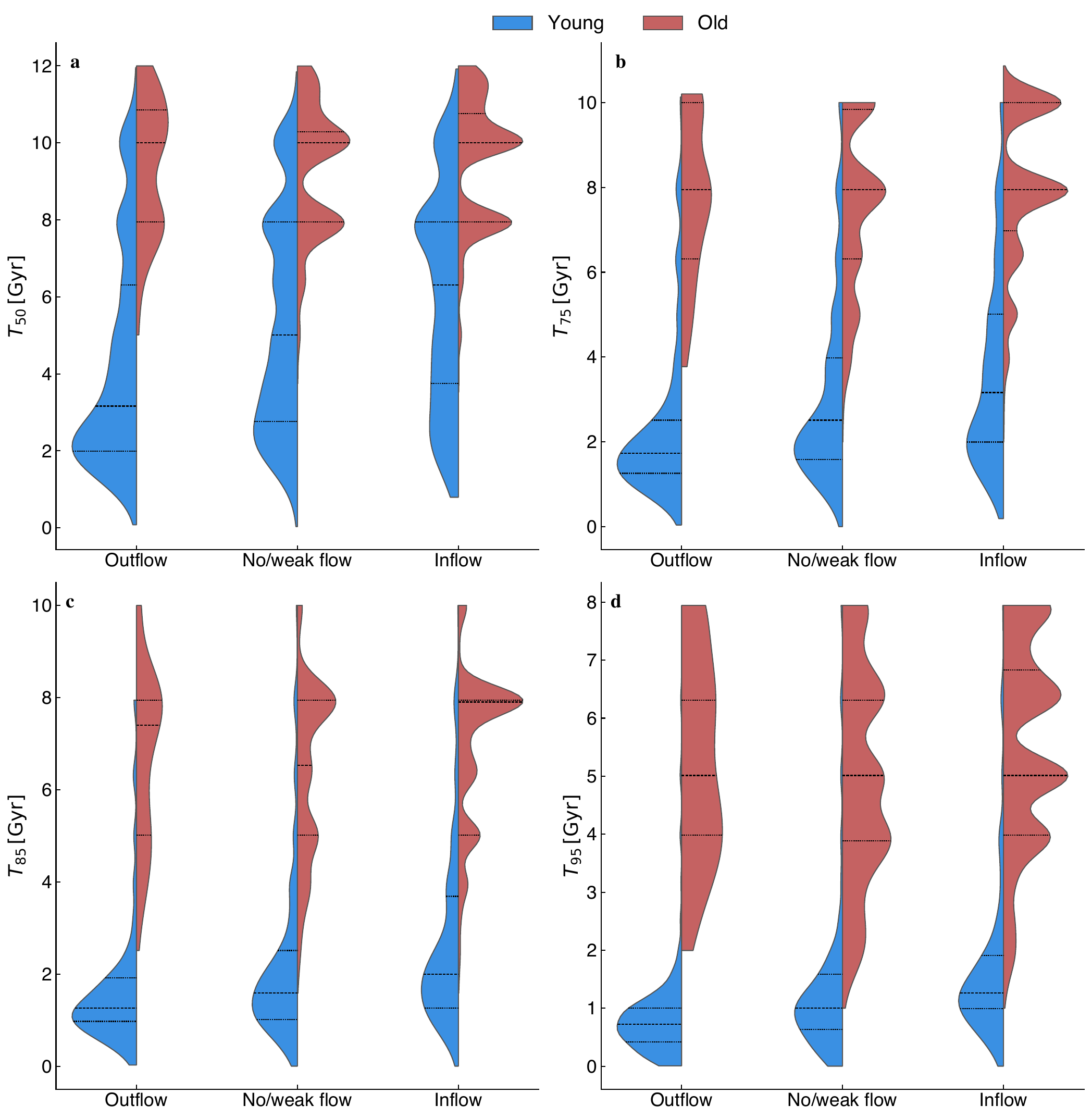}
\caption{\textbf{Stellar mass assembly times across gas-flow states in young and old galaxies.}
\textbf{a--d}, Lookback times at which galaxies assembled 50\%, 75\%, 85\%, and 95\% of their present-day stellar mass, denoted as $T_{50}$, $T_{75}$, $T_{85}$, and $T_{95}$, respectively.
In each panel, dashed lines indicate the 25th, 50th (median), and 75th percentiles of the distributions. The sample is restricted to galaxies with $\log\,(M_\star/\mathrm{M_\odot}) > 10$. Assembly-time estimates are available for all galaxies in the sample. However, owing to the small number of outflow hosts among the oldest galaxies, the corresponding distributions should be interpreted with caution.}
\end{figure}

\begin{figure}
 \includegraphics[width=0.49\textwidth]{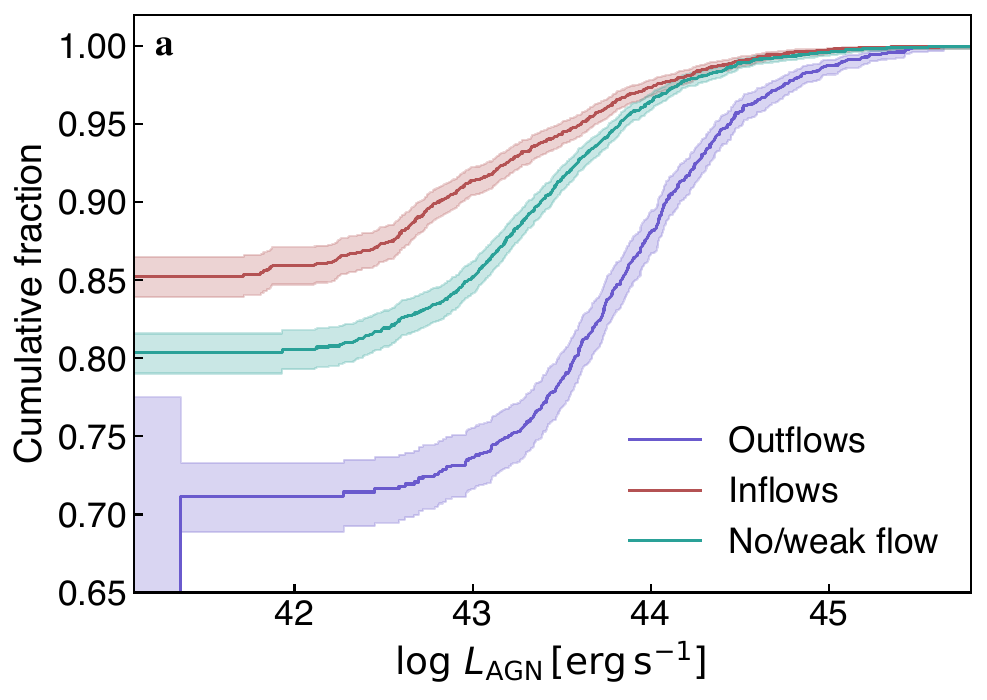}
  \includegraphics[width=0.49\textwidth]{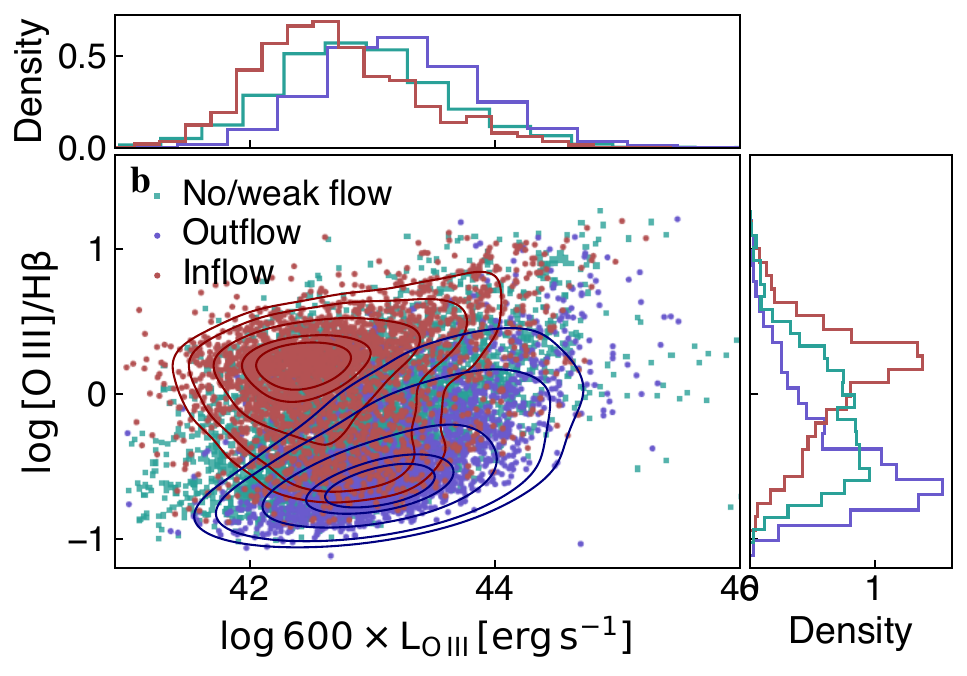}
 \includegraphics[width=0.49\textwidth]{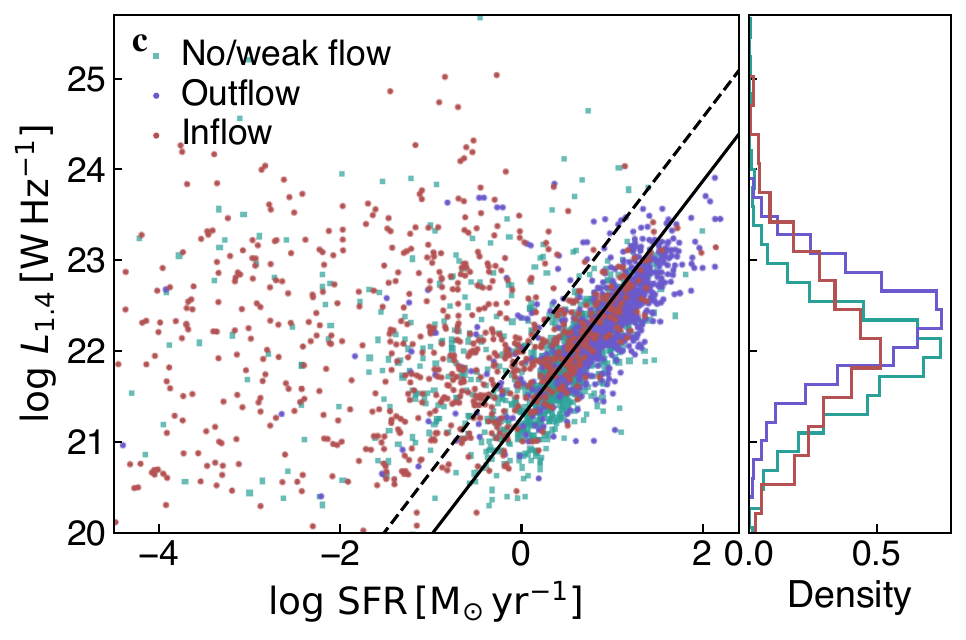}
 \includegraphics[width=0.49\textwidth]{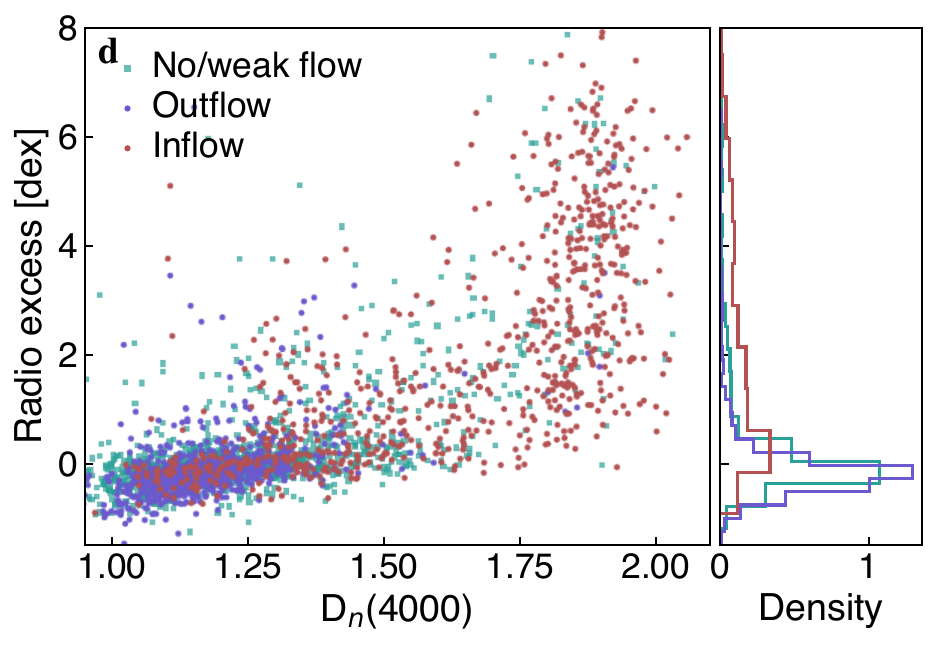}
\caption{\textbf{Association between gas flow state and weak AGN activity.}
In all panels, galaxies classified as inflow, no/weak flow, and outflow systems are shown in red, teal, and blue, respectively. \textbf{a}, Kaplan--Meier cumulative distributions of AGN bolometric luminosity derived from spectral energy distribution (SED) fitting, accounting for upper limits. \textbf{b}, AGN luminosity estimated from the dust-corrected [O\,III] luminosity as a function of the [O\,III]/H$\beta$ emission-line ratio. \textbf{c}, Star formation rate (SFR) versus 1.4\,GHz radio luminosity based on LOFAR and VLASS matched samples. Black lines indicate the best-fitting mean relation and a 0.7\,dex offset; the latter is adopted to identify radio-excess emission.\textbf{d}, 4000\,{\AA} break strength (light-weighted stellar age) as a function of radio excess.Together, these diagnostics show that although most flow galaxies host weak AGN or are inactive, observed inflow systems preferentially exhibit radio-excess emission, whereas outflow galaxies largely avoid the radio-excess regime despite including some of the most luminous AGN. This suggests distinct fueling and feedback pathways, with the observed inflows linked to low-level, radio-mode (hot) accretion and outflows associated with radiatively efficient (cold-mode) black hole growth.
}
\label{ExFig:AGN}
\end{figure}

\begin{figure}
\includegraphics[width=0.49\textwidth]{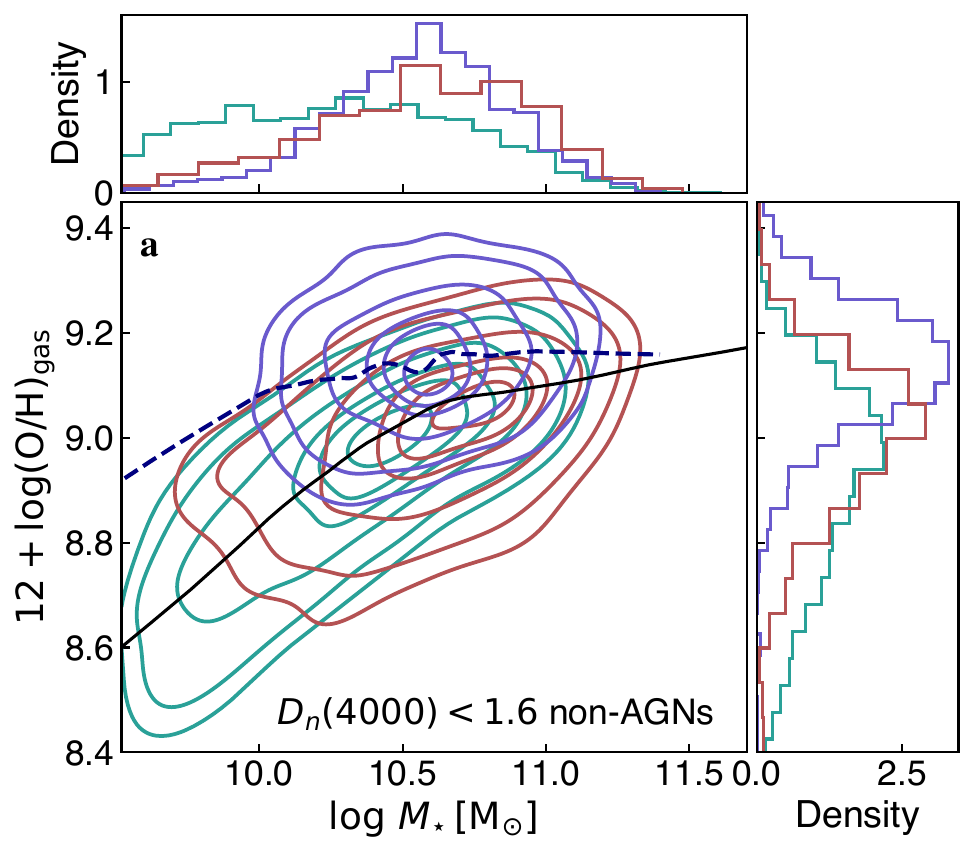}
\includegraphics[width=0.49\textwidth]{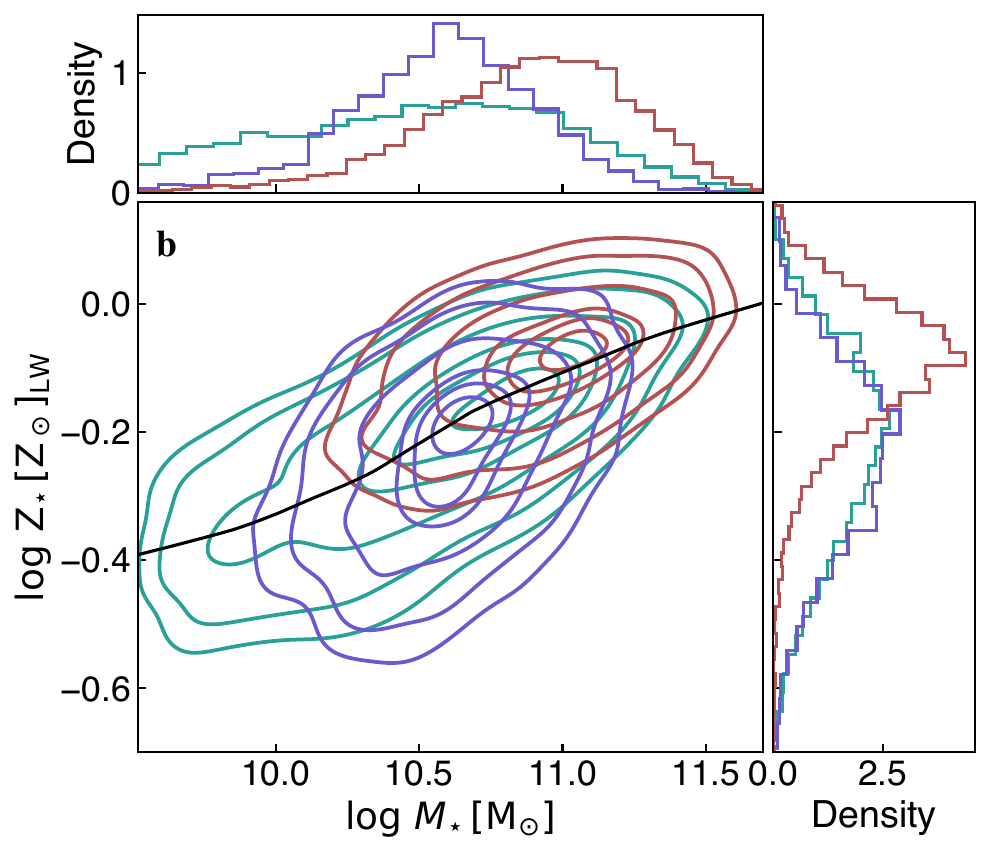}
\includegraphics[width=0.45\textwidth]{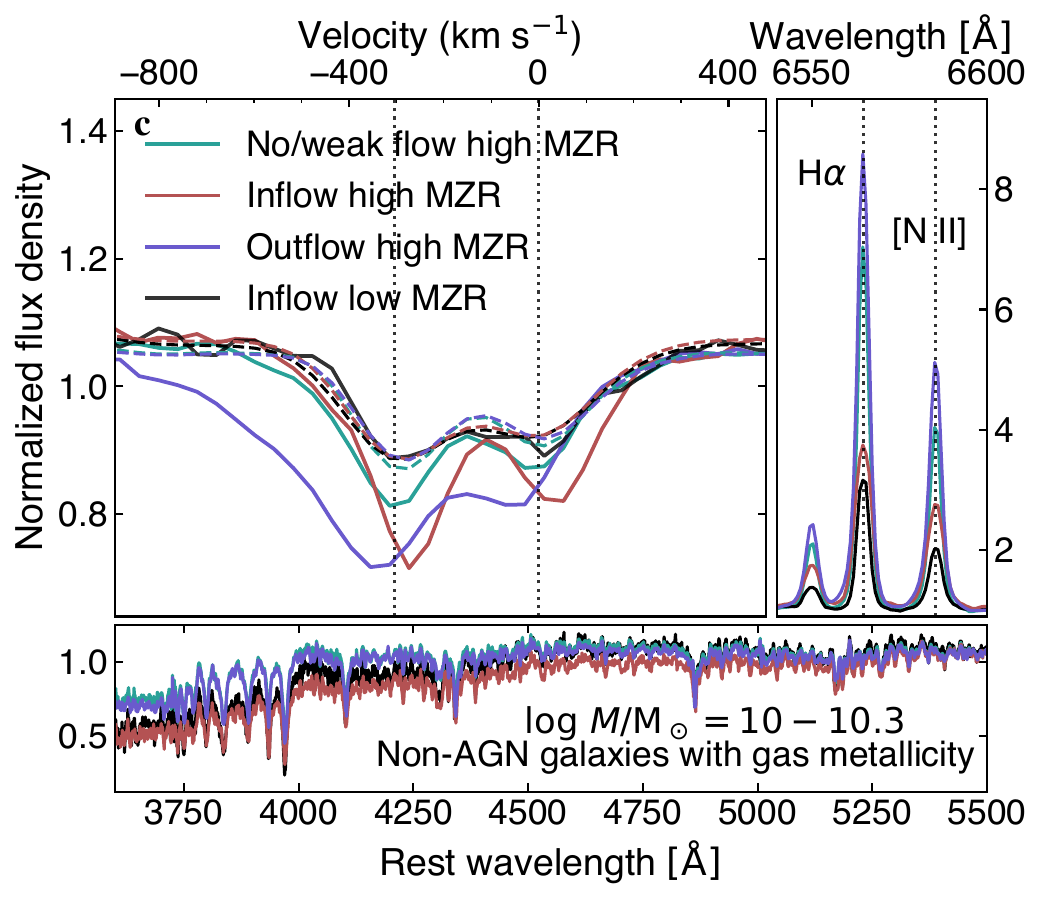}
\includegraphics[width=0.45\textwidth]{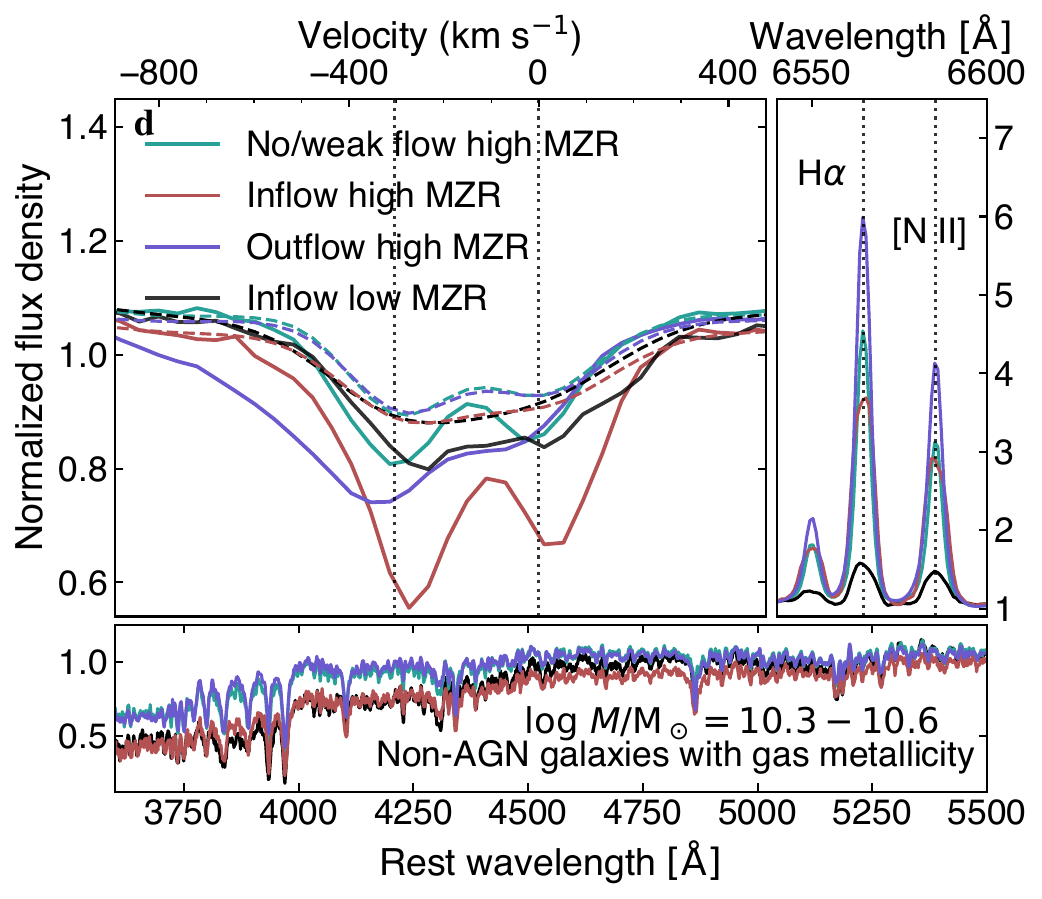}
\includegraphics[width=0.45\textwidth]{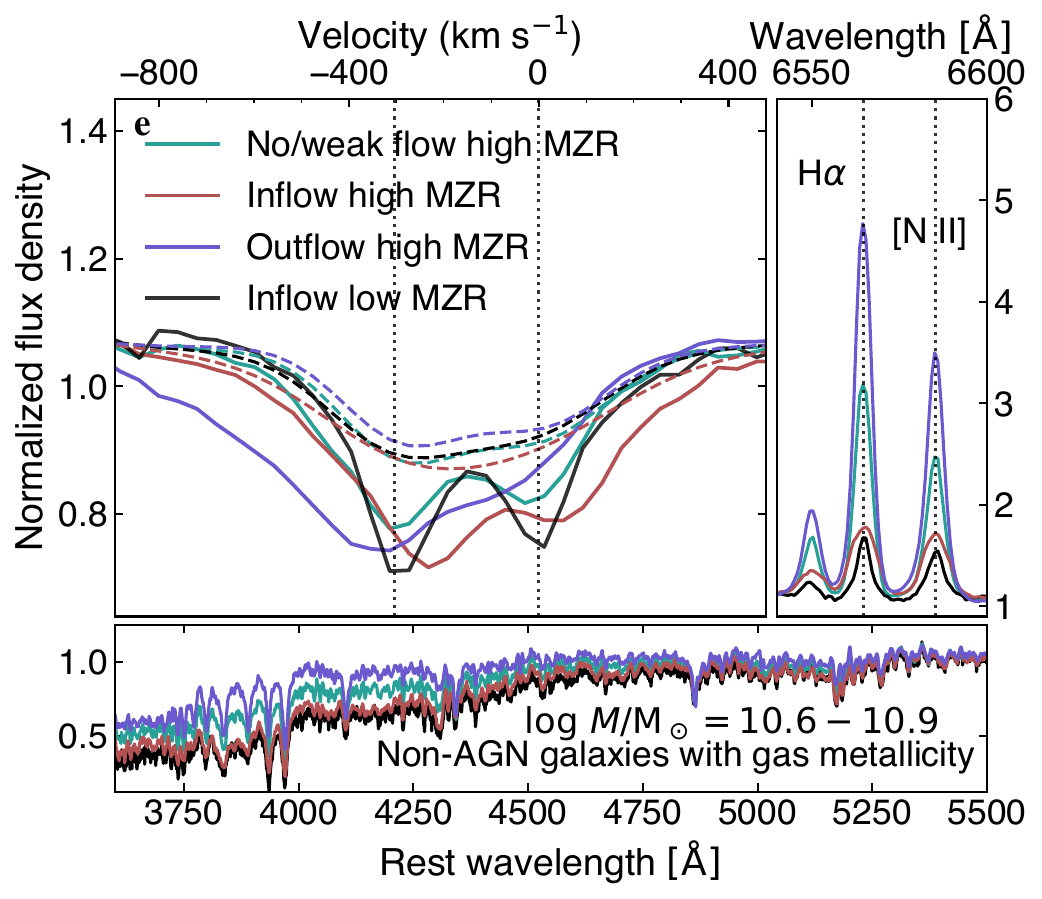}
\hfill
\includegraphics[width=0.45\textwidth]{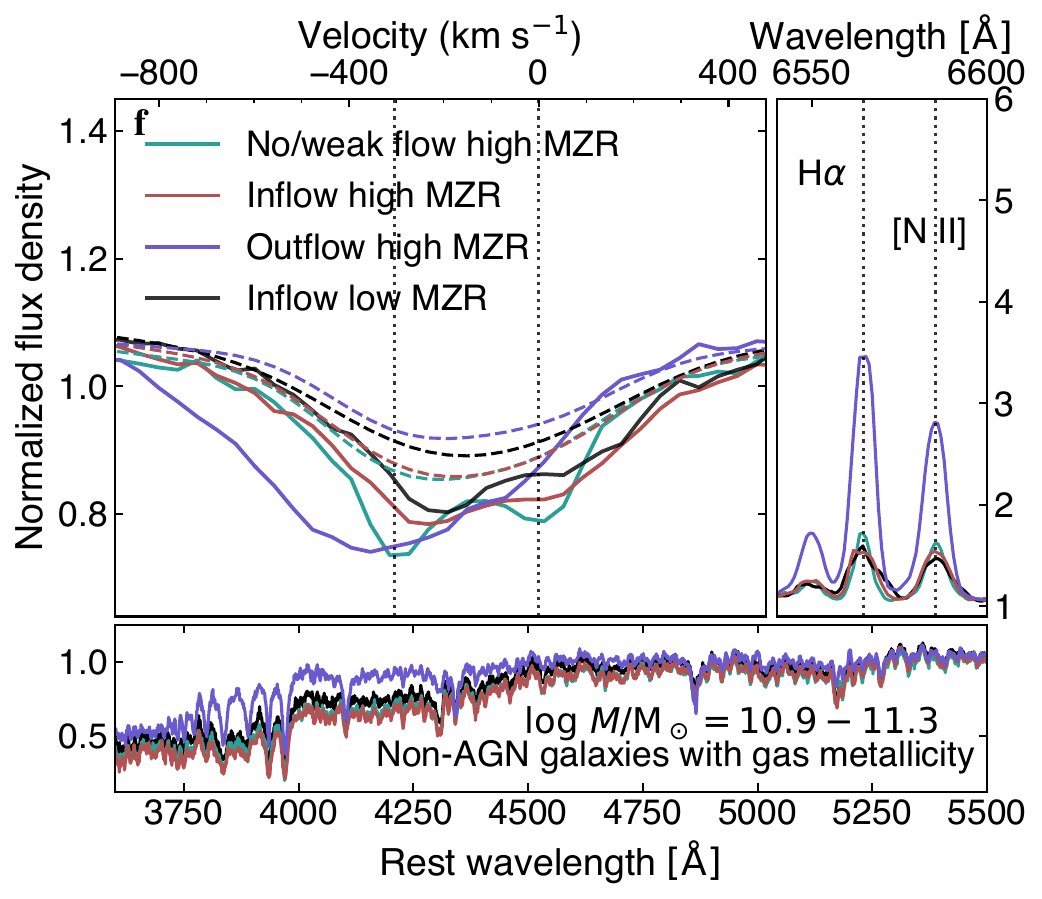}
\caption{\textbf{The mass–metallicity relation across gas flow states.}
Galaxies classified as inflow, no/weak-flow, and outflow systems are shown in red, teal, and blue, respectively.
\textbf{(a)} Stellar mass versus gas-phase metallicity derived from strong-line calibrations \citep{Dopita+16}.
The solid black curve shows the LOESS regression for the full sample, while the dashed navy curve shows the LOESS fit to outflow systems only.
\textbf{(b)} Stellar mass versus light-weighted stellar metallicity, with the LOESS regression for the full sample shown in black.
\textbf{(c–f)} Stacked spectra of galaxies selected to lie $+0.1$\,dex above and $-0.2$\,dex below the gas-phase mass–metallicity relation, illustrating systematic differences in absorption features at fixed mass. For gas-phase metallicities, only non-AGN emission-line galaxies are included, whereas the stellar metallicity analysis uses the full sample without this restriction.}
\end{figure}

\begin{figure}
\includegraphics[width=0.49\textwidth]{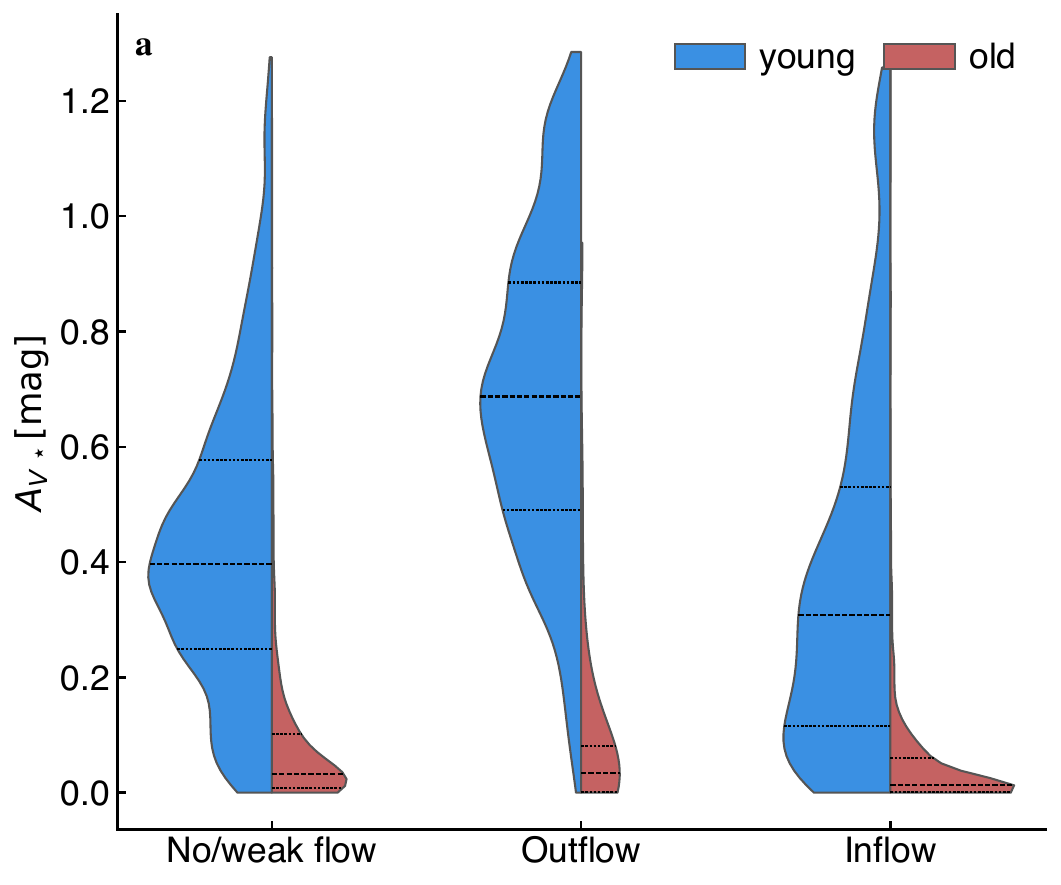}
\includegraphics[width=0.49\textwidth]{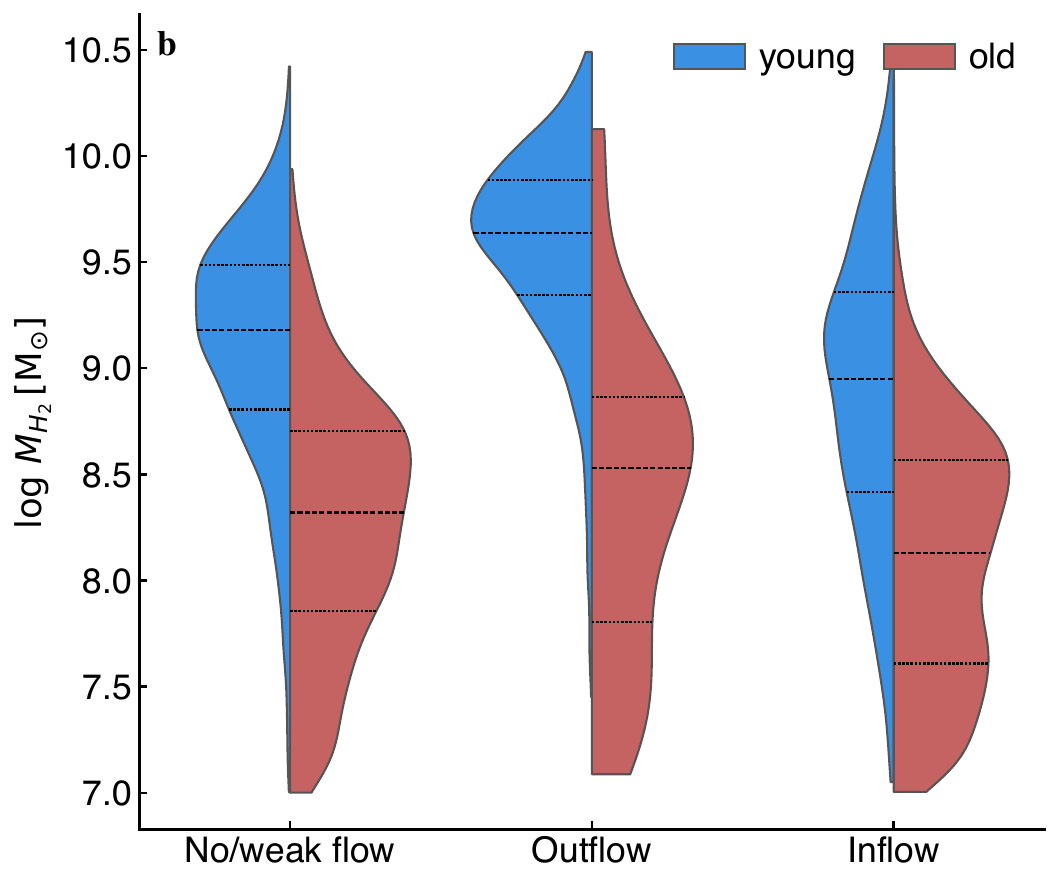}
\includegraphics[width=0.49\textwidth]{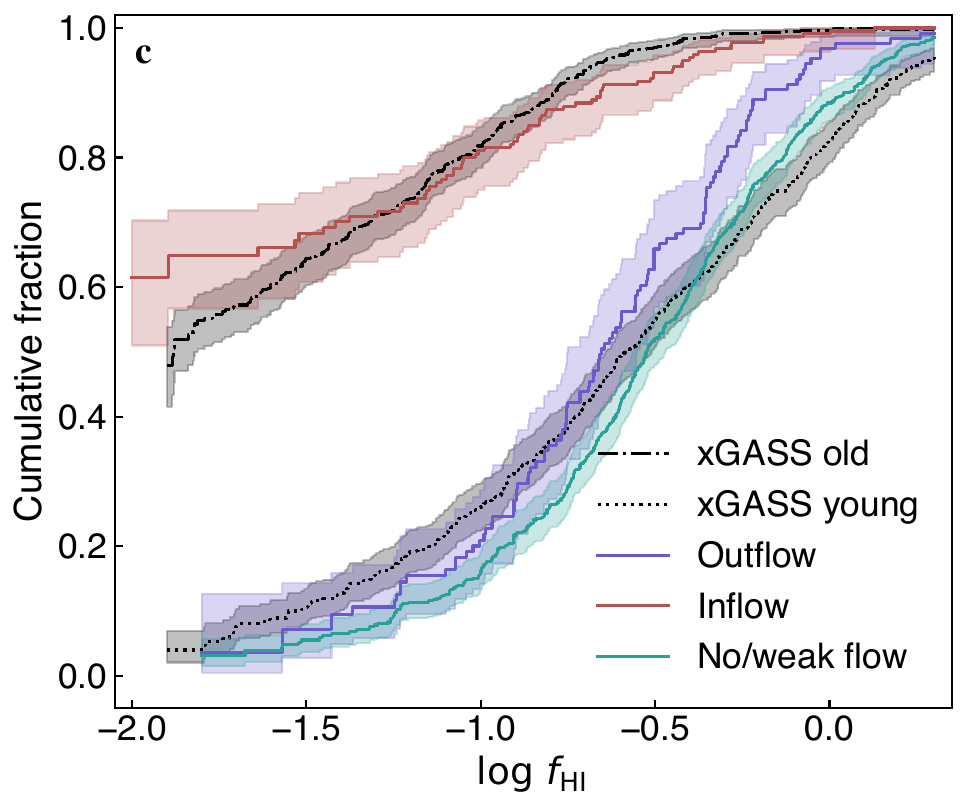}
\includegraphics[width=0.49\textwidth]{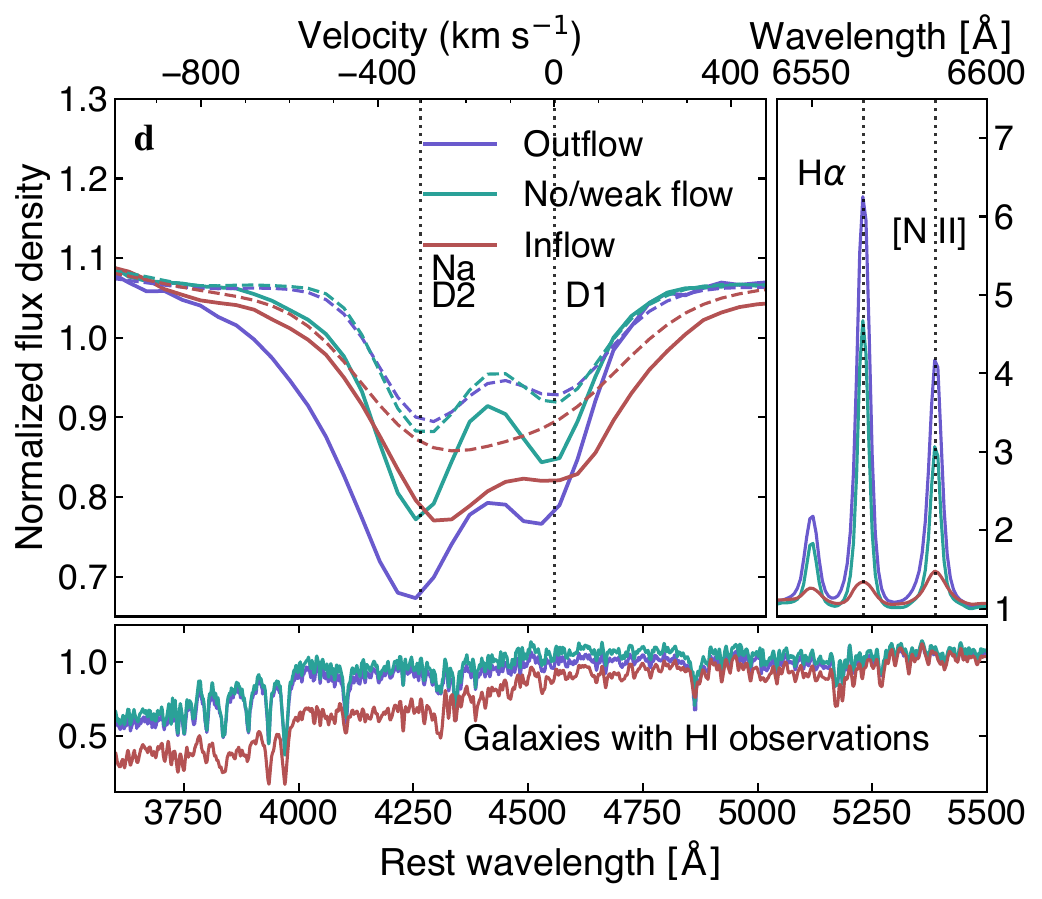}
\caption{ \textbf{Dust and gas content across gas-flow states in young and old galaxies.}
\textbf{(a)} Stellar dust attenuation in the $V$ band, $A_V$, derived from spectral energy distribution fitting. \textbf{(b)} Molecular hydrogen mass predicted from a scaling relation combining $A_V$, star formation rate, stellar mass, and galaxy radius \citep{YesufHo19}.
In \textbf{a} and \textbf{b}, dashed lines indicate the 25th, 50th (median), and 75th percentiles of the distributions. Owing to the limited number of outflow systems in old galaxies, their distributions should be interpreted with caution. \textbf{(c)} Kaplan--Meier cumulative distributions of the H\,\textsc{i} gas-to-stellar mass ratio, i.e, H\,\textsc{i} gas fraction. Galaxies classified as inflow, no/weak-flow, and outflow systems are shown in red, teal, and blue, respectively. For comparison, the black curves show young ($D_n(4000) < 1.6$; dotted) and old (dot--dashed) galaxies from the representative xGASS sample. \textbf{(d)} Stacked spectra of galaxies with available H\,\textsc{i} gas measurements.}
\end{figure}

\section*{Methods}

\subsection*{Parent sample selection}

We construct our parent sample from the DESI data release (DR) 1 \citep{DESI_DR1} value-added catalog of stellar masses and emission-line measurements for galaxies with reliable spectroscopic redshifts \citep{Zou+24}. From this catalog we select 30,416 galaxies with stellar masses $M_\star > 10^{9}\,M_\odot$ in the redshift range $0.005 < z < 0.16$. We further require high-quality spectroscopy, imposing a continuum signal-to-noise ratio ${\rm S/N} > 10$ per pixel to ensure robust stellar population modeling and absorption-line measurements.

DESI is an ongoing survey; the upper redshift limit is chosen to maximize spectroscopic completeness for environmental characterization by incorporating existing SDSS redshifts in the same volume \citep{Yesuf2022}.

The value-added catalog provides stellar masses derived from $g$, $r$, $z$, W1, and W2 photometry from the DESI Legacy Imaging Surveys. Strong optical emission lines in the catalog are measured using single-Gaussian fits, with stellar absorption corrected through continuum modeling with \textsc{starlight}. We adopt only a small subset of the provided measurements (e.g., spectral S/N and the 4000\,\AA\ break index).

For most key physical parameters used in this work, we perform independent measurements to ensure methodological consistency. In particular, we derive stellar masses and star formation rates through CIGALE spectral energy distribution fitting using an expanded photometric dataset spanning the ultraviolet to mid-infrared, while emission-line properties are independently remeasured using a separate spectral-fitting pipeline. These measurements are therefore independent of the catalog estimates, although we find good agreement for overlapping quantities. When available, the SED fitting incorporates UV observations from \textit{GALEX} (FUV and NUV) \citep{Martin+05}, optical photometry from SDSS ($u,g,r,i,z$) \citep{Abazajian+09}, DESI Legacy ($g,r,i,z$) \citep{Dey+19}, and Pan-STARRS ($g,r,i,z,y$) \citep{Chambers+16}, near-infrared measurements from 2MASS ($J,H,K_s$) \citep{Skrutskie+06} and UKIRT/WFCAM ($Y,J,H,K$) from UKIDSS \citep{Hewett+06,Lawrence+07}, and mid-infrared photometry from \textit{WISE} (W1--W4) \citep{Wright+10,Eisenhardt+20}, providing broad wavelength coverage to constrain star formation rate and dust attenuation.

\subsection*{Stellar population fitting of DESI spectra}

We derive stellar population parameters using full spectral fitting with \texttt{pPXF} \citep{Cappellari2023}. Observed spectra were corrected for Galactic extinction and shifted to rest-frame wavelengths using DESI pipeline redshifts. Fits were restricted to 3600--9700\,\AA, masking strong sky residuals and emission lines, including the Na\,\textsc{i}\,D doublet (5889.95, 5895.92\,\AA) and prominent sky lines (5577, 6300, 6363\,\AA) with velocity-based windows ($\pm$300 km\,s$^{-1}$ for sky lines, $\pm$800 km\,s$^{-1}$ for Na\,\textsc{i}\,D).

We used the default templates from Flexible Stellar Population Synthesis (FSPS) models \citep{Cappellari2023}, which were generated via the Python bindings \citep{Johnson+21} and FSPS v3.2 \citep{Conroy+09,Conroy+10}, covering 43 ages logarithmically spaced by 0.1\,dex from 1\,Myr to 15.85 Gyr and 9 metallicities, [Z/H] $= [-1.75, -1.5, -1.25, -1, -0.75, -0.5, -0.25, 0, 0.25]$, for a total of 387 SSP templates. Models adopt a Salpeter IMF (0.08--100 M$_\odot$) with Mesa Isochrones and Stellar Tracks (MIST) \citep{Choi+16} and default FSPS parameters, and spectra are computed using the MILES stellar library \citep{Falcon-Barroso+11}. For fitting, we restrict metallicities to [Z/H] $\ge -1$ and ages to $<14$\,Gyr.

In the \texttt{pPXF} modeling, multiplicative Legendre polynomials of order 8 were used to correct continuum mismatches, with no additive polynomials. Outliers were removed via robust 3$\sigma$ clipping. Emission lines were modeled simultaneously with single-Gaussian profiles. Balmer lines share a common velocity and dispersion, as do forbidden lines, though the two sets are allowed to differ. Both the templates and line profiles were convolved with the wavelength-dependent DESI instrumental resolution prior to fitting. A regularization value of $\texttt{regul}=50$ was used to generate smooth star formation histories.

Tests with X-shooter and E-MILES templates yielded consistent results. The resulting stellar kinematics ($V_\star$, $\sigma_\star$), light- and mass-weighted ages and metallicities, and emission-line fluxes form the basis for comparisons of galaxies with inflows, outflows, and weak or undetected gas flows.

\subsection*{Two-component modeling of sodium absorption}

We analyse continuum-normalized spectra, obtained by dividing the observed spectra by a \textsc{pPXF} stellar continuum model, in the Na\,\textsc{i}\,D region using a two-component absorption model that separates interstellar medium (ISM) gas from line-of-sight inflow or outflow. Each component is modeled with a Gaussian optical-depth profile \citep{Rupke+05}, and the resulting profile is convolved with a Gaussian kernel whose width is set by the DESI spectral resolution to account for the instrumental line-spread function (LSF). In galaxies with strong emission (H$\alpha$ EW $>20$\,\AA), the nearby He\,\textsc{i}\,$\lambda5876$ line is included, with its Doppler width tied to that of the ISM component; the average He\,\textsc{i}/H$\alpha$ ratio is only 0.03.

We infer model parameters using Bayesian analysis with dynamic nested sampling \citep{Speagle+20}. The free parameters are the local covering fractions ($C_f$, $C_g$), flow velocity centroid ($v_f$), Doppler widths ($b_f$, $b_g$, where $b=\sqrt{2}\sigma$), column densities ($N_f$, $N_g$), and He\,\textsc{i} amplitude ($A_{\rm HeI}$). The local covering fractions of the ISM and flow quantify the fraction of sightlines blocked by absorbing clouds, reflecting the clumpiness of the gas and incorporating effects of spectral resolution that can dilute absorption if multiple clouds are unresolved.

Uniform priors are adopted: $C_f, C_g = 0.0$–0.8; $v_f = -500$ to $+300~\mathrm{km\,s^{-1}}$; $b_f, b_g = 30$–450~$\mathrm{km\,s^{-1}}$; and $\log N_f, \log N_g = 11$–15~cm$^{-2}$. When the stellar velocity dispersion is measured, the ISM Doppler width is restricted to within $\pm100~\mathrm{km\,s^{-1}}$ of that value. The likelihood assumes independent Gaussian uncertainties per spectral pixel. This framework enables separation of static ISM absorption from kinematically shifted flow components.

We classify galaxies using both posterior flow probability and velocity offset. We define inflows as systems with posterior probability $P_{\rm in} > 0.85$ and redshifted Na\,\textsc{i}\,D absorption exceeding $+30~\mathrm{km\,s^{-1}}$ relative to systemic velocity. Outflows are defined as systems with $P_{\rm out} > 0.85$ and blueshifted absorption more negative than $-30~\mathrm{km\,s^{-1}}$. Systems with $|v_{\rm flow}| < 30~\mathrm{km\,s^{-1}}$ are classified as no/weak-flow galaxies.

\subsection*{Flow statistics and abundance}

Our parent sample contains 29,716 non-dwarf galaxies with $M_\star > 3 \times 10^{9}\,M_\odot$. 
Of these, 5,696 show no significant Na\,\textsc{i}\,D velocity shift ($-30 < v_{\rm flow} < 30~\mathrm{km\,s^{-1}}$), 6,575 exhibit high-probability inflows ($P>0.85$, $v_{\rm flow} > 30~\mathrm{km\,s^{-1}}$), and 2,215 show high-probability outflows ($P>0.85$, $v_{\rm flow} < -30~\mathrm{km\,s^{-1}}$).

Relaxing the probability threshold to $P>0.5$ increases the number of candidate inflows and outflows to 15,213 and 8,707, respectively. Stacking spectra in narrow stellar-mass bins shows that intermediate-probability systems ($0.5 < P \le 0.85$, $v_{\rm flow} > 30~\mathrm{km\,s^{-1}}$) display coherent velocity shifts on average, indicating genuine but lower signal-to-noise detections at least in some galaxies. 

Using a Poisson–binomial model to account for individual detection probabilities, we estimate that $\sim54\%$ of galaxies host inflows and $\sim35\%$ host outflows. To ensure robustness and minimize false positives, however, all primary analyses adopt a conservative threshold of $P>0.85$. Under this definition, $\sim20\%$ of nearby galaxies exhibit clear inflows and $\sim7\%$ show clear outflows. Including intermediate-probability systems suggests the intrinsic incidence likely spans $\sim20$–50\% for inflows and $\sim7$–30\% for outflows.

The mean Na\,\textsc{i}\,D equivalent width (EW), including both ISM and flow components, is $\sim2$–3\,\AA\ across all flow classes, and 90\% of galaxies have non-stellar ${\rm EW}_{\rm NaD} > 1$\,\AA. The residual Na\,\textsc{i}\,D absorption not reproduced by stellar population models correlates strongly with stellar age ($\rho = 0.7$), implying that quiescent galaxies retain substantial cool gas reservoirs, while in younger star-forming systems Na\,\textsc{i}\,D absorption is more affected by ionization and scattering. Galaxies with weak velocity shifts ($-30 < v_{\rm flow} < 30~\mathrm{km\,s^{-1}}$) typically have ISM Na\,\textsc{i}\,D EWs of $\sim1.7$\,\AA, with $\sim75\%$ showing $\lesssim0.5$\,\AA\ attributable to potential flow components. In short, residual Na\,\textsc{i}\,D absorption correlates strongly with stellar age, indicating substantial cool gas reservoirs in quiescent galaxies and more ionization-affected absorption in young star-forming systems.

The choice of $|v| = 30\,\mathrm{km\,s^{-1}}$ reflects instrumental, spectral-modelling, and physical considerations. This threshold lies well above the effective velocity (redshift) accuracy achievable for these data ($\sim10$--$20\,\mathrm{km\,s^{-1}}$) \citep{Lan+23}, while also being comparable to the typical turbulent velocity dispersion of cold interstellar gas in galaxy disks. Expected secular radial inflow velocities in disks ($\sim10\,\mathrm{km\,s^{-1}}$) are similarly at or below our effective velocity resolution \citep{Schmidt+16, DiTeodoroPeek21}. We therefore define a combined ``no/weak-flow'' class corresponding to systems without statistically significant detections of coherent cold-gas motion. To assess the robustness of this choice, we tested alternative classifications that separate weak and non-detected flows using different velocity and probability thresholds. We find that $\sim$95\% of galaxies with $|v_{\rm flow}| < 30~\mathrm{km\,s^{-1}}$ are low-probability detections ($P < 0.5$), with essentially no high-confidence cases, and that all key results remain unchanged under these alternative definitions.

Our conclusions are likewise insensitive to modest variations in the adopted velocity threshold: increasing the cut to $|v| = 50\,\mathrm{km\,s^{-1}}$ changes the inflow sample size by only $\sim$2\%, and 90\% of galaxies classified as inflows under our fiducial definition have velocities $\gtrsim70\,\mathrm{km\,s^{-1}}$. These inflow velocities are substantially larger than those expected from secular transport processes within galactic disks, such as viscous inflow or bar-driven streaming. This disfavors slow internal redistribution as the dominant origin of the detected absorption signatures. Instead, the measured velocities are more consistent with dynamically energetic processes, including recycled gas in galactic fountains, cooling halo gas, or accretion from the circumgalactic medium along non-circular and radially biased trajectories \citep{Fraternali17,Gaspari+18}.

\subsection*{Galaxy environments}

We quantify galaxy environments using multiscale stellar mass overdensities and nearest-neighbor statistics \citep{Yesuf2022}. Stellar mass overdensity is computed within fixed physical apertures of $x = 0.1$, 0.25, 0.5, 1, 2, 4, and $8\,h^{-1}\,{\rm Mpc}$ by summing the stellar mass of neighboring galaxies within a line-of-sight velocity window of $|\Delta v| < 1000\,\mathrm{km\,s^{-1}}$ and normalizing by the median density in redshift slices of $\Delta z = 0.02$. 

This velocity window is chosen to balance the inclusion of physically associated structures while minimising contamination from projection effects in redshift space, and is consistent with typical velocity dispersions of galaxy groups and clusters, as well as peculiar velocities along the line of sight. For neighbors with only photometric redshifts, a wider window of $|\Delta v| < 3000\,\mathrm{km\,s^{-1}}$ is adopted to account for larger redshift uncertainties. 

The normalization in narrow redshift slices removes redshift-dependent selection effects and ensures that overdensities are defined relative to the evolving survey completeness. Finally, we have verified that our results are insensitive to these choices, remaining unchanged when the velocity cut is increased by a factor of two to three.

To mitigate incompleteness in the neighbor sample, we incorporate spectroscopic redshifts from SDSS \citep{SDSS_DR16} and DESI DR1, as well as photometric redshifts from refs.~\citep{Zou+22,Zou+24}, in our environmental calculations.

Group halo masses are adopted from the DESI group catalogue \citep{YangX+20}, which extends the halo-based group finder to incorporate both spectroscopic and photometric redshifts. The catalogue also provides central and satellite classifications. The fractions of galaxies in different flow states do not show major differences between centrals and satellites; a detailed analysis is left for future work. This weak dependence is consistent with our broader finding that environment is not the primary driver of the gas flow state.

Filamentary structures are traced using the Tempel et al. catalog \citep{Tempel+14}, which employs a stochastic Bisous model to identify three-dimensional filament spines. For each galaxy, the catalog provides the distance to the nearest filament, filament length, and the galaxy number density along the filament spine.

After controlling for stellar age, we find that the three flow states occupy similar distances from cosmic filaments. Among young galaxies, inflow, outflow, and no/weak-flow systems all have mean filament distances of $\sim$2\,Mpc, with broad 15th--85th percentile ranges of $\sim$0.2--20\,Mpc. Older galaxies lie somewhat closer to filament spines, with median distances of $\sim$1\,Mpc. We therefore do not find compelling evidence that gas-flow state depends strongly on proximity to filaments. Although cosmic filaments have been proposed to influence galaxy gas content and star formation, and simulations suggest possible cold-mode accretion along filamentary structures \citep[][]{Ma+25}, the connection between directly detected gas flows and filament environment appears weak in our data and is not explored further here.

\subsection*{Star formation and stellar mass assembly history}

The star formation histories and stellar population properties provide a time-integrated complement to the instantaneous gas flow signatures traced by Na\,\textsc{i} (Extended Data Figs. 4 and 5). While inflow and outflow diagnostics probe the current state of baryon cycling, stellar age distributions encode the cumulative response of galaxies to sustained gas accretion and feedback over gigayear timescales. By linking flow state to age and stellar mass assembly histories, we can test whether inflow-, outflow-, and no-flow systems correspond to distinct evolutionary stages or instead reflect transient phases within a self-regulated baryon cycle. This approach directly probes the connection between gas exchange processes and long-term stellar mass growth in a mass- and time-dependent framework.

A Gaussian mixture model decomposition of the $D_n(4000)$ distribution yields two components with means $\mu_1 \approx 1.2$ and $\mu_2 \approx 1.76$, and similar dispersions ($\sigma \sim 0.13$), indicating overlapping young and old stellar populations as well as the presence of intermediate “green valley” systems. The formal midpoint between these components is $\sim 1.5$, reflecting the substantial overlap between the two populations.

We adopt $D_n(4000)=1.6$ as the division between young and old populations, placing the threshold toward the higher-$D_n(4000)$ side of the transition region. This choice provides a conservative separation of young stellar populations given the overlap between the distributions, and is consistent with commonly adopted divisions in the range $D_n(4000)\sim1.5$--1.6 in previous studies. We have verified that adopting $D_n(4000)=1.5$ does not affect our main results.

Using stellar mass assembly histories expressed as lookback formation times, we find that outflow-dominated young ($D_n(4000)<1.6$) galaxies assemble the bulk of their stellar mass at late epochs. The median outflow system reaches 50\%, 75\%, and 95\% of its present-day stellar mass only $\sim$3.5, $\sim$1.8, and $\sim$0.8\,Gyr ago, respectively. Even among the oldest 15\% of outflow galaxies, half-mass assembly typically occurs within $\lesssim$5\,Gyr. These timescales demonstrate that galaxies hosting strong neutral outflows have built a substantial fraction of their stellar mass recently, with star formation sustained over several gigayears rather than confined to a brief, one-time burst.

In contrast, inflow-dominated galaxies exhibit systematically earlier stellar mass assembly histories. Their median half-mass formation times are $\sim6$\,Gyr for young systems and $\sim10$\,Gyr for old systems, while the corresponding $T_{95}$ values are $\sim1.2$\,Gyr and $\sim5$\,Gyr, respectively. This indicates that inflow galaxies, on average, assembled the bulk of their stellar mass at significantly earlier cosmic epochs than outflow-dominated systems.

Focusing on the older population, $\sim85\%$ of inflow galaxies with $D_n(4000)>1.6$ show no evidence for substantial star formation or stellar mass growth over the past several gigayears and remain largely quenched. Nevertheless, their assembly histories are diverse, with a broad distribution of recent formation times and a non-negligible fraction experiencing stellar mass growth within the past $\sim1.5$--4\,Gyr.

This diversity is further reflected in a small but significant subset ($\sim12\%$) of galaxies with mass-weighted stellar ages $>6$\,Gyr that still exhibit relatively recent assembly, with median $T_{95}$ and $T_{98}$ values of $\sim2$\,Gyr and $\sim1$\,Gyr, respectively. Old galaxies with no or weak Na\,\textsc{i} absorption display assembly histories broadly similar to those of inflow-dominated systems.

Splitting the young galaxy sample into narrow stellar mass bins reveals that the relative timing of gas flows follows a simple and robust pattern. At fixed stellar mass, outflow-dominated galaxies are consistently the youngest, systems with no or weak Na\,\textsc{i} absorption occupy an intermediate evolutionary stage, and inflow-dominated galaxies are the most evolved. This ordering persists across all formation-time percentiles ($T_{95}$, $T_{85}$, $T_{75}$, and $T_{50}$), indicating that it reflects a coherent evolutionary sequence. The separation between flow classes remains nearly invariant with stellar mass, implying that the inflow–outflow cycle operates in a broadly self-similar manner across the mass range probed ($\log\,M/{\rm M_\odot}=10$--11.5).

Given typical cold-gas depletion times of $\sim$1--2\,Gyr, the extended star formation histories inferred for young galaxies require ongoing gas replenishment. The offset in formation times between inflow- and outflow-dominated systems, as quantified by $T_{95}$ and $T_{85}$, is modest ($\sim0.5$\,Gyr) and comparable to molecular gas depletion timescales \citep{SaintongeCatinella22}. This correspondence suggests a short gas recycling timescale, in which feedback-driven outflows are followed by the re-accretion of cool gas on similarly short intervals, while inflow and star formation persist over several gigayears.

Importantly, these results reveal a structured relationship between gas flow state and stellar population properties that is consistent with a two-regime baryon cycling framework. Among young star-forming galaxies, we find a coherent internal ordering in which outflow-dominated systems are systematically the youngest, no/weak-flow systems represent an intermediate stage, and inflow-dominated systems are more evolved. This sequence is consistent with a scenario in which feedback-driven outflows and subsequent gas accretion operate on comparable gigayear timescales within a self-regulated cycle. In contrast, inflows in older galaxies are systematically associated with early stellar mass assembly and quenched star formation, and are consistent with accretion of previously enriched halo gas. Taken together, these trends suggest that the observed flow classes reflect a combination of (i) feedback-driven gas circulation in star-forming galaxies and (ii) a distinct mode of gas accretion in quiescent systems, although other channels may also contribute.

\subsection*{S\'{e}rsic index and inclination measurements}

Structural parameters were obtained from the \texttt{Tractor} catalog of the DESI Legacy Imaging Survey, which provides parametric surface-brightness model fits for all sources \citep{Lang+16, Dey+19}. We adopt the S\'{e}rsic index $n$ from the best-fitting S\'{e}rsic profile as a measure of the central concentration of the galaxy light distribution. Larger values of $n$ indicate more centrally concentrated, bulge-dominated systems, while lower values correspond to disk-dominated morphologies.

To remove bulge-dominated galaxies, we classify systems with single-S\'{e}rsic index $n < 2.5$ as disk-dominated for the purpose of estimating disk inclination. Given the measurement uncertainties, this threshold is broadly consistent with the commonly adopted division at $n_{b} = 2$ \citep{FisherDrory16}, although those works use bulge S\'{e}rsic indices derived from bulge--disk decompositions. We have verified that adopting thresholds of $n = 2$ or $n = 3$ does not significantly affect the inferred inclination trends.

Galaxy inclinations were estimated from the measured ellipticity parameters (\texttt{shape\_e1}, \texttt{shape\_e2}) provided by \texttt{Tractor}. These components define the complex ellipticity $e = e_1 + i e_2$, from which the projected axis ratio is computed as $q = (1-|e|)/(1+|e|)$, where $|e| = \sqrt{e_1^2 + e_2^2}$. 
Assuming galaxies can be approximated as oblate spheroids, the inclination $i$ is derived from $\cos^2 i = (q^2 - q_0^2)/(1 - q_0^2)$, where $q_0$ is the intrinsic axial ratio (thickness) of an edge-on disk. These components define. We adopt a typical value of $q_0 = 0.2$ for disk galaxies, consistent with previous studies \citep{XuD+24}. This approach provides a statistical estimate of viewing angle suitable for population-level comparisons, while recognizing that individual inclinations may be uncertain for irregular or spheroid-dominated systems.

Edge-on disk galaxies generally exhibit higher line-of-sight extinction. That inflow-host galaxies show lower $A_V$ despite being preferentially edge-on indicates that their interstellar media contain, on average, less dust mass, consistent with inflows occurring in more quiescent, gas-poor hosts (Extended Data Fig.~8).

In addition to S\'{e}rsic index (Fig.~3d), we use our non-parametric concentration index measurement to confirm that no/weak flow galaxies are less concentrated than strong inflow and outflow galaxies (see Supplementary Information, Supplementary Fig.~8b). 

\subsection*{Spiral, bar, and merger classification}

We complement our analysis of Na\,\textsc{i}\,D-based gas-flow states with an investigation of internal morphological structures---bars, spiral arms, and merger signatures---which are known gravitational drivers of gas inflow. Our goal is to test whether such non-axisymmetric features show a preferential association with the inflow, outflow, or no/weak-flow states identified in this work (Fig.~5c), thereby linking large-scale structure to the nuclear gas kinematics traced by Na\,\textsc{i}\,D absorption. Interestingly, we find that these features are more strongly associated with outflow systems than with currently detected inflows. In the Supplementary Information, we discuss this result in terms of possible observational and physical effects, including viewing-angle effects, spatially extended inflow beyond the fiber aperture, low-velocity and spectroscopically unresolved flows ($|v| < 30\,\mathrm{km\,s^{-1}}$), and temporal offsets between the inflow event and the observed star formation or morphological disturbance.

To that end, we use galaxy morphological classifications (e.g., mergers, bars, and spirals) from the deep-learning catalog \citep{YeR+25}, which was trained on debiased labels from the Galaxy Zoo DECaLS 5 (GZD-5) project and covers the full DESI Legacy Imaging Surveys (DESI-LIS). The DESI-LIS comprise three imaging surveys: the Dark Energy Camera Legacy Survey (DECaLS), the Beijing–Arizona Sky Survey (BASS), and the Mayall $z$-band Legacy Survey (MzLS). The catalog provides morphological classifications for 593{,}737 galaxies in total, with probabilistic estimates for ten key features, including bar strength, spiral arm presence, and merger signatures. Its performance is comparable to previous Galaxy Zoo classifications \citep{Walmsley+23}, enabling a homogeneous morphological analysis across the DESI-LIS footprint.

\subsection*{Structural asymmetry measurements}

We measure galaxy structural asymmetries from DESI Legacy Imaging Survey $z$-band images using the \texttt{statmorph} package \citep{Rodriguez-Gomez+19}, which derives non-parametric morphological diagnostics from galaxy light distributions. Prior to measurement, we performed careful background estimation and subtraction, along with source detection and segmentation, using \texttt{photutils} and \texttt{SEP} \citep{Barbary+16,Bradley+20}. Neighboring sources were deblended and masked to avoid contamination from overlapping light. We applied additional masking for foreground stars using positions from the \textit{Gaia} catalog provided by the DESI Legacy Imaging Surveys, and for projected foreground and background galaxies using spectroscopic redshifts where available and photometric redshift estimates otherwise. This multi-step masking procedure reduces contamination from unrelated line-of-sight sources and improves the robustness of the asymmetry measurements compared to approaches that rely solely on image-based segmentation.

Source detection was performed with \textsc{sep} using a PSF-matched Gaussian filter corresponding to the $\sim1.1^{\prime\prime}$ Legacy Survey $z$-band seeing (kernel FWHM $\approx$ 4 pixels). We adopted a minimum detection area of 25 pixels and moderate multi-threshold deblending (\texttt{deblend\_nthresh}=32, \texttt{deblend\_cont}=0.005). Unrelated close neighbors were masked if their redshift differed by less than 1000\,km\,s$^{-1}$ for spectroscopic redshifts or 3000\,km\,s$^{-1}$ for photometric redshifts.

The asymmetry parameter $A$ quantifies deviations from 180$^\circ$ rotational symmetry in a galaxy’s light distribution \citep{Conselice+00}. It is computed by rotating the galaxy image by 180$^\circ$ about its center, subtracting the rotated image from the original, summing the absolute value of the residuals, and normalizing by the total galaxy flux after background correction. A correction term derived from blank sky regions is subtracted to account for noise-induced asymmetry. Larger values of $A$ indicate increasingly disturbed or asymmetric morphologies, commonly associated with galaxy interactions (including minor mergers), lopsided or asymmetric spiral structure, and enhanced star formation \citep{Yesuf+21, Bottrell+24}.

In Fig.~5d we present asymmetry measurements for young galaxies only. Repeating the same analysis for quiescent systems, we find that both inflow and no/weak-flow hosts exhibit low asymmetry, whereas quiescent galaxies with outflows show significantly higher asymmetry than inflow hosts (KS test: $D = 0.3$, $p \approx 10^{-5}$). 

We further computed the outer asymmetry, measured beyond one effective radius to reduce the influence of central structures. The same qualitative trends persist, indicating that the excess disturbance is not confined to galaxy centers and is likely driven by external processes. 

Within the young population, inflow hosts remain the least asymmetric even when restricting to narrow stellar-mass bins. No/weak-flow galaxies are less symmetric than outflow hosts at high stellar masses ($M_\star > 2\times 10^{10}\,\mathrm{M_\odot}$), while at lower masses they show asymmetries comparable to those of outflow systems. Their asymmetry levels are also similar to those of inflow hosts among quiescent galaxies.

For a subset of our sample, we verified that our asymmetry measurements are consistent with independent measurements derived from the higher-quality imaging of the Subaru Hyper Suprime-Cam (HSC) Wide Survey \citep{Bottrell+24}.

Mergers are expected to drive strong gas inflows, while also producing elevated morphological asymmetries and increasing the likelihood of close companions. However, disturbed morphologies are observed in only a minority of inflow hosts ($\sim$8--20\%), and inflow systems are, overall, more morphologically regular than outflow hosts. Likewise, comparable fractions ($\sim$10--20\%) of old galaxies with inflows and with no or weak flows possess nearby companions within $\lesssim$100\,kpc. Together, these results indicate that while mergers, tidal interactions, or external stripping may contribute in individual cases, they are unlikely to dominate the observed inflow population.

We note, however, that the identification of close companions is limited by redshift incompleteness and projection effects. A more definitive assessment will require improved spectroscopic completeness and more uniform spatial sampling to secure redshifts for faint satellites and robustly identify physical companions. Such advances will be essential for quantifying the relative roles of environment, galaxy interactions, and halo cooling in regulating the gas flow cycle.

\subsection*{AGN classification}

As shown in Fig.~3a, we classify galaxies as star-forming (SF), composite, or AGN-dominated using the [N\,\textsc{ii}]/H$\alpha$ versus [O\,\textsc{iii}]/H$\beta$ BPT diagram \citep{Baldwin+81}, adopting the demarcation lines of refs \citep{Kauffmann+03} and \citep{Kewley+01}. Only galaxies with signal-to-noise ratios greater than 2 in all four emission lines are included in the classification. This approach allows robust quantification of the AGN fraction within each gas flow category.

Among 2,655 inflow galaxies with reliable BPT classifications, 62\% are AGN-dominated, 14\% are SF-dominated, and 24\% are composite, indicating that more than 80\% of inflow galaxies likely host weak AGN or exhibit ionization from shocks or evolved stars. By contrast, among 2,022 outflow galaxies, only 12\% are AGN-dominated while 50\% are SF-dominated. No/weak-flow systems are similarly SF-dominated (57\% SF, 19\% AGN).

Restricting to young galaxies ($D_n4000 < 1.6$), the fractions shift modestly: no/weak-flow galaxies are 64\% SF and 12\% AGN; outflows are 51\% SF and 11\% AGN; and inflows are more evenly distributed across BPT classes (31\% AGN, 32\% SF, remainder composite), indicating that weak AGN remain common in inflow systems even among young stellar populations.

In contrast, more than 85\% of old galaxies with inflows exhibit weak H$\alpha$ (equivalent width $<3$\,{\AA}), suggesting that photoionization by evolved stars may dominate their line emission. Nevertheless, the BPT diagram remains effective for identifying AGN when compared with X-ray–selected AGN. The substantial fraction of non–star-forming BPT inflow galaxies is therefore consistent with genuine AGN activity, further supported by radio excess.

\subsection*{AGN luminosity distributions}

To compare the AGN luminosity distributions of inflow, outflow, and no-flow galaxies while properly accounting for non-detections (Extended Data Fig.~6a), we applied non-parametric survival analysis using the Kaplan--Meier (KM) estimator as implemented in the \textsc{lifelines} package. Upper limits on $\log L_{\rm AGN}$ were treated as left-censored data and incorporated using the \texttt{fit\_left\_censoring} formalism.

AGN bolometric luminosities were obtained from \textsc{CIGALE}. Objects with uncertainties exceeding 0.5\,dex in $\log L_{\rm AGN}$ were classified as non-detections. For these systems, we adopted an upper limit of $\log L_{\rm AGN} + 0.3$, where 0.3\,dex corresponds to the typical uncertainty of sources with well-constrained AGN luminosities (i.e., those with $\sigma_{\log L_{\rm AGN}} < 0.5$). This choice provides a conservative and physically motivated bound on the true AGN luminosity of poorly constrained sources while preserving the integrity of the censored distribution. Cumulative distribution functions were constructed for each flow class, and characteristic percentiles were derived directly from the KM CDFs. We verified that adopting a more conservative bound of $\log L_{\rm AGN}+0.4$ for censored sources produces no qualitative change to our conclusions.

To provide physical context, we adopt $L_{\rm bol} \sim 10^{43}\,\mathrm{erg\,s^{-1}}$ as the boundary between weak, radiatively inefficient AGN and radiatively efficient accretion, corresponding to an Eddington ratio of $\sim10^{-2}$ for a typical $M_{\rm BH} \sim 10^7\,M_\odot$ black hole.

Because the AGN fraction in nearby galaxies is typically only $\sim$10--30\%, the median AGN luminosity of all three flow classes lies at the adopted upper-limit threshold, $\log L_{\rm AGN}\approx41$, reflecting that the majority of systems host either very weak or undetected AGN activity. However, the high-luminosity tail of the distribution differs markedly among the populations. While inflow galaxies remain dominated by weak AGN activity even at the 90th percentile ($\log L_{\rm AGN}^{90}\approx42.7$), outflow galaxies exhibit a substantially more luminous tail ($\log L_{\rm AGN}^{80}\approx43.7$, $\log L_{\rm AGN}^{90}\approx44.1$), compared with no-flow systems ($\log L_{\rm AGN}^{80}\approx41.9$, $\log L_{\rm AGN}^{90}\approx43.4$). Thus, although most flow galaxies are not powered by luminous AGN, the presence of strong AGN activity is preferentially associated with systems hosting outflows, as expected, because these galaxies are gas- and dust-rich, providing fuel for both black hole accretion and star formation.

For galaxies with detectable emission lines, the dust-corrected [O\,\textsc{iii}] $\lambda5007$\,{\AA} luminosity can serve as an additional tracer of AGN activity, although it may be contaminated by star formation in starburst galaxies. In Extended Data Fig.~6b, we pair $L_{\rm [O\,III]}$ with the [O\,\textsc{iii}]/H$\beta$ line ratio to help distinguish AGN-dominated and star formation-dominated systems. A bolometric conversion factor of 600 is adopted to convert $L_{\rm [O\,III]}$ to $L_{\rm AGN}$ \citep{KauffmannHeckman09}. This figure confirms that, independent of flow state, the majority of galaxies host weak AGN or are inactive: $\sim$85\% of inflow galaxies have $L_{\rm AGN} < 3\times10^{43}\,\mathrm{erg\,s^{-1}}$, whereas $\sim$60\% of outflow galaxies fall below this threshold. The [O\,\textsc{iii}]/H$\beta$ ratio is less than 2 in 90\% of outflow galaxies, with only $\sim$3--4\% satisfying the [O\,\textsc{iii}]/H$\beta > 3$ criterion commonly used to identify Seyfert galaxies.

\subsection*{Radio emission and radio excess}

We examine the radio properties of our sample using data from the VLASS \citep{Lacy+20} and LOFAR LoTSS \citep{Shimwell+22} surveys. VLASS provides 2–4\,GHz continuum imaging of the northern sky ($\delta > -40^\circ$) at $\sim2.5\,\arcsec$ resolution. We use the Epoch~2 catalog from the Canadian Initiative for Radio Astronomy Data Analysis (CIRADA), which delivers uniformly calibrated source catalogs optimized for reliable detection. The higher observing frequency makes VLASS particularly sensitive to compact radio emission associated with AGN activity.

LoTSS surveys the northern sky at 144\,MHz with $\sim 6\,\arcsec$ resolution and high surface-brightness sensitivity. We use optical identifications from the LoTSS second data release \citep{Hardcastle+23}, based on DESI Legacy Imaging Survey counterparts.

Approximately 5,500 galaxies in our parent sample have radio coverage from one or both surveys. We combine the measurements by converting luminosities to rest-frame 1.4\,GHz assuming a power-law spectrum ($S_\nu \propto \nu^\alpha$). When both VLASS and LoTSS detections are available, $\alpha$ is measured directly; otherwise we adopt $\alpha = -0.7$. This conversion includes both frequency scaling and K-correction.

Extended Data Fig.~6c shows radio luminosity as a function of star formation rate (SFR) for different flow classes. We adopt the empirical radio–SFR relation from ref.~\citep{Heesen+24} and define radio excess as emission exceeding this relation by 0.7\,dex. We further distinguish systems with strong radio excess as those lying an additional 0.2\,dex above this threshold. 

We find that $\sim$27\% of outflow galaxies and $\sim$45\% of no/weak-flow galaxies show radio excess, but only $\sim$15\% and $\sim$20\% of these populations, respectively, exceed the strong-excess boundary. In contrast, $\sim$83\% of inflow galaxies exhibit radio excess, with $\sim$75\% lying more than 0.2\,dex above the excess threshold; the median radio excess of the inflow population is $\sim$1.4\,dex. This trend is consistent with inflows being preferentially detected in older, more quiescent galaxies (Extended Data Fig.~6d), where radio AGN activity is more prevalent. Even among galaxies with young stellar populations ($D_n(4000) < 1.6$), the radio-excess fraction remains higher in inflow systems than in outflow systems ($\sim$65\% versus $\sim$25\%).

These results suggest that cold outflows are more closely associated with star formation or radiatively efficient AGN activity, whereas radio-mode AGN activity is more commonly linked to systems hosting cold inflows.

\subsection*{Metallicity indicators and photoionization models}

We primarily use $N2S2 \equiv \log\,([\mathrm{N\,II}]\lambda6583 / [\mathrm{S\,II}]\lambda\lambda6717,6731)$ to trace relative gas-phase metallicity trends (Fig.~4b) for the N2S2 results. Because these lines are close in wavelength, the ratio is largely insensitive to dust attenuation and flux-calibration uncertainties. Both species arise from low-ionization zones with similar ionization potentials, reducing sensitivity to ionization parameter variations. Moreover, nitrogen transitions from primary to secondary nucleosynthetic production at moderate-to-high metallicity, causing N/O—and thus $N2S2$—to increase with overall chemical enrichment, while sulfur closely tracks oxygen as an $\alpha$-element \citep{PerezMontero17}. 

Empirical comparisons with direct electron-temperature ($T_e$) abundances show that diagnostics incorporating $N2S2$ correlate strongly with direct-method metallicities and are less sensitive to ionization parameter than commonly used indicators such as $N2$ or $O3N2$ \citep{Easeman+24,Brazzini+24}. These properties make $N2S2$ well suited for comparative studies of metallicity variations across galaxy populations.

To express abundances on an absolute oxygen scale, we also adopt the calibration of ref. \citep{Dopita+16}, which combines $N2S2$ with $N2 \equiv \log\,([\mathrm{N\,II}]\lambda6583/\mathrm{H}\alpha)$ to estimate $12+\log\,(\mathrm{O/H})$. This conversion is used only to place measurements on a physically motivated abundance scale in the Methods and Supplementary analyses; our primary conclusions rely on relative abundance trends traced directly by $N2S2$.

As with all strong-line methods, absolute metallicities carry systematic uncertainties \citep{MaiolinoMannucci19}, but these do not affect the differential comparisons between gas-flow classes that are central to this work.

Furthermore, we assess gas-phase metallicities using MAPPINGS~V photoionization and shock models \citep[][see Supplementary Information]{Flury+25}. Among common strong-line ratios, [N\,\textsc{ii}]/[S\,\textsc{ii}] provides the most reliable monotonic tracer of metallicity across both stellar- and AGN-dominated regimes, consistent with MAPPINGS~V predictions.

The [N\,\textsc{ii}]/[S\,\textsc{ii}] ratio exhibits a moderate positive correlation with stellar mass ($\rho \approx 0.5$), which remains essentially unchanged after controlling for AGN activity (partial $\rho \approx 0.45$--$0.5$). In contrast, its correlation with AGN strength—traced by log\,$L_{\rm [O\,III]}$ or log\,$L_{\rm AGN}$ from \textsc{CIGALE}—is weak ($\rho \approx 0.2$--$0.4$) and is further reduced after controlling for stellar mass (partial $\rho \approx 0.15$--$0.2$). These results indicate that the dominant driver of the [N\,\textsc{ii}]/[S\,\textsc{ii}] variation is stellar mass, with AGN activity contributing only a secondary effect.

In our analysis of the mass–metallicity relation, we remove pure AGN based on the BPT classification, and further excluding galaxies with $L_{\rm AGN}$ above the sample median yields qualitatively similar trends: galaxies hosting outflows continue to exhibit systematically higher [N\,\textsc{ii}]/[S\,\textsc{ii}] ratios. In addition, the vast majority ($85\%$) of non-AGN galaxies show [O\,\textsc{i}]/H$\alpha$ ratios below 0.06, well below the values characteristic of shock-excited gas, which typically has [O\,\textsc{i}]/H$\alpha > 0.1$ \citep{Allen+08}. Furthermore, the [N\,\textsc{ii}]/[S\,\textsc{ii}] ratio shows only a weak anti-correlation with the shock-sensitive diagnostic [O\,\textsc{i}]/H$\alpha$ ($\rho \approx -0.2$), further disfavoring shocks as the primary origin of the enhanced [N\,\textsc{ii}]/[S\,\textsc{ii}] ratios.

\textbf{Acknowledgments}
We are sincerely grateful to the referees for their careful reading of the manuscript and for their constructive comments, which have helped us improve the manuscript, both in its scientific content and in the clarity and presentation. We are grateful to Ms. Xuejie Dai for assistance in producing schematic Fig.~2 under the guidance of H.M.Y, H.L., and R.J. C.B. gratefully acknowledges support from the Forrest Research Foundation and from the Pawsey Supercomputing Research Centre’s Setonix Supercomputer (https://doi.org/10.48569/18sb-8s43) and Acacia Object Storage (https://doi.org/10.48569/nfe9-a426), with funding from the Australian Government and the Government of Western Australia.

\medskip
\textbf{Author contributions}

H.M.Y. conceived and led the study, designed and performed the observational analysis and measurements, developed the interpretation, and wrote the manuscript. Z.Y. and F.Y. performed the simulation analysis, and co-wrote the sections comparing simulations and observations. R.J., L.C.H., L.L., H.L., S.S., C.B., F.G., and J.D.S. contributed to the interpretation of the data, provided technical and conceptual input, and contributed to manuscript revision. C.B. also provided independent structural measurements based on the HSC Wide survey using a separate analysis software, confirming the H.M.Y. measurements derived from the DESI Legacy Survey. All authors discussed the results and approved the final manuscript.

\medskip
\textbf{Competing interests}
The authors declare no competing interests.

\medskip
\textbf{Data availability} 
All observational data used in this study are publicly available through the survey data releases cited below. The catalog of $\sim$30{,}000 galaxies, including derived gas-flow measurements and related physical properties, will be made publicly available upon publication of this Article. The MACER simulation data supporting this work are available from Y.Z. upon reasonable request.
\begin{itemize}
    \item DESI Legacy survey:\href{https://www.legacysurvey.org/dr10}{https://www.legacysurvey.org/dr10}
    \item DESI DR1: \href{https://data.desi.lbl.gov/doc/releases/dr1/}
    {https://data.desi.lbl.gov/doc/releases/dr1/}
    \item Sloan Digital Sky Survey (SDSS): \href{https://skyserver.sdss.org/casjobs/}{https://skyserver.sdss.org/casjobs/}
    \item Pan-STARRS: \href{https://catalogs.mast.stsci.edu/panstarrs/}{https://catalogs.mast.stsci.edu/panstarrs/}
    \item 2MASS: \href{https://irsa.ipac.caltech.edu/Missions/2mass.html}{https://irsa.ipac.caltech.edu/Missions/2mass.html}
    \item UKIDSS: \href{http://wsa.roe.ac.uk/}{http://wsa.roe.ac.uk/}
    \item WISE: \href{https://irsa.ipac.caltech.edu/Missions/wise.html}{https://irsa.ipac.caltech.edu/Missions/wise.html}
\end{itemize}

\clearpage

\renewcommand{\thefigure}{\arabic{figure}}  
\renewcommand{\figurename}{Supplementary Fig.}  

\setcounter{figure}{0}

\section*{Supplementary Information}

\subsection*{Sample statistics}

Table~\ref{tab:subsamples} summarizes the number of galaxies classified as inflow, outflow, and no/weak-flow systems across various subsamples. For the machine-learning-based morphological classifications, we adopt two probability thresholds to assess the robustness of the sample statistics. Likewise, for disk galaxies we use two S\'{e}rsic index thresholds to illustrate the sensitivity of sample statistics to the adopted selection criteria. Structural measurements such as S\'{e}rsic index, concentration, and asymmetry are available for all or $\gtrsim98\%$ of the parent sample, while environmental-density and group-halo-mass estimates are available for 93\% and 48\% of galaxies, respectively.

The inflow, outflow, and no/weak-flow samples do not sum to the full parent sample in Table~\ref{tab:subsamples} because we adopt a conservative flow-classification threshold ($P>0.85$). Consequently, a substantial number of galaxies with candidate flow signatures remain unclassified. For example, lowering the threshold to ($P>0.5$) increases the number of classified inflow and outflow galaxies to 15,213 and 8,707, respectively, yielding a total of 29,616 classified galaxies. The main conclusions of this work are insensitive to reasonable variations in the adopted threshold.

\begin{table*}
\centering
\caption{Number of galaxies in different subsamples with inflow, outflow, and no/weak-flow Na\,\textsc{i}\,D classifications.}
\label{tab:subsamples}
\begin{tabular}{lccc}
\hline
Subsample & Inflow & Outflow & No/weak flow \\
\hline
Total & 6575 & 2215 & 5696 \\
Young ($D_n(4000) < 1.6)$ & 1450 & 2147 & 4241 \\
Old ($D_n(4000) \ge 1.6)$ & 5125 & 68 & 1455 \\
BPT AGN & 1634 & 240 & 773 \\
Bar $P > 0.8$ & 83 & 195 & 126 \\
Bar $P > 0.5$ & 291 & 378 & 413 \\
Young and bar $P > 0.5$ & 104 & 373 & 350 \\
Spiral $P > 0.5$ & 254 & 511 & 646 \\
Young and Spiral $P > 0.5$ & 136 & 508 & 607 \\
Merger $P > 0.5$ & 100 & 72 & 94 \\
Merger $P > 0.8$ & 29 & 16 & 22 \\
Young $n < 2.5$ & 331 & 457 & 1704 \\
Young $n < 2.0$ & 188 & 280 & 1289 \\
$\log\,(M_\star/M_\odot)=10-10.3$ & 316 & 303 & 895\\
$\log\,(M_\star/M_\odot)=10.3-10.6$ & 955 & 702 & 1156\\
$\log\,(M_\star/M_\odot)=10.6-10.9$ & 1853 & 748 & 1240\\
Group mass & 2464 & 1220 & 2854\\
\hline
\end{tabular}
\end{table*}

\subsection*{MACER disk galaxy simulations}

To aid interpretation of the observations, we compare with state-of-the-art hydrodynamic simulations of disk galaxies performed with the MACER framework \citep{Yuan2018}. A full description of the models will be presented in ref.~\citep[][]{ZouYX+26}; here we summarize the key ingredients relevant for understanding the results discussed below.

MACER builds on a long lineage of studies of galaxy–AGN feedback \citep[e.g.,][]{2001Ciotti,2009Ciotti,2011Novak,2014Gan} and is specifically designed to resolve gas dynamics on galactic scales while simultaneously capturing the essential physics of black-hole accretion. Its inner radial boundary is placed well inside the Bondi radius, so that the mass inflow rate can be measured directly at the boundary. This setup avoids the potentially large biases—both overestimates and underestimates—in inferred accretion rates \citep{Negri2017,Hopkins_2016} that can arise from approximate Bondi-based prescriptions \citep{Springel05,Weinberger2017}. Coupled with black hole accretion theory, this provides a physically grounded estimate of the accretion rate, which determines the AGN power and feedback strength. This high-resolution treatment distinguishes MACER from cosmological simulations, where subgrid prescriptions typically regulate black hole growth and feedback.

AGN feedback is modeled self-consistently through direct coupling of radiation, winds, and jets with the interstellar medium, rather than via parameterized energy injection. The simulations are two-dimensional and axisymmetric, performed with the parallel ZEUS-MP/2 code \citep{Hayes_2006} in spherical polar coordinates $(r,\theta,\phi)$. The grid for the disk galaxy simulation contains 72 uniform angular zones and 280 logarithmically spaced radial zones extending from 100\,pc to 500\,kpc, giving a minimum cell size of $\sim4$\,pc at the inner boundary.

The simulated galaxy represents a Milky Way–like disk embedded in a dark matter halo with $(M_{\rm halo}, c) = (1.6\times10^{12}\,M_\odot, 12)$. The baryonic components have masses $(M_{\rm bulge}, M_\star, M_{\rm gas}) = (1.5\times10^{10},\,4.7\times10^{10},\,9\times10^{9})\,M_\odot$. The stellar disk follows scale lengths $(h_\star, z_\star) = (3.0\,{\rm kpc},\,0.3\,{\rm kpc})$. The central black hole begins with a mass of $5\times10^{7}\,M_\odot$. The ISM and circumgalactic medium (CGM) are initialized at $10^4$\,K and $10^6$\,K, respectively. The system is evolved for $\sim0.6$\,Gyr to reach quasi-equilibrium before being followed for 12\,Gyr.

Cosmological gas supply is modeled through both hot- and cold-mode accretion. Hot-mode accretion is implemented as a quasi-spherical inflow at $500\,\mathrm{kpc}$, representing shock-heated halo gas, while cold-mode accretion enters as filamentary streams at $100\,\mathrm{kpc}$. The total inflow rate is fixed at $100\,M_\odot\,\mathrm{yr}^{-1}$ throughout the simulation and is partitioned in a 60:40 ratio between the cold and hot modes. For the cold-mode inflow, a constant radial velocity of $v_r=-200\,\mathrm{km\,s^{-1}}$ (negative denotes inflow) is adopted.

Angular momentum transport in the disk is parameterized using a spatially uniform kinematic viscosity $\nu$. By varying $\nu$, the simulations mimic different efficiencies of angular momentum redistribution driven by processes such as gravitational torque or turbulence. Four values are explored: $\nu = 0.25,\ 0.55,\ 0.75,$ and $1.1$.

Star formation follows the prescription used in earlier MACER studies but is restricted to gas with $T < 4\times10^4$\,K and number density $n > 1\,{\rm cm^{-3}}$. Supernova feedback provides both thermal and momentum input, with the effective coupling scale decreasing at higher ambient densities.

AGN feedback operates in two modes regulated by the accretion rate: a radiatively efficient cold (quasar) mode at high accretion rates and a kinetic hot (radio) mode at low accretion rates. The transition occurs at a luminosity of 2\% of the Eddington limit \citep{Yuan2014}. The AGN outputs from the two modes as a function of accretion rate, namely radiation, jet, and wind, are directly taken from small-scale MHD simulations of black hole accretion or observations.

\subsection*{Results from MACER galaxy simulations}

The MACER simulations indicate that a galaxy’s star formation history and cold gas content are governed primarily by the interplay between AGN feedback, angular momentum transport within the disk, and the angular momentum carried by cosmological inflows. Among the disk models explored, only the Fiducial run with AGN feedback and intermediate viscosity ($\nu = 0.55$) undergoes rapid quenching. The other models do not quench efficiently, providing insight into why most present-day galaxies sustain star formation over extended periods.

The simulations show that diversity in gas accretion histories and star formation activity can arise from variations in internal angular momentum transport, the strength and intermittency of AGN feedback, and the momentum of cold inflows. Crucially, these differences occur at fixed halo mass (only one galaxy with $M_h=10^{12}\,M_\odot$ is simulated), consistent with the observational finding that short-timescale changes in gas flow state are not primarily driven by halo mass or large-scale environment. Instead, the models demonstrate that internal processes regulate the transition between inflow-dominated, outflow-dominated, and weak-flow phases. Variants of the Fiducial model with reduced angular momentum, as well as the $\nu=0.75$ and noAGN0.75 runs, are therefore particularly relevant for slowly evolving, star-forming disk galaxies in the nearby Universe.

\subsection*{Gas kinematics and flow geometry}

The simulated velocities, orientations, and overall flow geometries of both inflows and outflows are broadly consistent with established observational trends. In particular, the models offer a straightforward explanation for why inflows in actively star-forming galaxies are so challenging to detect: gas accretion proceeds preferentially within (or close to) the disk plane, whereas hot, feedback-driven outflows are launched approximately perpendicular to the disk. As a result, along many common lines of sight the outflowing material can displace, obscure, or otherwise mask the inflowing component.

\begin{figure*}
\centering
\includegraphics[width=\textwidth]{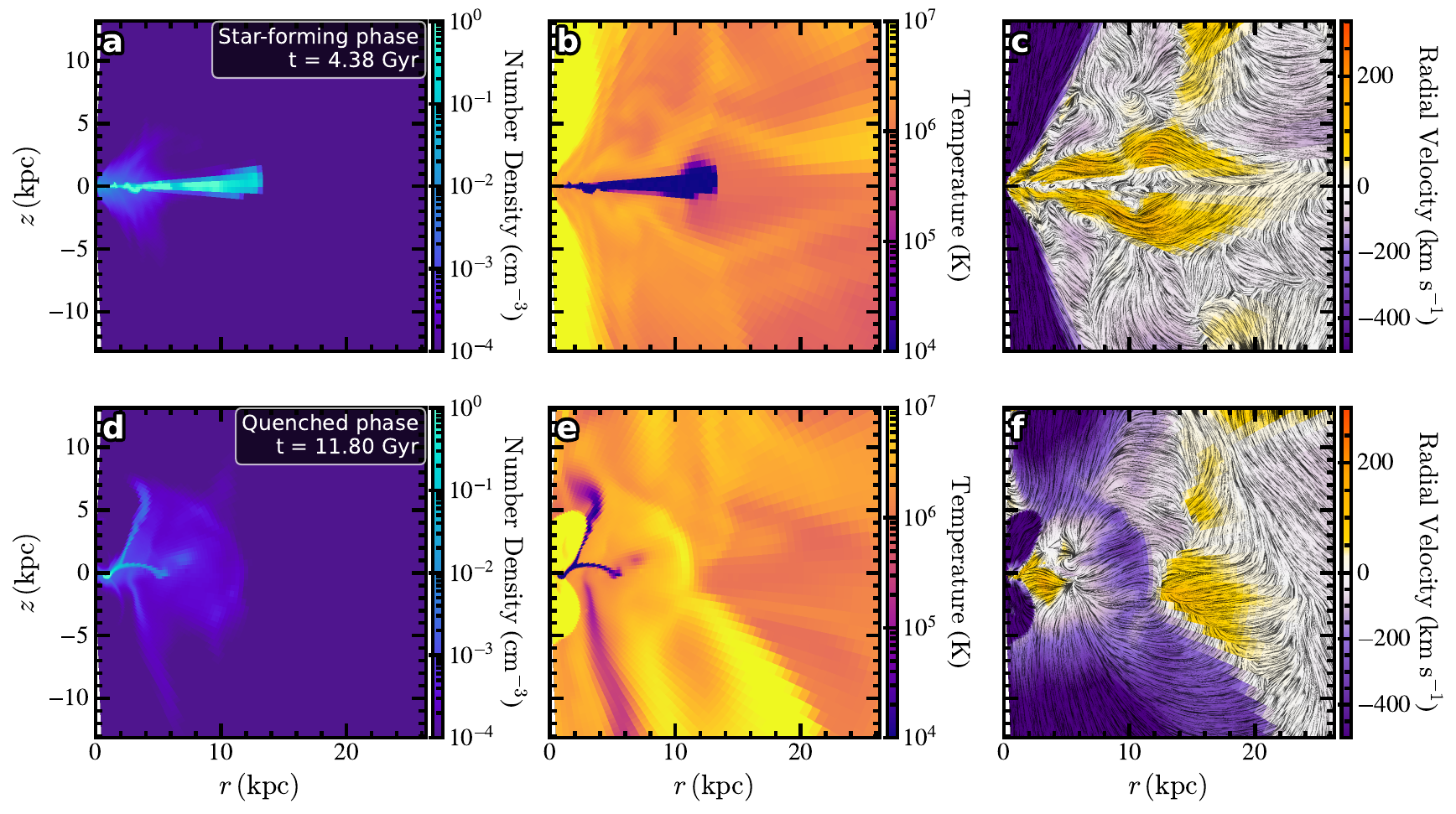}
\caption{\textbf{Simulated gas properties in star-forming and quenched phases.} Snapshots of the Fiducial simulation at $t = 4.38$\,Gyr (\textbf{a--c}), corresponding to a star-forming phase, and at $t = 11.80$\,Gyr (\textbf{d--f}), corresponding to a quenched phase. From left to right, the panels show gas number density, temperature, and radial velocity, with positive values indicating inflow. In the radial-velocity panels, black line segments are tangential to the local velocity field and indicate the direction of the flow.}\label{fig:Fiducial1}
\end{figure*}

\begin{figure*}
\centering
\includegraphics[width=\textwidth]{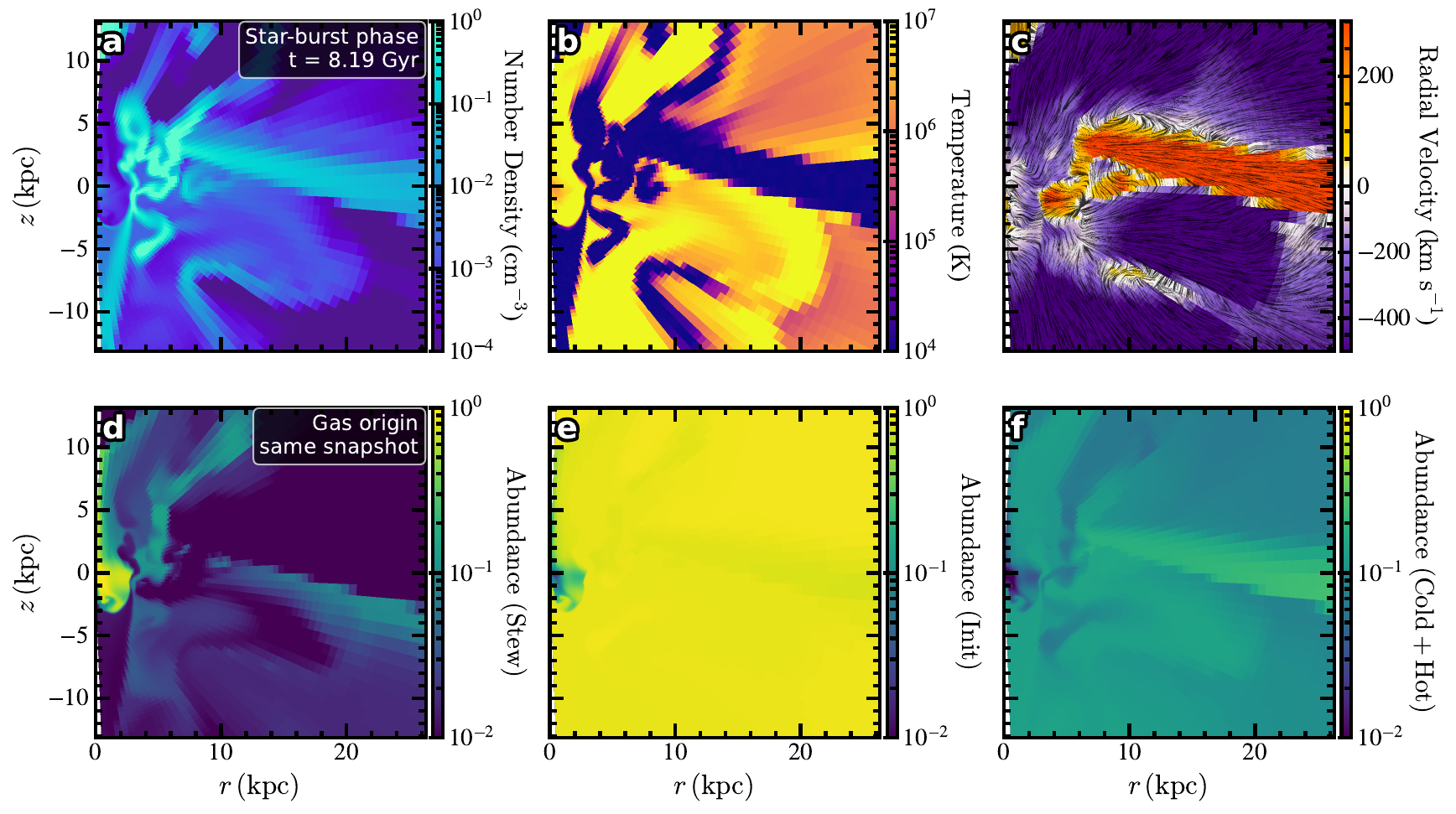}
\caption{\textbf{Gas properties and origin during a starburst phase.} Snapshot of the Fiducial simulation at $t = 8.19$\,Gyr, corresponding to a starburst phase. Panels \textbf{a--c} show gas number density, temperature, and radial velocity (positive values denote inflow). In \textbf{c}, short black line segments, drawn tangentially to the local velocity field, indicate the flow direction. The bottom panels illustrate the origin of the gas supply: \textbf{d}, stellar-wind material; \textbf{e}, residual initial gas; and \textbf{f}, cosmological accretion (including both cold- and hot-mode components). The snapshot highlights disk-aligned inflow coexisting with a roughly bipolar hot outflow, while the gas supply reflects a combination of recycled material, remaining initial gas, and ongoing cosmological accretion.}
\label{fig:Fiducial2}
\end{figure*}

\begin{figure*}
\centering
\includegraphics[width=\textwidth]{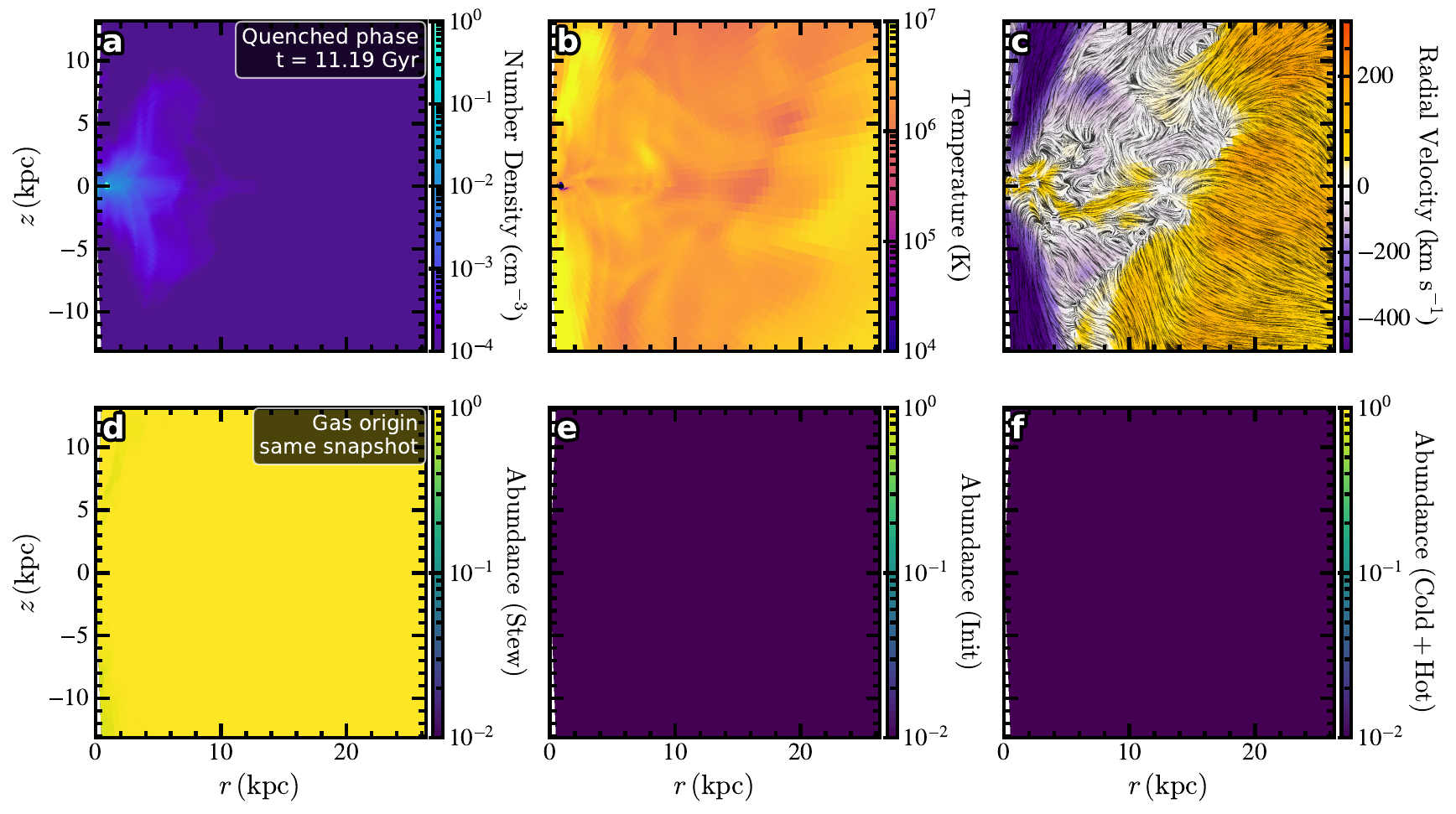}
\caption{\textbf{Gas properties and origin in a quenched phase.} Snapshot of the Fiducial simulation at $t = 11.19$\,Gyr, corresponding to a quenched phase. Panels \textbf{a--c} present the gas number density, temperature, and radial velocity (positive values denote inflow). In \textbf{c}, short black line segments, drawn tangentially to the local velocity field, indicate the flow direction. The bottom panels show the origin of the gas supply: \textbf{d}, stellar-wind material; \textbf{e}, residual initial gas; and \textbf{f}, cosmological accretion.}
\label{fig:Fiducial3}
\end{figure*}

Here we present snapshots from the Fiducial simulation during star-forming and quiescent phases (Supplementary Figs.~\ref{fig:Fiducial1}--\ref{fig:Fiducial3}). These illustrate the key qualitative features discussed above, namely disk-aligned inflow coexisting with a roughly bipolar hot outflow oriented perpendicular to the disk.

In the Fiducial model, episodes of elevated star formation coincide with enhanced cold gas accretion. In the central regions, however, cold streams are frequently disrupted or pushed back by hot outflows, typically stalling at radii of $\sim$5\,kpc (Supplementary Fig.~\ref{fig:Fiducial2}). During phases of low star formation, both inflows and outflows are dominated by hot gas, although occasional cold clumps and narrow streams can still penetrate inward. In quiescent stages, the gas supply is dominated by stellar mass loss, with only a minor contribution from ongoing cosmological accretion (Supplementary Fig.~\ref{fig:Fiducial3}d--f).

The simulations further show that inflows and outflows coexist over extended periods, sustaining star formation on observed depletion timescales. The predicted cold gas masses and flow rates are consistent with observational constraints. A large fraction (up to $\sim$90\%) of the accreting material is recycled---originating from stellar ejecta and previously ejected ISM----while the contribution from cosmological inflow is typically below 10\% and increases during high-SFR phases. Feedback-driven outbursts repeatedly expel gas to large radii (tens to hundreds of kpc, and occasionally beyond 500\,kpc), much of which subsequently cools and re-accretes on timescales of $\sim$0.5\,Gyr.

Furthermore, we select four representative models from ref.~\citep[][]{ZouYX+26} to summarize the time evolution of the SFR and the sSFR (see their Figures 9--10). In the Fiducial model, a pronounced starburst drives the SFR to about \(100\,M_\odot\,\mathrm{yr^{-1}}\) around \(t \approx 7.8-8.2\,\mathrm{Gyr}\), followed by a rapid decline between \(\approx 8.2-9.2\,\mathrm{Gyr}\) and quenching by \(\approx 9\,\mathrm{Gyr}\), with the system remaining quenched up to \(12\,\mathrm{Gyr}\); the sSFR remains in the range \(\sim 10^{-12}-10^{-13}\,\mathrm{yr^{-1}}\) once quenched. ModelLow exhibits a similar qualitative pattern but with a somewhat delayed peak near \(t \approx 8.0\,\mathrm{Gyr}\) and a high-SFR phase lasting longer (\(\mathrm{SFR} \approx 50\,M_\odot\,\mathrm{yr^{-1}}\)), with quenching around \(9-10\,\mathrm{Gyr}\) and the sSFR sustaining at \(\sim 10^{-13}\,\mathrm{yr^{-1}}\) through \(12\,\mathrm{Gyr}\). The noAGN0.55 case, lacking AGN feedback, shows no global quenching; the sSFR stays near \(\sim 10^{-10}\,\mathrm{yr^{-1}}\), and after \(\approx 9\,\mathrm{Gyr}\) the cold CGM filaments fuel additional ISM SFR, boosting it by roughly a factor of \(3-5\). The noAGN0.75 model behaves similarly in that no sustained quenching occurs; the sSFR remains around \(\sim 10^{-10}\,\mathrm{yr^{-1}}\), with low-amplitude oscillations between \(\sim 10^{-10}\) and above \(10^{-11}\) after \(\approx 7\,\mathrm{Gyr}\).

\subsection*{Star formation and AGN co-evolution}

The simulations reveal a close coupling between star formation and AGN activity, arising from their shared dependence on a common gas supply regulated by feedback. Episodes of enhanced gas inflow simultaneously fuel black hole growth and elevated star formation. In particular, starburst events are triggered by the accretion of cold filaments that condense out of the circumgalactic medium (CGM) through radiative cooling.

AGN-driven outflows play a dual role in this process. Although they can temporarily expel gas from galactic centres, they also deposit mass and energy into the CGM, promoting gas accumulation and the eventual formation of massive cold filaments. These filaments later fall back toward the galaxy, fueling both a surge in star formation and enhanced AGN activity \citep{ZouYX+26}. AGN feedback therefore does not simply suppress star formation; it helps set the stage for a major accretion episode by creating the conditions required for the formation and infall of cold streams. In the simulations, models with the strongest cumulative AGN feedback do not quench. Instead, long-term quenching only occurs when gas is rapidly consumed during a single, intense starburst --- an event enabled by AGN-driven outflows and perturbations that promote the formation and inward transport of massive cold filaments.

The overall impact of AGN activity depends sensitively on angular momentum transport within the galaxy, which regulates how efficiently gas can reach the nucleus and thus sets the AGN duty cycle. The formation of cold filaments is ultimately governed by the thermal state of the CGM: if AGN heating is too strong, cooling is suppressed and filament formation is inhibited. This delicate balance between heating and cooling establishes a self-regulated cycle that unifies feedback, star formation history, and the multiphase structure of gas in both the ISM and CGM.

\subsection*{Recently quenched and post-starburst galaxies in the MACER simulation}

The Fiducial disk model has clear analogs among recently quenched and post-starburst galaxies. These systems frequently show either inflows or outflows, and in some cases both, consistent with a transitional phase in the feedback cycle. Quantitatively, the Fiducial disk model describes an evolutionary sequence in which a starburst is ignited, followed by a post-starburst phase and eventual quenching. Cold filaments condense and feed gas into the ISM, boosting the star formation rate to roughly $100\,M_\odot\,\mathrm{yr}^{-1}$ and triggering a luminous AGN \citep{ZouYX+26}. Within about $1\,\mathrm{Gyr}$, AGN feedback rapidly quenches the system, and the gas fraction $f_{\rm gas}=M_{\rm gas}/M_{*}$ declines from $\sim 1$ to approximately $10^{-3}$ (see details in Zou et al. 2026c, in prep.). Qualitatively, the Fiducial model correctly predicts the co-existence of cold outflow and inflow in PSBs, their rapid changes in star formation within $1\,\mathrm{Gyr}$ and their gas fraction. However, rapid quenching is rare in the nearby Universe, and only a small fraction of galaxies are post-starbursts. The Fiducial model therefore represents a special evolutionary pathway rather than the dominant mode of galaxy evolution today.

\subsection*{Direct evidence for refueling in post-starburst galaxies}

PSBs are traditionally interpreted as systems that have recently quenched after expelling or consuming most of their gas reservoirs. However, a significant minority ($\sim$10--30\%) retain substantial cold gas masses despite little ongoing star formation, posing a long-standing puzzle: whether this gas is simply leftover from before quenching and why it remains inefficient at forming stars. Our detection of Na\,{\sc i}\,D inflows provides direct insight into this question by revealing that some of this gas is being newly accreted or recycled, rather than solely being residual fuel from the pre-quenching phase.

Extended Data Fig.~2 presents six PSBs with clear Na\,{\sc i}\,D inflow signatures spanning a range of stellar masses, environments, and structural properties. Two of these galaxies are already confirmed to host substantial cold gas reservoirs. For example, TARGETID~2852167325057024 has $\log M_\star = 10.2$ and contains a large atomic gas reservoir ($\log M_{\rm HI} = 9.4$) but comparatively little molecular gas ($\log M_{\rm H_2} = 8.3$; refs.~\cite{French2015,Ellison2025}). The resulting atomic-to-molecular gas ratio ($M_{\rm HI}/M_{\rm H_2} \approx 10$) indicates that much of the gas remains in a diffuse atomic phase rather than in the dense molecular form required for star formation. This imbalance is naturally expected if the gas has been recently accreted and has not yet fully condensed into molecular clouds. The simultaneous detection of Na\,{\sc i}\,D inflow provides direct evidence that gas acquisition is actively ongoing in this system, linking its substantial atomic reservoir to recent or continuing accretion.

Other examples show complementary signatures consistent with recent gas acquisition, including merger remnants, nearby gas-rich companions, and group environments. These systems demonstrate that cold-gas inflows occur in at least some PSBs across a wide range of environments and structural states, indicating that refueling is not limited to a single evolutionary pathway.

The direct detection of inflows provides a natural explanation for the presence of substantial cold gas in a subset of PSBs (refs.~\cite{French+15, YesufH17, Baron+23}). In contrast to the traditional view in which this gas represents residual fuel left over from the pre-quenching phase, our observations indicate that at least some PSBs are actively reacquiring gas after quenching. This suggests that these galaxies are observed during an early stage of renewed gas accretion, before molecular clouds and sustained star formation have fully re-established.

Future observations combining Na\,{\sc i}\,D inflow measurements with sensitive H\,{\sc i} and molecular gas data will be critical for determining how commonly refueling occurs in PSBs and for establishing its role in governing their long-term evolution.

\subsection*{Details on examples of post-starburst galaxies with detected inflows}

Extended Data Fig.~2 presents Na\,{\sc i}\,D absorption profiles for six  post-starburst galaxies with detected inflows, diverse gas content, environments, and structural properties.

TARGETID~2852167325057024 has $\log\,(M_\star/M_\odot)=10.2$, with substantial atomic and molecular gas reservoirs ($\log\,(M_{\mathrm{HI}}/M_\odot)=9.4$ and $\log\,(M_{\mathrm{H_2}}/M_\odot)=8.3$; refs.~\cite{French2015,Ellison2025}). The galaxy is edge-on, structurally undisturbed ($A=0.03$), exhibits radio emission, and resides in a group environment without close companions within several hundred kpc, consistent with gradual gas accretion or recycling.

TARGETID~2842634825498625 ($\log M_\star=10.6$) contains $\sim10^9\,M_\odot$ of molecular gas but lacks dense gas tracers (HCN) in ALMA observations \cite{French2015,French+23}. Its disturbed morphology and merger signatures ($A=0.11$, $C=4.3$), combined with elevated N2S2 ratios indicating chemical enrichment, suggest recent external gas acquisition.

TARGETID~39627932301266709 ($\log M_\star=10.6$) has an upper limit of $M_{\mathrm{H_2}}<1.5\times10^9\,M_\odot$ \cite{French2015}, appears largely isolated, and exhibits faint tidal features ($A=0.05$, $C=3.9$), consistent with minor interactions or recent accretion.

TARGETID~2842467036561408 has a close star-forming companion within $\sim30$\,kpc but otherwise appears morphologically undisturbed ($A=0.03$, $C=3.1$) and lacks strong radio emission. Although direct gas measurements are unavailable, the nearby companion provides a plausible external gas supply.

TARGETID~39628530157360232 is structurally smooth, exhibits radio excess emission, and resides in a group environment, although its nearest companion lies several hundred kpc away. It is undisturbed ($A=0.04$, $C=3.7$) weak bar may facilitate inward gas transport.

TARGETID~39628517691883730 is isolated, and morphologically undisturbed ($A=0.03$, $C=3.0$), with no strong radio emission or prior gas measurements, suggesting that inflow may arise from low-level external accretion or stellar mass recycling.

Together, these examples demonstrate that cold-gas inflows occur in post-starburst galaxies spanning a wide range of gas content, environments, and structural states, consistent with ongoing gas accretion or recycling during the post-starburst phase.

Extended Data Fig.~3 presents six illustrative post-starburst galaxies with strong outflow signatures, exhibiting median velocities of $\sim 400$\,km\,s$^{-1}$---about twice typical cold-gas inflow velocities in such systems \citep{YesufH17} These high velocities suggest that a substantial fraction of the outflowing gas may escape, particularly if post-starburst galaxies undergo repeated inflow–outflow cycles, and may reflect enhanced feedback linked to recent AGN activity.

\subsection*{Quiescent galaxies and chaotic cold accretion}

Most present-day quiescent galaxies quenched at earlier cosmic epochs, yet many still exhibit evidence for ongoing gas accretion. These systems are more naturally compared with MACER elliptical-galaxy models \citep[e.g.,][]{Yuan2018,zhu_2023} than with the disk-dominated simulations discussed above. In this regime, inflows are primarily associated with hot halo cooling and maintenance-mode feedback, consistent with the observational picture of low-level, long-lived gas cycling in quiescent systems.

Recent theoretical models of chaotic cold accretion (CCA) provide a natural framework to interpret inflows in quiescent galaxies. In the CCA paradigm \citep[e.g.,][]{Gaspari+18}, turbulent hot halos, stirred by outflow or mergers, condense into warm ionized and neutral filaments, which then rain toward the central black hole as cold clouds, driving self-regulated AGN activity and associated outflows. The ensemble velocity shifts and dispersions predicted by these models are $|v| \sim 10$--200~km~s$^{-1}$ and $\sigma \sim 100$--200~km~s$^{-1}$ for cold/warm gas, while pencil-beam measurements of individual clouds can display $\sigma$ as low as $\sim 10$~km~s$^{-1}$ \citep{Gaspari+18}. These model predictions are consistent with our observations of cool gas inflows in massive quiescent galaxies, supporting a scenario in which cooling, precipitation, and black hole feedback form a coupled, self-regulating gas cycle \citep{Gaspari+20}.

It is important to note that current CCA simulations, like most idealized galaxy-scale feedback models, are not embedded in a full cosmological context. Our observational census of gas flows across a wide range of stellar masses and galaxy types therefore provides a critical benchmark for future cosmological simulations. In particular, it can inform models in areas where direct predictions remain lacking, including the relationship between gas flow states and halo mass, the influence of galaxy substructure, and the duty cycles of inflow and outflow over cosmic time.

\subsection*{Rarity of cold outflows in quiescent galaxies}

Cold outflows are an order of magnitude less common than cold inflows in quiescent galaxies in our sample. This asymmetry is not unexpected. At low accretion rates, AGN feedback is predicted to operate predominantly in a radiatively inefficient, hot mode, in which most of the released energy emerges mechanically through jets and hot winds rather than through cool gas outflows \citep[][]{Yuan2014,Yuan2018}. Gas driven from such hot accretion flows is expected to be highly ionized, with temperatures $\gtrsim10^{6}$--$10^{7}$\,K, making it difficult to detect using the low-ionization absorption tracers employed here.

Although the overall prevalence of such winds remains uncertain, observational evidence for highly ionized outflows has been reported in low-luminosity and radio-loud AGN through X-ray spectroscopy, including blueshifted Fe\,K absorption features associated with fast, highly ionized winds \citep{Tombesi+14, ShiFangzheng+25}. The scarcity of cool outflow signatures in quiescent systems is therefore consistent with feedback that couples to the surrounding medium primarily through hot, volume-filling phases.

This interpretation is further supported by the low incidence of resolved outflows in nearby quiescent galaxies. Integral-field surveys identify compact AGN-like ionized outflow structures (``red geysers’’) in only $\sim2$--3\% of quiescent galaxies \citep{Roy+21, Ilha+24}, indicating that large-scale cool or warm outflows are not ubiquitous in this population. At the same time, the frequent detection of cold inflows does not imply an absence of feedback activity. Dense, clumpy inflowing gas can survive in the presence of hot winds because coupling between hot flows and cold clouds is inefficient, interactions can promote rapid cooling in mixed gas, and feedback is both geometrically variable and temporally intermittent \citep[e.g.,][]{Gaspari+18}.

Taken together, these results reinforce a two-mode picture of gas accretion and feedback. Star-forming galaxies commonly host cold, multiphase outflows driven by stellar feedback, which may later reaccrete and, in some cases, contribute to rapid quenching following starbursts. In contrast, quiescent galaxies are more often characterized by low-level accretion and hot-mode AGN feedback that heats halo gas without producing prominent cold outflow signatures, potentially alternating with episodes of cool gas inflow \citep{guo18}. The observed excess of cold inflow relative to cold outflow in quiescent systems thus reflects phase-dependent detectability of gas flows rather than a lack of feedback activity.

In passive galaxies, the detected inflows may originate from stellar mass loss, residual disk gas, or the cooling of hot halo gas, although the relative contribution of each channel cannot be disentangled with the present data. The observed low gas fractions, near-solar metallicities, and association with low-level AGN activity are broadly consistent with the slow cooling of previously enriched halo material, potentially supplemented by stellar ejecta. The general absence of strong outflows in these systems disfavors merger-driven triggering, which would typically be accompanied by elevated star formation and stronger feedback signatures. A dominant contribution from pristine cosmological accretion is likewise disfavored by the inferred chemical enrichment.

\subsection*{Energetic impact of low-power radio AGN}

To place our results in a broader physical context, we first note that the weak AGN activity commonly observed in galaxies with inflows requires only a minute gas supply to operate, yet is energetically sufficient to influence gas in and around galaxies. Most inflow galaxies have radiative $L_{\rm AGN} \sim 10^{41-42}\,\mathrm{erg\,s^{-1}}$ (Extended Data Fig.~6). Using standard radiatively inefficient accretion models \citep[e.g.,][]{XieYuan12}, such low-luminosity AGN correspond to mass accretion rates of $\dot{M} \sim 10^{-5}$--$10^{-3}\,M_\odot\,\mathrm{yr^{-1}}$, equivalent to only $\sim 10^{-5}$--$10^{-3}$ times the Eddington accretion rate for typical $10^8\,M_\odot$ black holes, and implying total gas consumption of $M_{\rm acc} \sim 10^2$--$10^4\,M_\odot$ over a $10^7$~yr duty cycle. This is negligible compared with typical cold gas reservoirs of $10^{8-9}\,M_\odot$, demonstrating that even extremely inefficient AGN activity is fully compatible with ongoing inflow and recycling.

We next compare AGN energy output with the gravitational and thermal energies of galactic gas. Typical radio luminosities of $L_{1.4} \sim 10^{22}$--$10^{23}\,\mathrm{W\,Hz^{-1}}$ (Extended Data Fig.~6) correspond to mechanical jet powers $P_{\rm jet} \sim 10^{42}$--$10^{43}\,\mathrm{erg\,s^{-1}}$ \citep{Cavagnolo+10, HeckmanBest14}. For a representative galaxy with $\sigma_\star \approx 150\,\mathrm{km\,s^{-1}}$ and $M_{\rm gas} \sim 10^9\,M_\odot$, the ISM binding energy is $E_{\rm bind} \sim M_{\rm gas}\sigma_\star^2 \sim {\rm few} \times 10^{55}\,\mathrm{erg}$. Over a conservative AGN lifetime of $t_{\rm AGN} \sim 10^7$~yr, even modest jets inject $E_{\rm jet} \sim 10^{57}$--$10^{58}\,\mathrm{erg}$, one to two orders of magnitude above $E_{\rm bind}$, sufficient to heat or perturb the ISM while radiative losses and imperfect coupling prevent wholesale gas removal.

The same energy is comparable to the thermal content of hot circumgalactic gas in halos of $M_{\rm halo} \sim 10^{12}$--$10^{13}\,M_\odot$ (Fig.~3), with $T_{\rm vir} \sim 10^{6}$--$10^{7}$~K and $E_{\rm CGM} \sim 10^{58}\,\mathrm{erg}$ \citep{DonahueVoit22}, showing that repeated low-power episodes can offset cooling losses and maintain hot halos \citep{Fabian12}. At a few kpc, gas densities of $n \sim 10^{-2}$--$10^{-3}\,\mathrm{cm^{-3}}$ and temperatures $T \sim 10^{6}$--$10^7$~K yield cooling times $t_{\rm cool} \sim 10^8$--$10^9$~yr \citep{DonahueVoit22}, supporting maintenance-mode regulation.

Observationally, inflow galaxies typically show compact, core-dominated radio emission, with extended jets being rare. This morphology is expected for low-Eddington, radiatively inefficient AGN accreting primarily from hot gas, where energy is released via small-scale jets, shocks, bubbles, and hot winds rather than large radio lobes \citep{Fabian12, HeA+25}. The rarity of strong cold outflows and extended jets indicates that feedback operates primarily in a gentle maintenance mode, heating and pressurizing the halo to prevent rapid cooling \citep{Fabian12, Gaspari+18}.

In summary, low-power radio AGN require only a tiny fraction of the available gas to operate and are energetically capable of perturbing both ISM and circumgalactic gas. Their primary role is long-term thermal regulation, consistent with the compact radio cores and ongoing inflows observed in our sample.

\subsection*{Matched comparison of inflow and outflow host galaxies}

\begin{figure}
    \centering
    \includegraphics[width=0.95\linewidth]{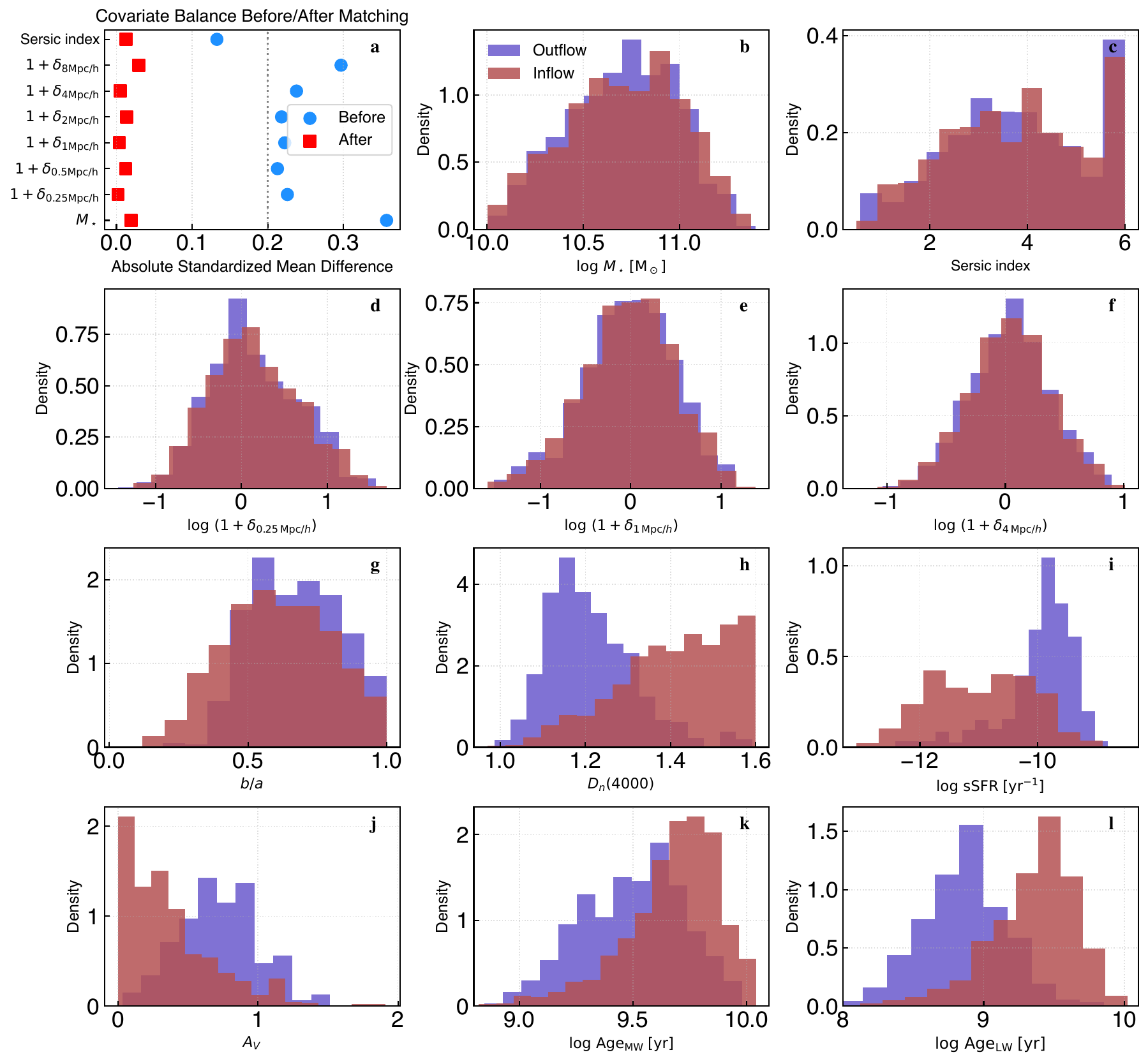}
    \caption{\textbf{Matched comparison of young inflow and outflow host galaxies}.\textbf{a}, the absolute standardized mean differences between the inflow and outflow samples before (blue circles) and after (red squares) matching, with the vertical dashed line indicating the adopted balance criterion of 0.2. \textbf{b--f}, the distributions of the variables used in the matching procedure: stellar mass, S\'{e}rsic index, and projected mass overdensities measured on scales of 0.25, 1, and 4 Mpc\,$h^{-1}$. \textbf{g--l}, the distributions of galaxy properties not used in the matching: axis-ratio ($b/a$), $D_n(4000)$ index, specific star formation rate ($\log\,\mathrm{sSFR}\,[\mathrm{yr}^{-1}]$), visual extinction ($A_V$), mass- and light-weighted ages. Outflow hosts are shown in slate blue and inflow hosts in dark red. The matched samples are well balanced in stellar mass, morphology, and environment, enabling differences in stellar populations, star-formation activity, and dust attenuation to be assessed independently of these host-galaxy properties.}
    \label{fig:Suppl_LP_young_matched_var}
\end{figure}

If inflows and outflows are linked through short-timescale gas cycling, as envisioned in galactic fountain and recycling scenarios, the gas-flow state should evolve more rapidly than global galaxy properties such as stellar mass, morphology, and environment, while quantities that respond directly to recent gas supply and feedback, including star-formation activity and dust content, may vary substantially. Because individual galaxies cannot be observed at multiple stages of their evolution, statistical comparisons between galaxies with similar long-lived properties provide a means of testing whether different flow states are associated with distinct recent evolutionary histories. At the same time, gas-flow signatures are known to depend on host-galaxy properties and viewing geometry, raising the possibility that differences between inflow and outflow hosts may reflect differences in the underlying galaxy populations. To isolate the connection between gas-flow state and recent galaxy evolution, and to test whether the observed inflow and outflow populations are consistent with expectations from short-timescale gas cycling, we construct matched samples of young galaxies ($D_n(4000) < 1.6$) hosting inflows and outflows that are closely matched in stellar mass, morphology, and environment (Supplementary Fig.~\ref{fig:Suppl_LP_young_matched_var}).

We matched each inflow galaxy to its nearest outflow counterpart in a standardized parameter space defined by stellar mass, S\'{e}rsic index, and projected mass overdensity measured on scales of 0.25--8\,Mpc. To compare galaxies with similar host properties, we required matches to lie within a maximum separation threshold in this multidimensional parameter space. We performed the matching with replacement, allowing an outflow galaxy to serve as the best match for more than one inflow galaxy when appropriate. The matching threshold was selected to retain the majority of the sample ($N=1,192$) while maintaining good balance in all matching variables; only $\sim$10\% of galaxies lacking sufficiently similar counterparts were excluded. After matching, all covariates had absolute standardized mean differences below 0.2, indicating good balance in stellar mass, morphology, and environment between the inflow and outflow samples (Supplementary Fig.~\ref{fig:Suppl_LP_young_matched_var}a--f). Note also that the axis ratio distributions (b/a; Supplementary Fig.~\ref{fig:Suppl_LP_young_matched_var}g) are broadly similar. The inferred trends are insensitive to reasonable variations in the matching threshold and to whether matching is performed with or without replacement. 

Despite the close agreement in stellar mass, morphology, and environment, inflow and outflow host galaxies exhibit systematically different stellar population and dust properties (Supplementary  Fig.~\ref{fig:Suppl_LP_young_matched_var}h--l). Inflow hosts have larger $D_n(4000)$ values, older stellar populations, lower galaxy-wide specific star-formation rates, and lower dust attenuation than their matched outflow counterparts. The median (16th, 84th percentile) $D_n(4000)$ values are 1.2\,(1.1, 1.3) for outflow hosts and 1.4\,(1.3, 1.55) for inflow hosts. This difference is reflected in the stellar population ages derived from pPXF spectral fitting: the light-weighted $\log\,(\mathrm{Age/yr})$ values are 8.9\,(8.5, 9.2) and 9.4\,(9.0, 9.6), respectively. The corresponding $\log\,(\mathrm{sSFR/yr^{-1}})$ values are $-9.8\,(-10.4, -9.4)$ and $-10.9\,(-11.9, -10.0)$, placing a substantially larger fraction of inflow hosts in the green-valley regime. Similarly, the stellar-continuum dust attenuation decreases from $A_V=0.7\,(0.4, 1.0)$ in outflow hosts to $A_V=0.3\,(0.1, 0.7)$ in inflow hosts. 

\begin{figure}
    \centering
    \includegraphics[width=0.95\linewidth]{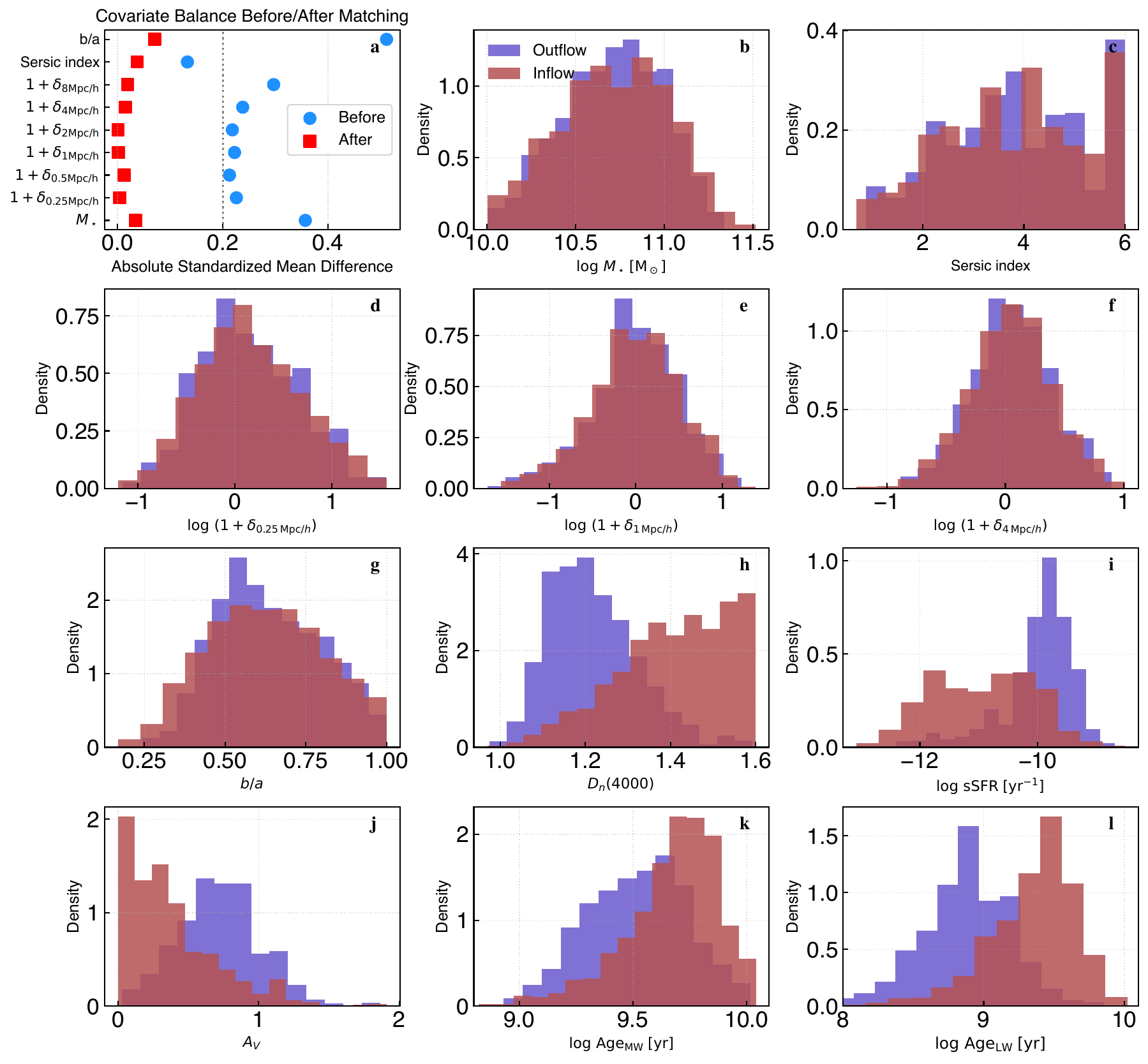}
    \caption{\textbf{Matched comparison of young inflow and outflow host galaxies with inclination matched}. The matching procedure now includes axis ratio ($b/a$) as an additional matching variable. Outflow hosts are shown in slate blue and inflow hosts in dark red. The matched samples demonstrate that differences in stellar populations, star-formation activity, and dust attenuation persist after controlling for galaxy inclination.}
    \label{fig:Suppl_LP_young_matched_incl}
\end{figure}

To assess the impact of viewing geometry, we repeat the matched-sample analysis by additionally matching galaxies in axial ratio ($b/a$), a proxy for inclination. The resulting matched sample ($N=1,181$) exhibits the same trends as the original analysis (Supplementary Fig.~\ref{fig:Suppl_LP_young_matched_incl}): inflow hosts continue to exhibit older stellar populations, lower specific star-formation rates, and lower dust attenuation than their matched outflow counterparts. These results indicate that projection effects influence the detectability of gas-flow signatures but cannot, by themselves, account for the observed differences between the two flow populations.

Although our snapshot observations do not allow us to follow the evolution of individual galaxies or measure transition times accurately between flow states, the observed differences in the matched young galaxies ($D_n(4000) < 1.6$) in stellar age, star-formation activity, and dust content indicate that gas-flow signatures are closely linked to a galaxy's recent evolutionary state than to stellar mass, environment, and structure. The results are broadly consistent with galactic fountain and gas-recycling scenarios in which outflows are preferentially associated with active, gas-rich phases, while inflows are found in systems with more evolved stellar populations and lower levels of ongoing star formation. We therefore interpret the observed inflow and outflow populations as reflecting different phases of a connected gas-cycling process linking gas supply and feedback. However, the data do not uniquely determine the evolutionary pathways connecting these states or exclude alternative interpretations.

\subsection*{Matched comparison of flow and no/weak-flow host galaxies}

For completeness, we repeat the matched-sample analysis using no/weak-flow galaxies as the comparison population. Unlike the inflow–outflow analysis, we do not impose a $D_n(4000)$ cut, allowing the full range of stellar population ages represented by the three flow classes to be examined. We match galaxies in stellar mass, S\'{e}rsic index, projected mass overdensity, and axial ratio to minimize differences arising from host-galaxy demographics and viewing geometry.

\begin{figure}
    \centering
    \includegraphics[width=0.95\linewidth]{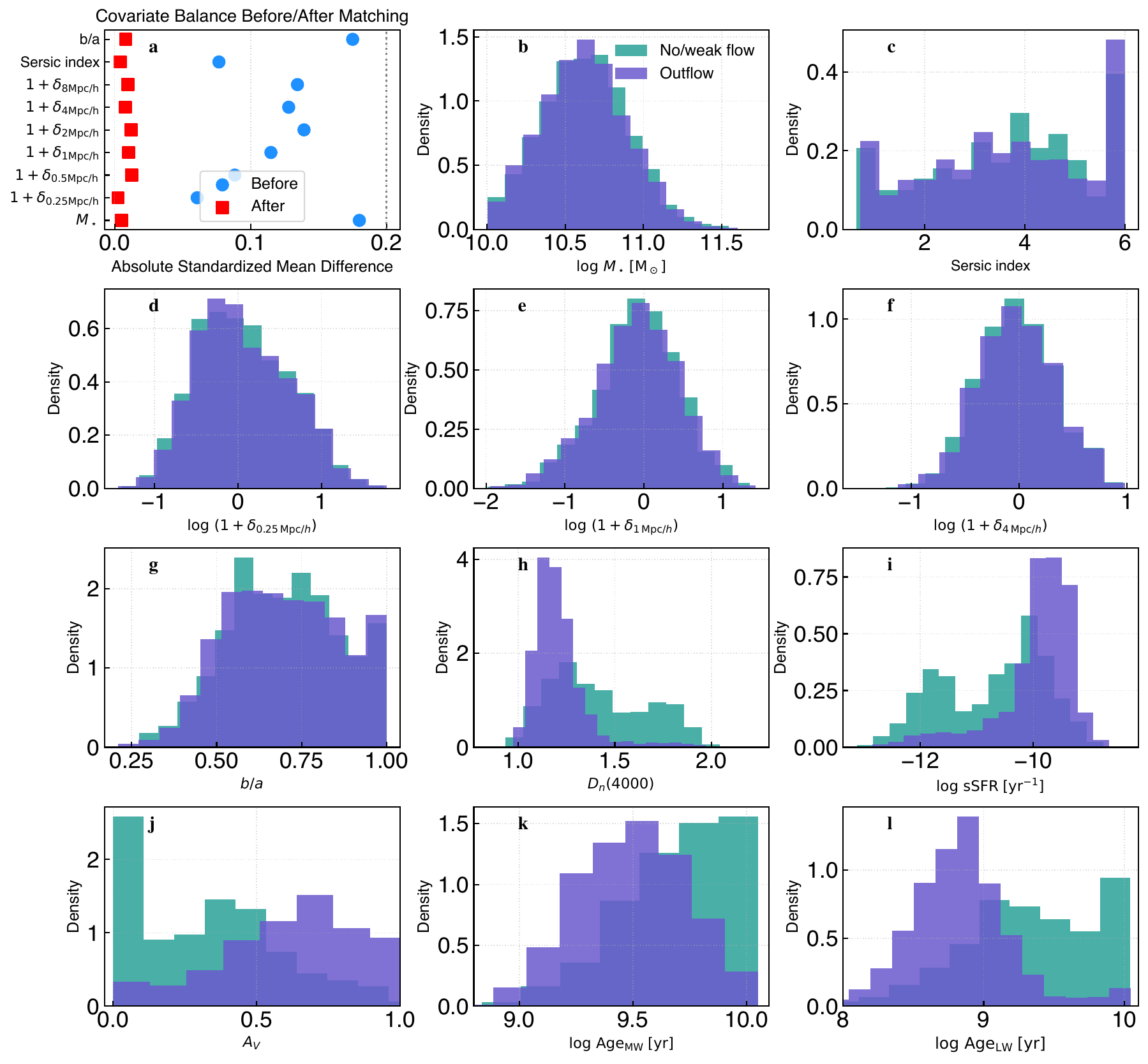}
    \caption{\textbf{Matched comparison of outflow to no/weak flow host galaxies}. Outflow hosts are shown in slate blue and no/weak flow hosts are shown in green.}
    \label{fig:Suppl_LP_outflow_noflow_matched}
\end{figure}

\begin{figure}
    \centering
    \includegraphics[width=0.95\linewidth]{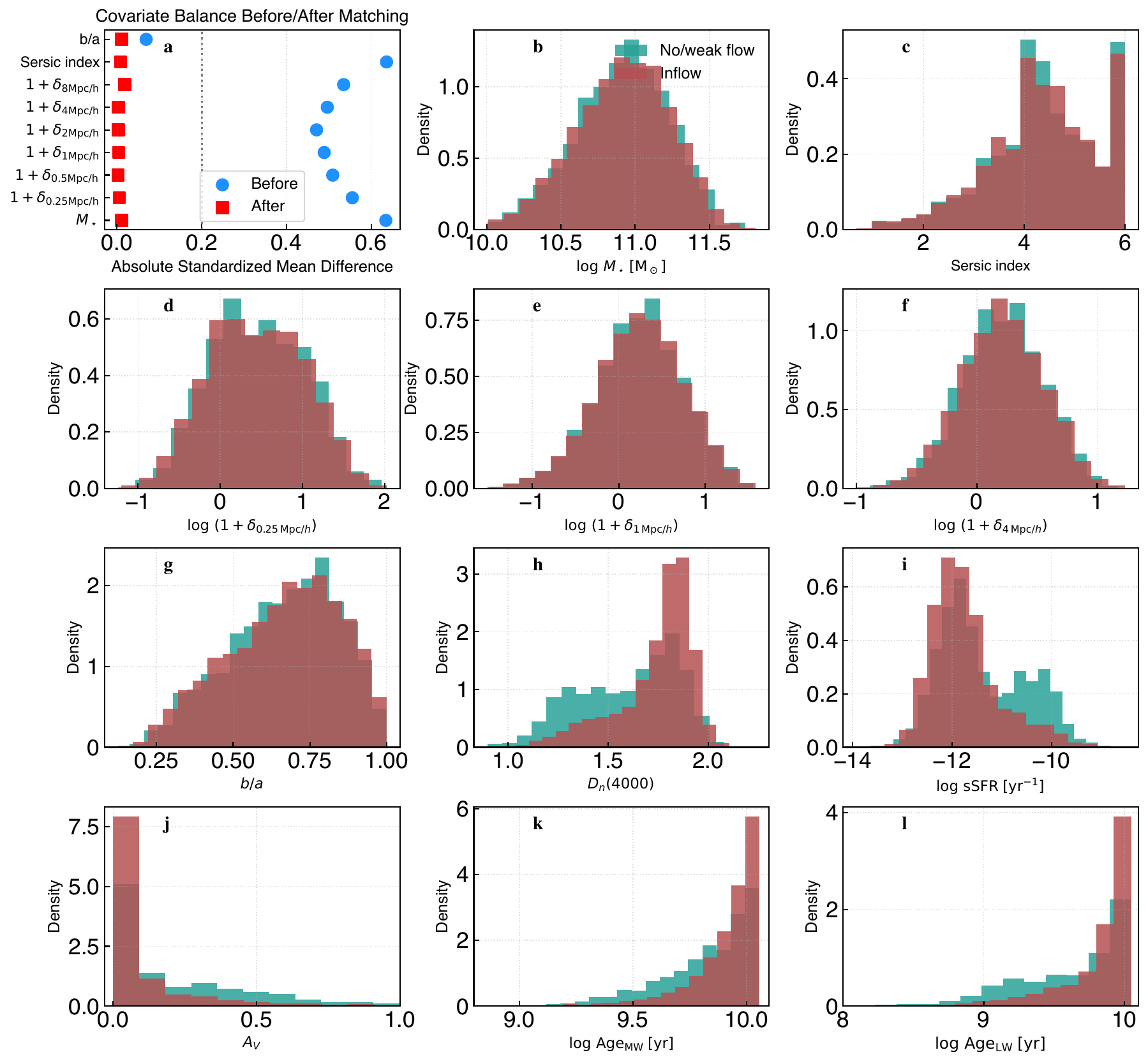}
    \caption{\textbf{Matched comparison of inflow to no/weak flow host galaxies}. Inflow hosts are shown in dark red and no/weak flow hosts are shown in green.}
    \label{fig:Suppl_LP_inflow_noflow_matched}
\end{figure}

Supplementary Fig.~\ref{fig:Suppl_LP_outflow_noflow_matched} shows that most outflow galaxies ($N=1,806$) can be closely matched to no/weak-flow galaxies in these global properties. Despite this matching, outflow hosts exhibit systematically younger stellar populations, higher specific star-formation rates, and larger dust attenuation than their no/weak-flow counterparts. Similarly, Supplementary Fig.~\ref{fig:Suppl_LP_inflow_noflow_matched} shows that most inflow galaxies ($N=5,647$) can be matched to no/weak-flow galaxies with comparable stellar mass, morphology, environment, and inclination. Inflow hosts remain systematically older, less actively star-forming, and less dust attenuated than the matched no/weak-flow population.

The matched distributions reveal a broad continuity among the three flow classes. Outflow hosts preferentially occupy the youngest, most actively star-forming, and dust-rich end of the distributions, whereas inflow hosts are skewed toward older stellar populations and lower levels of star formation. No/weak-flow galaxies span a wide range of properties and substantially overlap both populations. These results suggest that gas-flow signatures are more strongly associated with stellar population age, star-formation activity, and dust content than with global host-galaxy properties such as stellar mass, morphology, and environment. Future theoretical models that simultaneously predict gas-flow signatures and host-galaxy properties will provide a quantitative framework for interpreting these empirical distributions and for testing the physical interpretation we put forward here.

\subsection*{Star formation rate surface density and outflow connection}

We derived central star formation rate surface densities ($\Sigma_{\rm SFR}$) from dust-corrected H$\alpha$ luminosities measured within the 0.75$^{\prime\prime}$ DESI fibre radius. Strong AGNs are excluded using BPT diagram and requiring $S/N > 2$ for H$\alpha$ flux and other lines. The H$\alpha$ fluxes were corrected for dust attenuation using the visual extinction, $A_V$, derived from the Balmer decrement. For disk-dominated systems (S\'{e}rsic index $n<2.5$), the projected area was inclination-corrected by scaling with the photometric axis ratio $b/a$, assuming an intrinsically thin disk, such that $A=\pi R_{\rm fibre}^2/(b/a)$ for disks and $A=\pi R_{\rm fibre}^2$ otherwise. We converted H$\alpha$ luminosity surface density to $\Sigma_{\rm SFR}$ using the calibration of \cite{Kennicutt2012}, assuming a Kroupa initial mass function, $\Sigma_{\rm SFR}\, (\mathrm{M_\odot\,yr^{-1}\,kpc^{-2}})=5.37\times10^{-42}\,\Sigma_{L_{\rm H\alpha}}$. 

\begin{figure}
    \centering
    \includegraphics[width=0.85\linewidth]{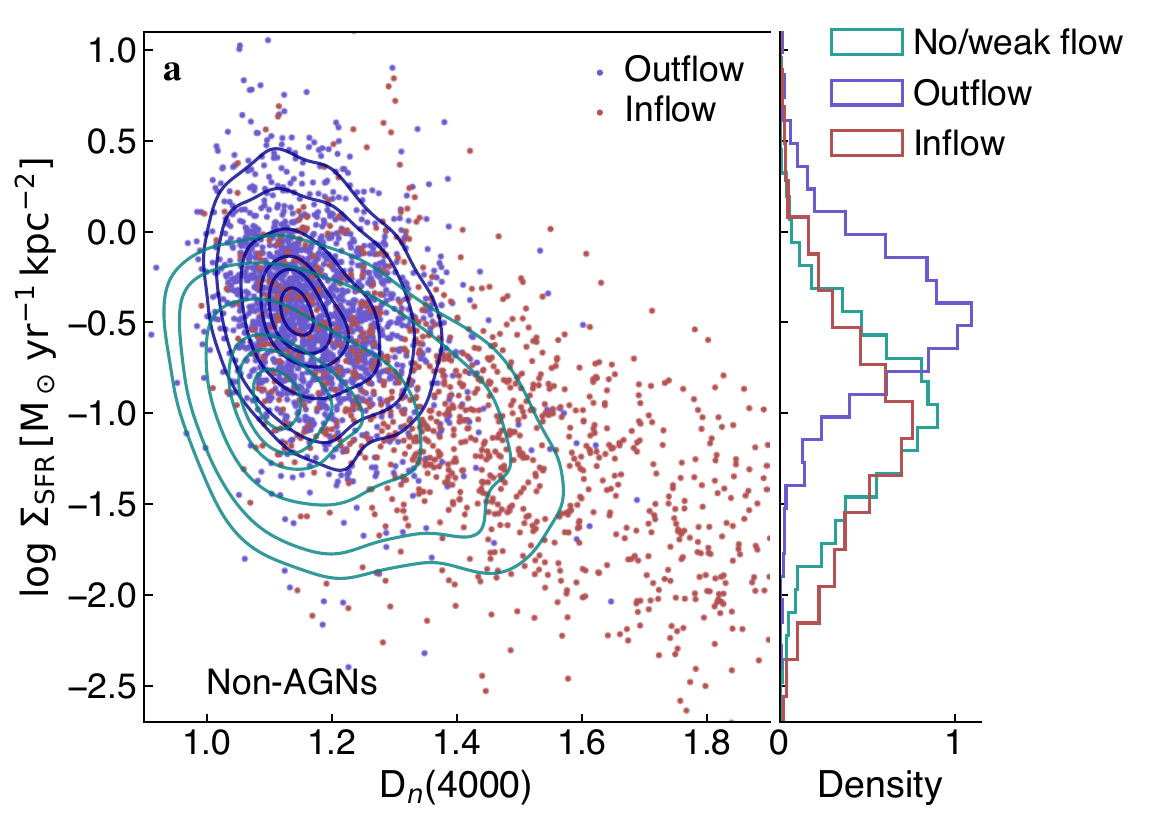}
    \caption{\textbf{Dependence of gas flows on central star formation rate surface density ($\Sigma_{\rm SFR}$) based on H$\alpha$}. Inflow ($N=1,028$), outflow ($N=1,750$), and no/weak-flow ($N=3,343$) non-AGN galaxies with detectable emission lines are shown in red, blue, and teal, respectively. Contours mark the 15th, 25th, 50th, 75th, and 85th percentiles of the distribution. Outflows are preferentially associated with elevated $\Sigma_{\rm SFR}$, exceeding the wind-launching threshold of $\Sigma_{\rm SFR}\,\approx$ 0.1 M$_{\odot}$ yr$^{-1}$ kpc$^{-2}$, underscoring their direct connection to enhanced central star formation.}
    \label{fig:sigmasfr}
\end{figure}

We find that the $\Sigma_{\rm SFR}$ distribution of outflow hosts differs significantly from those of no/weak-flow and inflow hosts (KS test, $p<10^{-6}$; Supplementary Fig.~\ref{fig:sigmasfr}). The median (16th–84th percentile) $\Sigma_{\rm SFR}$ of outflow hosts is $0.30\,(0.13,\,0.77)\,\mathrm{M_\odot\,yr^{-1}\,kpc^{-2}}$, compared to $0.10\,(0.03,\,0.27)$ and $0.07\,(0.02,\,0.22)\,\mathrm{M_\odot\,yr^{-1}\,kpc^{-2}}$ for no/weak-flow and inflow hosts, respectively.

The requirement for reliable dust-corrected H$\alpha$ measurements and the exclusion of AGN hosts reduces the sample size relative to the parent sample, particularly for inflow and no/weak-flow galaxies, which contain a larger fraction of systems with weak or undetected star formation. Consequently, this selection does not explain the elevated central star-formation surface densities observed in outflow hosts and, if anything, likely makes the measured differences conservative (Supplementary Fig.~\ref{fig:sigmasfr}).

The elevated $\Sigma_{\rm SFR}$ observed in outflow hosts places $\sim$90\% of them near or above the canonical threshold of $\Sigma_{\rm SFR}\sim0.1\,\mathrm{M_\odot\,yr^{-1}\,kpc^{-2}}$, widely identified as the regime where stellar feedback transitions from localized turbulence to large-scale galactic winds \citep{Heckman2000}. Above this limit, the collective energy and momentum injection from supernovae and massive stars can overcome gravitational confinement, driving multiphase outflows. In contrast, no/weak-flow systems predominantly lie below this threshold (with $\lesssim$50\% above it), consistent with star formation intensities insufficient to sustain coherent winds. The systematic separation in $\Sigma_{\rm SFR}$ between outflow and non-outflow populations therefore supports a physically regulated connection between central star formation intensity and wind launching, as expected in feedback-driven models of the baryon cycle.

Although observational selection effects, such as wind geometry and dust obscuration, do influence detectability, the intrinsic difference in star formation surface density appears comparably important. Structural features such as bars and spiral arms can funnel gas toward galactic centres, enhancing central star formation, chemical enrichment, and dust production, thereby both facilitating wind launching and increasing the likelihood of detecting dust-shielded outflows.

\begin{figure}
    \includegraphics[width=0.48\linewidth]{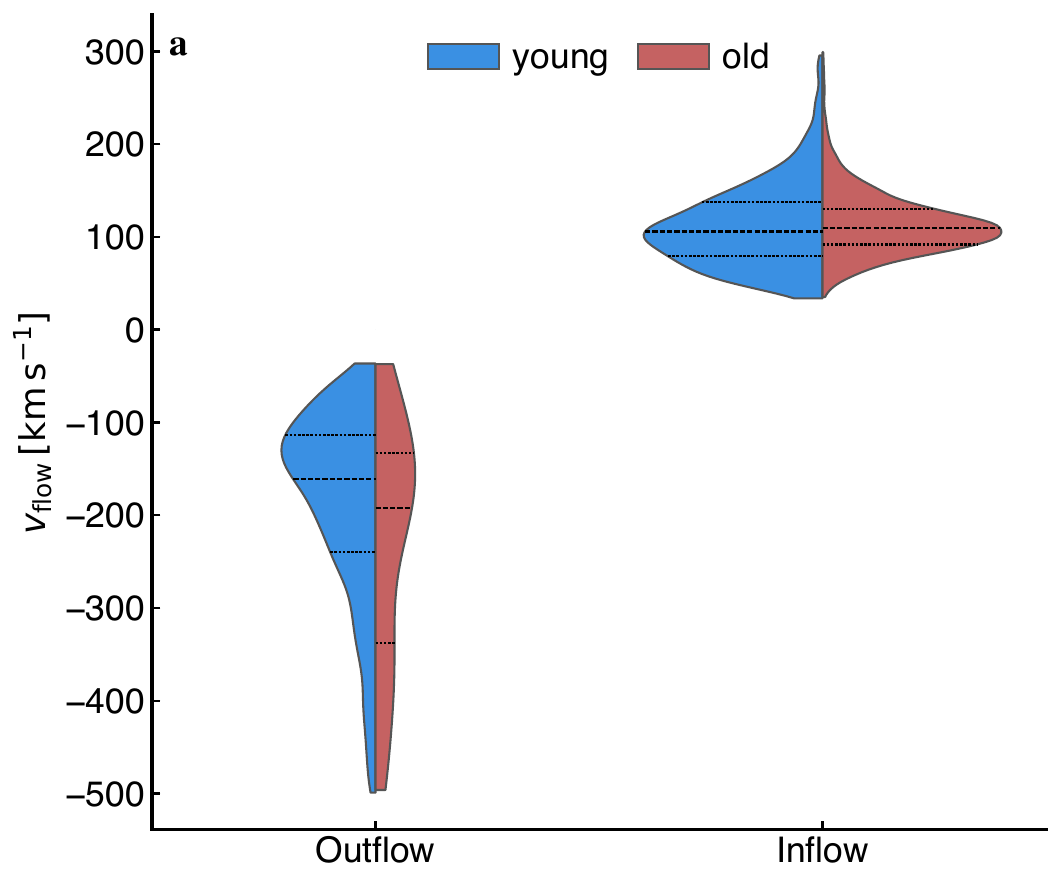}
    \includegraphics[width=0.48\linewidth]{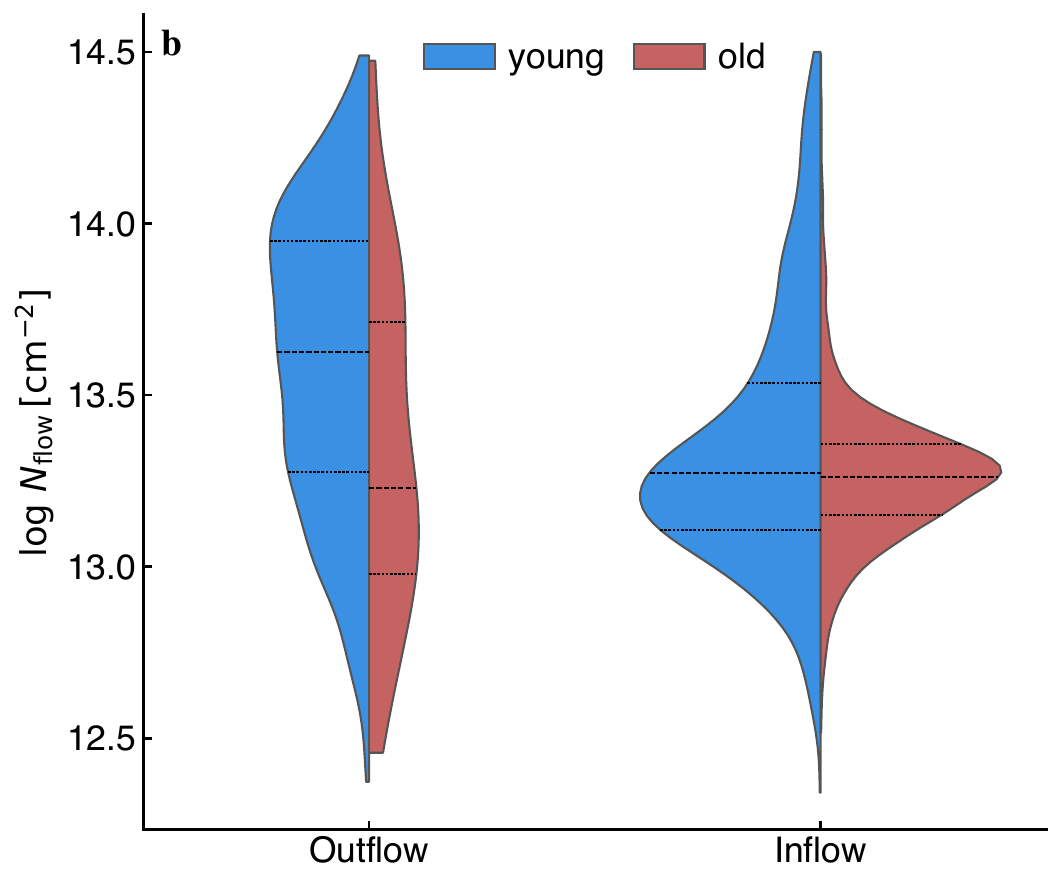}
    \includegraphics[width=0.48\linewidth]{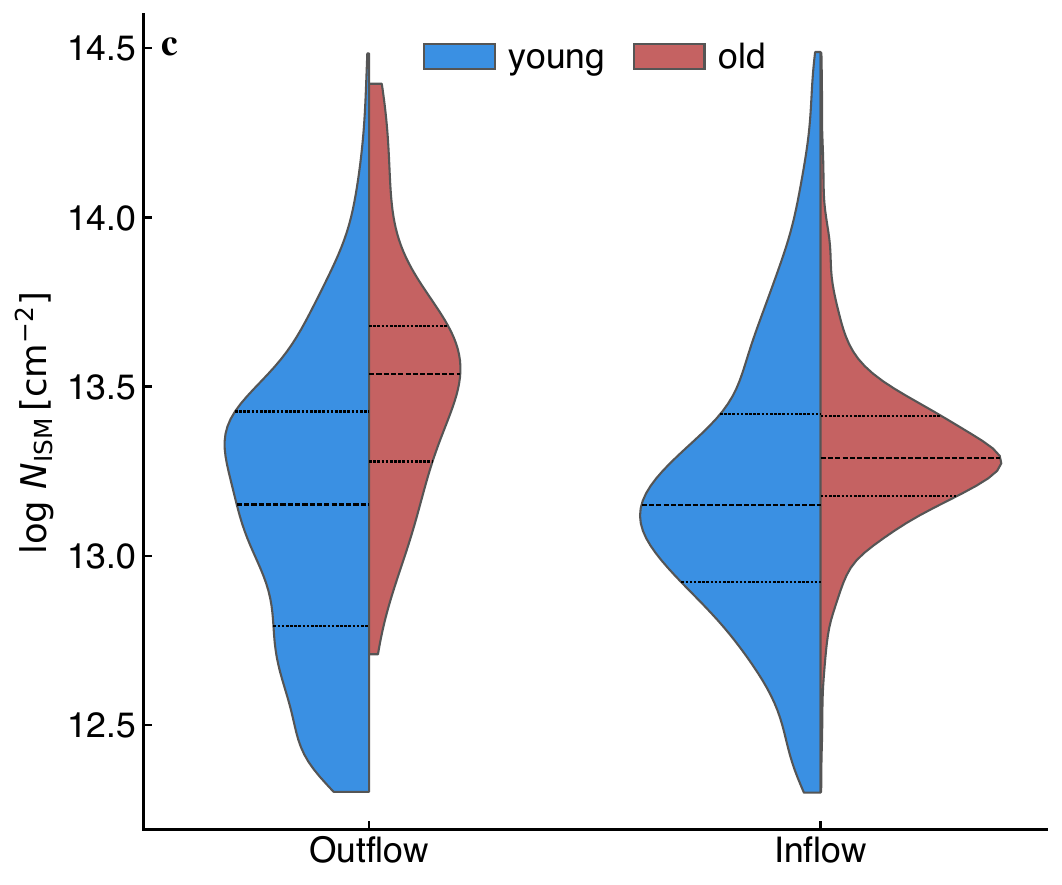}
    \includegraphics[width=0.48\linewidth]{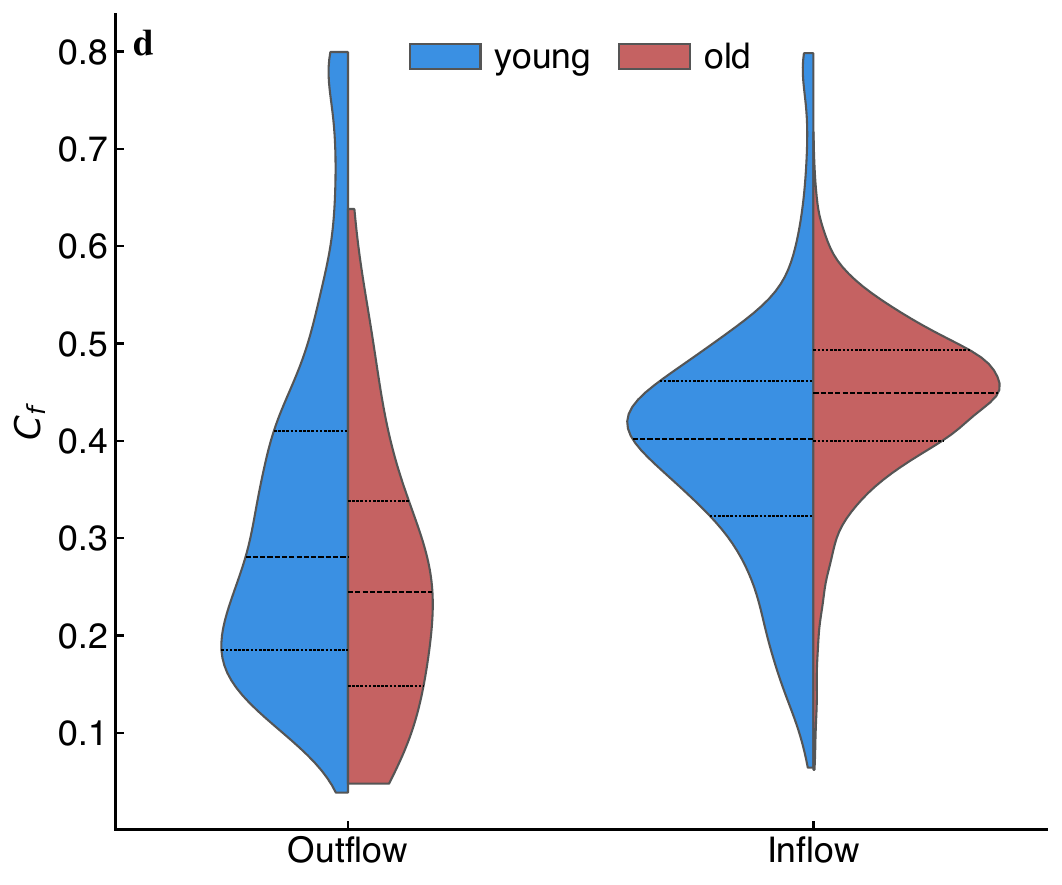}
    \includegraphics[width=0.48\linewidth]{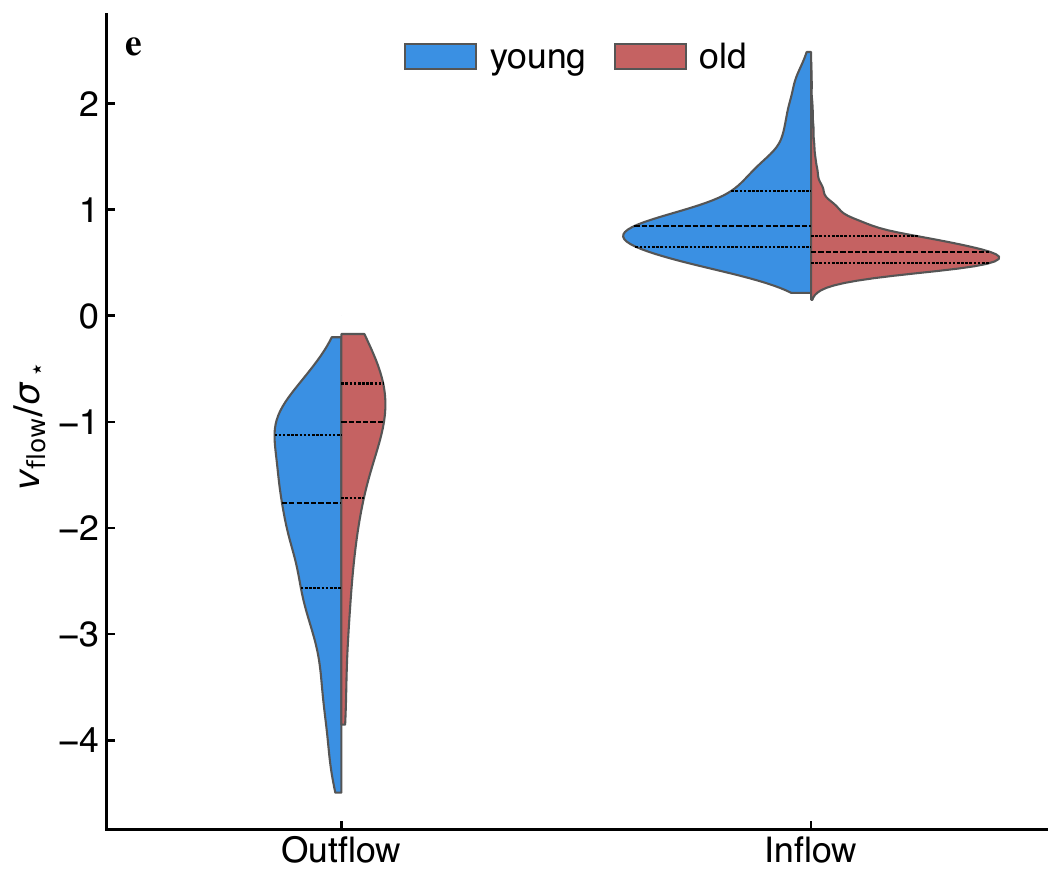}
    \includegraphics[width=0.48\linewidth]{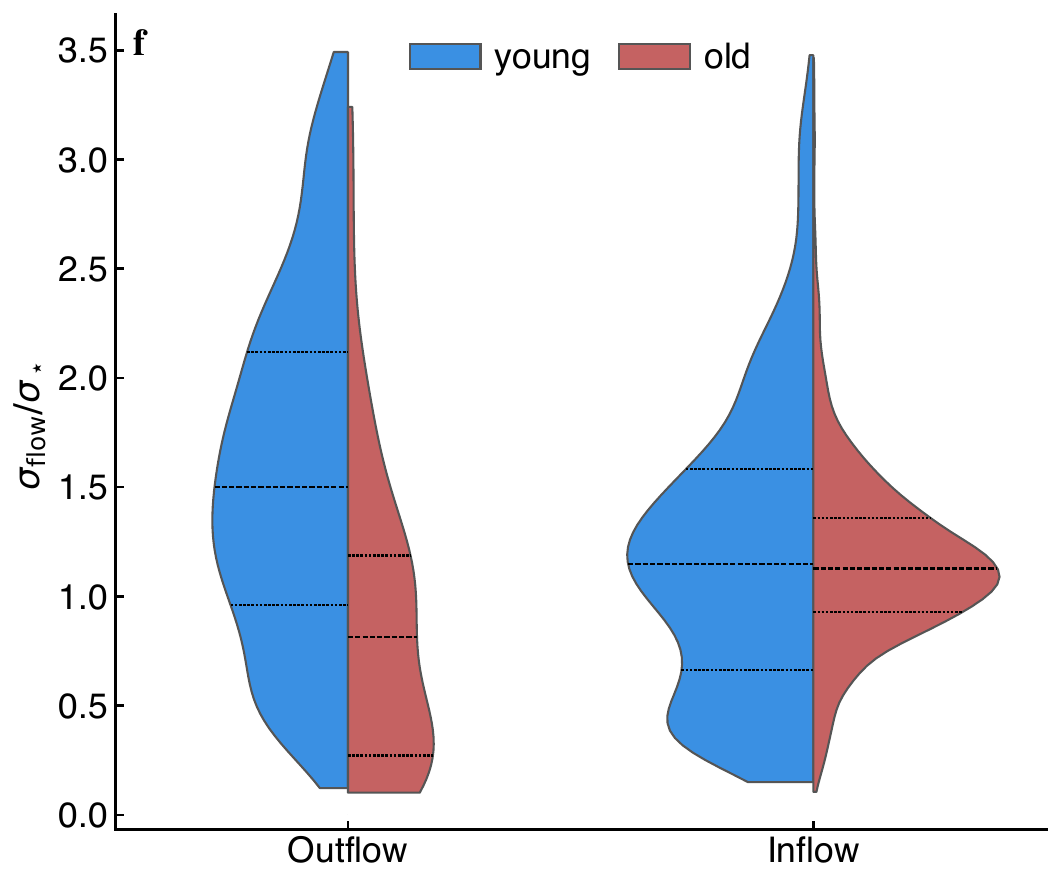}
    \caption{\textbf{Distributions of Na\,D-derived flow and ISM properties for young and old galaxies.} Violin plots show the distributions of the full sample ($\log M_\star/\mathrm{M_\odot} > 9.5$) separately by age. 
    \textbf{a}, Flow velocity ($v>0$ for inflow). \textbf{b}, Column density of the flow component. \textbf{c}, Column density of the ISM component. \textbf{d}, Local covering fraction of the flow component. \textbf{e}, Ratio of flow velocity to stellar velocity dispersion. \textbf{f}, Ratio of flow velocity dispersion to stellar velocity dispersion.}
    \label{fig:flowprop}
\end{figure}

\subsection*{Escape velocity and recycling}

Given that the median (85\%, 15\%) of outflow velocity of -165\,(-93, -297)\,$\mathrm{km\,s^{-1}}$ and median $|V_{\rm flow}|/\sigma_{\star}$ of $\sim2$ (Supplementary Fig.~6), most outflowing gas in our sample does not approach the escape velocity of a typical bulge potential.

For reference, the central velocity dispersion of the Milky Way bulge reaches $\sigma_{\star}\approx120-140\,\mathrm{km\,s^{-1}}$ in spectroscopic measurements of bulge stars \citep{Valenti+18}, and the escape velocity of Milky Way is estimated to be within $\sim500$--$600\,\mathrm{km\,s^{-1}}$ range \citep[][]{Roche+24}. 

Alternatively, we estimate the maximum circular velocity, $V_{\rm max}$, for each halo assuming a Navarro–Frenk–White (NFW) density profile \citep{Navarro+97} with fixed concentration $c=10$, consistent with $\Lambda$CDM expectations for $M_{200}\sim10^{12-13}\,M_\odot$ halos at $z\sim0$. Halo masses $M_{200}$ are taken from the group catalog \citep{YangX+20} and converted to $R_{200}$ assuming a spherical overdensity of 200 times the critical density. For an NFW halo, $V_{\rm max}$ is computed directly from $M_{200}$ and $c$.

We adopt $v_{\rm esc} \approx 2\,V_{\rm max}$ as a characteristic escape velocity. For NFW halos with realistic concentrations, the ratio between escape velocity and $V_{\rm max}$ is expected to lie in the range $\sim1.5$--$3$, depending on radius within the halo. Varying the proportionality factor within $1.5$--$3.0$ does not qualitatively affect our conclusions.

This rough estimate yields a median escape velocity of $\sim 470\,\mathrm{km\,s^{-1}}$ with 16\%–84\% percentiles of $(300, 850)\,\mathrm{km\,s^{-1}}$. Supplementary Fig.~7 shows the ratio of outflow velocity to escape velocity. Even when considering the $v_{85}$ velocity (defined as the sum of the median outflow velocity and its velocity dispersion), the median (16\%, 84\%) ratio is $0.6\,(0.3, 1.0)$, indicating that most of the outflows remain bound to the halo and only a small fraction enrich the intergalactic medium. The corresponding summary statistics for the ratio using median outflow velocity is 0.3\,(0.2, 0.6).

Our results suggest that galactic outflows in nearby star-forming galaxies are typically unable to escape their dark matter halos, instead remaining gravitationally bound and potentially contributing to later recycled accretion. However, a bound outflow does not necessarily imply directly observable return flows: expelled gas may mix with the hot halo, transition between gas phases, or become spatially diffuse before re-accreting onto the galaxy. We therefore do not assume a one-to-one correspondence between observed outflow and inflow phases, but instead interpret halo-bound outflows as a long-lived reservoir capable of sustaining subsequent recycled inflow.

This picture of bound, low-velocity feedback is also supported by observations at high redshift. Ref.~\citep{Bevacqua+26} recently reported neutral gas inflow into a massive, quiescent post-starburst galaxy at $z \approx 2.7$, with Na\,\textsc{i}\,D inflow velocities of $\sim270$~km~s$^{-1}$, suggesting that gas recycling may be a generic feature of galaxy evolution rather than a peculiarity of the local Universe. Many other studies likewise report low-velocity outflows that remain gravitationally bound to galaxy halos at earlier cosmic epochs \citep[e.g.,][]{Rubin+14}. Future high-spectral-resolution observations of edge-on transition systems---including post-starburst and quiescent disks---will be crucial for establishing the prevalence of recycled inflowing gas in massive distant galaxies and for directly testing the gas recycling paradigm.

\begin{figure}
\includegraphics[width=0.85\linewidth]{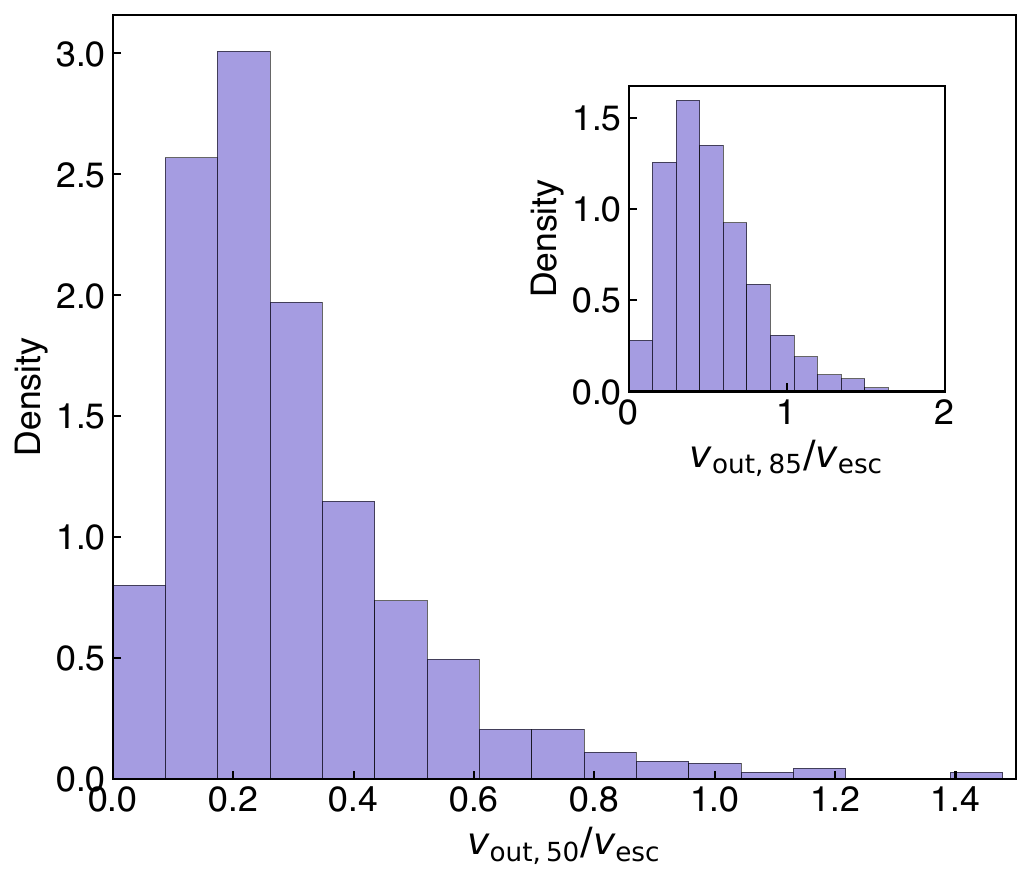}
\caption{\textbf{Comparison of outflow and escape velocities}. The main panel shows the distribution of the ratio between the median outflow velocity and the escape velocity. The inset shows the distribution of the ratio between the $v_{85}$ velocity (defined as the median outflow velocity plus the velocity dispersion) and the escape velocity, highlighting clouds moving significantly above the median ($1\sigma$).}
\end{figure}

\subsection*{Qualitative comparison with fountain and recycling models}

Several aspects of our observations are naturally interpreted within galactic fountain and gas recycling frameworks, in which gas expelled by stellar or AGN-driven outflows cools and later re-accretes onto galaxies over timescales $\sim 0.1-1$\,Gyr \citep[e.g.,][]{Bregman80, Oppenheimer+10, AnglesAlcazar2017,Fraternali17,Grand+19}. Stellar population analysis shows that inflow galaxies are offset toward older light-weighted ages by $\sim0.5$\,Gyr (Fig.~3, Extended Data Fig.~4 and 5) and commonly exhibit lower present-day star formation rates than matched control outflow hosts. This pattern is qualitatively consistent with a delayed gas return cycle operating in galaxies of otherwise similar global properties. The characteristic inflow velocities of $\sim100\,\mathrm{km\,s^{-1}}$ are also consistent with expectations for recycled gas moving within the halo potential and resemble velocities observed for Milky Way high- and intermediate-velocity clouds interpreted as fountain material \citep{Fraternali17, Marasco+22}.

Young galaxies in our sample exhibit both inflow and outflow signatures and predominantly reside in halos of $M_{\rm halo}\sim10^{12}$–$10^{13}\,M_\odot$, with a mean near $3\times10^{12}\,M_\odot$. This mass range corresponds to the regime where fountain-driven condensation of coronal gas is predicted to be most effective. Simulations show that metal-rich fountain clouds mixing with the hot circumgalactic medium can stimulate cooling and accretion in halos below a few $\times10^{12}\,M_\odot$, whereas this process becomes progressively less efficient and largely suppressed above $\sim10^{13}\,M_\odot$ \citep[e.g.,][]{Armillotta+16, Fraternali17}.

In these models, most of the accreting gas is not pristine but recycled, metal-enriched material that has mixed with the circumgalactic medium, resulting in metallicities slightly below solar yet well above the intergalactic medium \citep{Fraternali17, Muratov+17, Christensen+18, Grand+19}. The central ISM metallicities of our observed galaxies are near solar within a factor of two (Extended Data Fig.~7a; Supplementary Fig.~9), consistent with the return of enriched gas and the absence of substantial dilution by pristine inflows, which may instead occur at larger radii in the disk.

The trajectories of fountain material are unlikely to follow simple ballistic up-and-down paths near their launch sites. On their return, cool fountain clouds can interact with hot outflows, which may deflect their motion and redistribute them over larger halo volumes. After mixing with the ambient CGM, portions of this gas can cool, condense, and join circulating cold streams before re-accreting onto the disk at locations that may be offset from their original launch regions. In addition, the metallicity structure of the disk–halo interface and inner CGM is expected to be anisotropic \citep[e.g.,][]{ChoiB+16}, with enhanced enrichment along preferred outflow channels. These processes naturally lead to a broad spread in the metallicities of recycled inflowing gas. Gas-rich minor mergers may further contribute to this diversity.

\subsection*{Column densities of inflowing, outflowing, and ISM gas}

The mean Na\,\textsc{i} column densities measured from the two-component fits are
$\log N_{\rm Na\,I} = 13.6$ for outflows,
$\log N_{\rm Na\,I} = 13.3$ for inflows, and
$\log N_{\rm Na\,I} \simeq 13.1$ for the static ISM component (see also Supplementary Fig.~6b and c)
Following standard practice \citep[][]{Rupke+05,YesufH+17}, we convert Na\,\textsc{i} to total hydrogen column density using
\begin{equation}
N_{\rm H}
\approx
2.5\times10^{20}
\left(\frac{N_{\rm Na\,I}}{10^{13}\,{\rm cm^{-2}}}\right)
\left(\frac{0.2}{f_{\rm Na\,I}}\right)
\left(\frac{10^{-1}}{10^{\delta_{\rm Na}}}\right)
\left(\frac{2\times10^{-6}}{\rm (Na/H)_\odot}\right)
\ {\rm cm^{-2}}.
\end{equation}

where $f_{\rm Na\,I}$ is the fraction of neutral sodium, $\delta_{\rm Na}$ is the depletion factor onto dust of sodium, and (Na/H)$_\odot = 2\times10^{-6}$ is solar abundance of sodium. Adopting reasonable values, $f_{\rm Na\,I}=0.2$, $\delta_{\rm Na}=-1$ \citep{Murray+07, SavageSembach96} we obtain
\[
\log N_{\rm H} \approx 21 \quad \text{(outflow)}, \qquad
\log N_{\rm H} \approx 20.8 \quad \text{(inflow)}, \qquad
\log N_{\rm H} \approx 20.4 \quad \text{(ISM)}.
\]
Thus, both inflowing and outflowing material trace classical neutral ISM column densities, with only modest enhancement in the outflowing component.
These values are consistent with direct H\,\textsc{i} measurements in nearby galaxies, where column densities of order $N_{\rm H}\sim10^{20-21}\,{\rm cm^{-2}}$ are commonly observed \citep{Serra+12}. 

\subsection*{Estimating the global covering factor of gas flows}

The global covering factor, $C_\Omega$, quantifies the fraction of $4\pi$ steradians subtended by inflowing or outflowing gas around galaxies. Along a given sightline, the probability of detecting a flow is approximately the product of the global covering factor and the line-of-sight (local) covering fraction, $\langle C_f \rangle$, estimated from absorption-line modeling (Supplementary Fig.~6d).

\begin{equation}
P_{\rm detect} = C_\Omega \, C_f.
\end{equation}

Galaxies are divided into edge-on ($i \ge i_{\rm thresh}=60^\circ$) and face-on sub-samples. For each flow type, we define the parent sample and measure the observed detection fractions:

\begin{equation}
f_{\rm edge} = \frac{N_{\rm edge}^{\rm detect}}{N_{\rm edge}^{\rm parent}}, \qquad
f_{\rm face} = \frac{N_{\rm face}^{\rm detect}}{N_{\rm face}^{\rm parent}},
\end{equation}

where $N_{\rm edge}^{\rm detect}$ and $N_{\rm face}^{\rm detect}$ are the numbers of galaxies with detectable absorption in the edge-on and face-on sub-samples, respectively, and $N_{\rm edge}^{\rm parent}$ and $N_{\rm face}^{\rm parent}$ are the total numbers of galaxies in each orientation bin.

The overall detection probability in the full sample is then the weighted mean of the two orientation bins:

\begin{equation}
P_{\rm detect, total} = 
\frac{ N_{\rm edge}^{\rm parent} f_{\rm edge} + N_{\rm face}^{\rm parent} f_{\rm face} }{ N_{\rm edge}^{\rm parent} + N_{\rm face}^{\rm parent} }.
\end{equation}

Assuming that the local covering fraction $\langle C_f \rangle$ is independent of orientation, the global covering factor can be estimated as:

\begin{equation}
\boxed{
C_\Omega \approx \frac{ N_{\rm edge}^{\rm parent} f_{\rm edge} + N_{\rm face}^{\rm parent} f_{\rm face} }{ \langle C_f \rangle \, (N_{\rm edge}^{\rm parent} + N_{\rm face}^{\rm parent}) } }.
\end{equation}

This formulation reduces to the simple average, $C_\Omega \approx (f_{\rm edge} + f_{\rm face}) / (2 \langle C_f \rangle)$, when the edge-on and face-on parent samples are equal. In the special case of randomly oriented galaxies with equivalent sightlines, the detection probability is independent of inclination. Denoting the total detection fraction as 
\(
f_{\rm total} = \frac{N_{\rm detect}}{N_{\rm total}},
\) 
the weighted formula reduces to 
\(
C_\Omega \approx \frac{f_{\rm total}}{\langle C_f \rangle},
\) 
i.e., the global covering factor is simply the overall fraction of detections divided by the mean line-of-sight covering fraction.

For our sample, in young disk galaxies ($\rm D_n(4000) < 1.6$), inflows are detected preferentially along edge-on sightlines ($f_{\rm edge} \simeq 0.2$, $f_{\rm face} \simeq 0.1$) with a mean line-of-sight covering fraction $\langle C_f \rangle \simeq 0.4$, yielding $C_\Omega \simeq 0.3$. Outflows, in contrast, are more frequently detected along face-on sightlines ($f_{\rm edge} \simeq 0.1$, $f_{\rm face} \simeq 0.2$; $\langle C_f \rangle \simeq 0.3$), giving $C_\Omega \simeq 0.56$. In old disk galaxies ($\rm D_n(4000) > 1.6$), inflows appear more isotropic ($f_{\rm edge} \simeq 0.7$, $f_{\rm face} \simeq 0.6$), producing $C_\Omega \simeq 1.4$, while outflows are rare ($f_{\rm edge} \simeq 0$, $f_{\rm face} \simeq 0.03$; $C_\Omega \simeq 0.03$). These trends indicate that inflows in old galaxies are distributed more isotropically, consistent with stellar mass-loss-driven or clumpy cooling flows, whereas young galaxies show planar inflows and polar outflows associated with ongoing star formation and feedback.  

We emphasize that these global covering factor estimates are approximate. The values above are based on strong flows only; including flows with lower detection probability (e.g., $>0.6$) increases $C_\Omega$ to $\sim 0.6$ for inflows and $\sim 0.9$ for outflows in young disk galaxies. In addition, inclination estimates from axis ratios are uncertain for quiescent galaxies, which typically lack disks, and the mean line-of-sight covering fraction, $\langle C_f \rangle$, may be underestimated or fail to capture its variance. For example, using the 84th percentile of the posterior distribution for $C_f$ gives a mean of $\sim 0.7$, which would bring the inferred $C_\Omega$ closer to unity ($\sim 0.94$).

\subsection*{Consistency with 21\,cm hydrogen gas observations}

A complementary constraint on the product of the local and global covering factors, $C_f C_\Omega$, can be obtained by comparing the Na\,\textsc{i}–inferred column densities with those implied by 21\,cm observations. Using the HI mass–size relation \citep{WangJ+16}, 
\[
\log D_{\rm HI}/{\rm kpc} = 0.506 \log M_{\rm HI}/M_\odot - 3.293,
\] 
we can compute characteristic HI radii for a given mass. For early–type galaxies with $M_{\rm HI} \sim 10^8$–$10^9\,M_\odot$, this yields $R_{\rm HI} \sim 3$–10\,kpc and area-averaged neutral column densities of order $N_{\rm H}^{\rm (21cm)} \sim 5\times10^{20}\,{\rm cm^{-2}}$. Star–forming spirals with larger HI reservoirs ($M_{\rm HI} \sim 3\times10^{9}$–$10^{10}\,M_\odot$) have $R_{\rm HI} \sim 15$–30\,kpc, giving similar area-averaged columns $\langle N_{\rm H}\rangle_{\rm 21cm} \sim 5\times10^{20}\,{\rm cm^{-2}}$. These estimates are broadly consistent with deep, resolved 21\,cm measurements from the ATLAS$^{\rm 3D}$ survey \citep[][their Fig.~10]{Serra+12}, which report slightly lower mean columns for early–type galaxies (by roughly a factor of two), with the upper range of $N_{\rm H}^{\rm (21cm)} \sim 3-5\times10^{20}\,{\rm cm^{-2}}$ in agreement with our scaling.

Comparing with the column densities inferred from Na\,\textsc{i} absorption along detected sightlines 
($N_{\rm H}^{\rm (Na)}\sim 5\times10^{20}\,{\rm cm^{-2}}$ for early–type inflows, and $\sim 10^{21}\,{\rm cm^{-2}}$ for star–forming outflows), the ratio
\[
C_f\,C_\Omega \;\simeq\; \frac{N_{\rm H}^{\rm (21cm)}}{N_{\rm H}^{\rm (Na)}}
\]
provides an independent estimate of the product of the local and global covering factors. For early–type (old inflow) galaxies, this ratio is $\sim0.1$–$0.6$, in good agreement with the covering factors derived independently from inclination statistics and $C_f$ measurements. For star–forming (outflow) galaxies, the area-averaged H\,\textsc{i} columns imply $C_f\,C_\Omega \sim 0.1$–$1$, potentially larger than the Na\,\textsc{i}–derived covering product from absorption statistics ($\sim0.17$). This suggests that a large fraction of the neutral gas may reside in more diffuse or geometrically extended structures not efficiently traced by Na\,\textsc{i} absorption. Alternatively, the discrepancy may arise from systematic uncertainties in the Na–to–H column conversion (e.g., higher Na\,\textsc{i} fractions or weaker depletion than assumed in Eq.~1). These estimates are statistical in nature and broadly reasonable given the observed galaxy-to-galaxy variations.

\subsection*{Gas content and gas flow states}

To examine how atomic gas content relates to gas-flow classifications, we cross-matched our sample with major H\,\textsc{i} surveys: ALFALFA \citep{Haynes+18}, xGASS \citep{Catinella+18}, and H\,\textsc{i}-MaNGA \citep{Masters+19}. For galaxies classified as outflow or no/weak-flow systems, H\,\textsc{i} measurements from all three surveys are used, while inflow galaxies are primarily drawn from xGASS and H\,\textsc{i}-MaNGA owing to their greater depth and more uniform stellar-mass selection. The final cross-matched sample comprises 622 no/weak-flow galaxies, 217 high-confidence inflow hosts, and 126 high-confidence outflow hosts.

Extended Data Fig.~8c presents Kaplan–Meier cumulative distribution functions of the H\,\textsc{i} gas fraction ($M_{\rm HI}/M_\star$), incorporating non-detections as upper limits. For reference, we also show the full xGASS sample divided into young and old stellar populations. Outflow and no/weak-flow galaxies are predominantly H\,\textsc{i}-rich, with gas-fraction distributions similar to those of young xGASS galaxies. In contrast, inflow galaxies exhibit systematically lower H\,\textsc{i} fractions, comparable to those of older, more quiescent systems in xGASS.

Nevertheless, a non-negligible subset of inflow hosts remains gas rich: approximately 20\% have $M_{\rm HI}/M_\star > 0.1$, similar to the fraction observed among old xGASS galaxies. This fraction rises to $\sim$30\% when including lower-significance inflow candidates and to nearly $\sim$50\% when incorporating additional ALFALFA detections among high-confidence inflow systems. Because ALFALFA is a relatively shallow, flux-limited survey, it preferentially detects the most H\,\textsc{i}-rich galaxies at fixed stellar mass and provides fewer constraining upper limits for gas-poor systems \citep{Haynes+18}. Its inclusion therefore biases the detected subsample toward higher gas fractions and amplifies the apparent gas-rich tail among inflow hosts.

Overall, these results indicate that although inflow hosts are typically H\,\textsc{i}-poor on average, a substantial minority retain significant atomic gas reservoirs. This is consistent with a scenario in which some inflow systems continue to accrete or recycle gas while their global star formation activity is low.

\subsection*{Mass inflow and outflow rates}

We estimate the mass flux of neutral gas in inflows and outflows using the Na\,\textsc{i} column densities derived from our two-component absorption fits, the characteristic radius of the absorbing gas, and the line-of-sight and global covering factors. Following \citep{Rupke+05}, the mass outflow (or inflow) rate can be expressed as
\begin{equation}
\dot{M} = 1.4 \, m_{\rm H} \, N_{\rm H} \, v \, R\,4 \pi\, C_f \, C_\Omega,
\end{equation}
where $v$ is the flow velocity, $R$ is the characteristic radius of the absorbing gas.

\begin{equation}
\dot{M} \;\simeq\; 1\,M_\odot\,{\rm yr^{-1}}
\left(\frac{N_{\rm H}}{10^{20}\,{\rm cm^{-2}}}\right)
\left(\frac{v}{100\,{\rm km\,s^{-1}}}\right)
\left(\frac{R}{2\,{\rm kpc}}\right)
\left(\frac{C_f}{0.4}\right)
\left(\frac{C_\Omega}{1}\right),
\end{equation}

For old inflow galaxies with $\rm D_n(4000)>1.6$, we adopt $R\sim 2$\,kpc (corresponding to the 1.5$\arcsec$\ fiber radius), $N_{\rm H} \sim 1-5\times10^{20}\,{\rm cm^{-2}}$, mean $v\sim100$\,km\,s$^{-1}$, $C_f \sim 0.4$ and $C_\Omega \sim 0.3-1$, yielding a median inflow rate $\dot{M}_{\rm inflow} \sim 0.3$–$5\,{\rm M_\odot\, yr^{-1}}$. This is broadly comparable to the expected stellar mass-loss rate from evolved stars in massive galaxies \citep[][]{LeitnerKravtsov11,Pellegrini12}. This consistency is suggestive of a scenario in which recycled stellar ejecta contributes significantly to the cold gas reservoir associated with the observed inflow signatures---a picture supported by MACER simulations (Supplementary Figs.~2 and 3). We note, however, that additional gas sources (e.g., minor mergers and intergalactic medium accretion) may also contribute, and the observational constraints on the fraction of stellar ejecta that joins the hot atmosphere versus remains in the galaxy are uncertain.

In young, star-forming galaxies with $\rm D_n(4000)<1.6$, outflows are observed with mean velocities ($v\sim 200$\,km\,s$^{-1}$) and slightly larger column densities ($N_{\rm H} \sim 0.5-1 \times 10^{21}\,{\rm cm^{-2}}$). Adopting $R\sim2$\,kpc, $C_f \sim 0.3$, and $C_\Omega \sim 0.5$, we find average mass outflow rates $\dot{M}_{\rm outflow} \sim 4$–$8\,{\rm M_\odot\, yr^{-1}}$, consistent with modest star-formation-driven winds in these systems. 

We note that the mass fluxes for both inflows and outflows are rough estimates and are uncertain by factors of a few due to systematic uncertainties in $N_{\rm H}$, $R$, $C_f$, and $C_\Omega$, as well as assumptions about the flow geometry. Nevertheless, the derived rates indicate that inflows in quiescent galaxies are largely consistent with stellar mass-loss feeding, whereas outflows in young galaxies are plausibly driven by a combination of star formation and AGN feedback.

\begin{figure}
\includegraphics[width=0.49\linewidth]{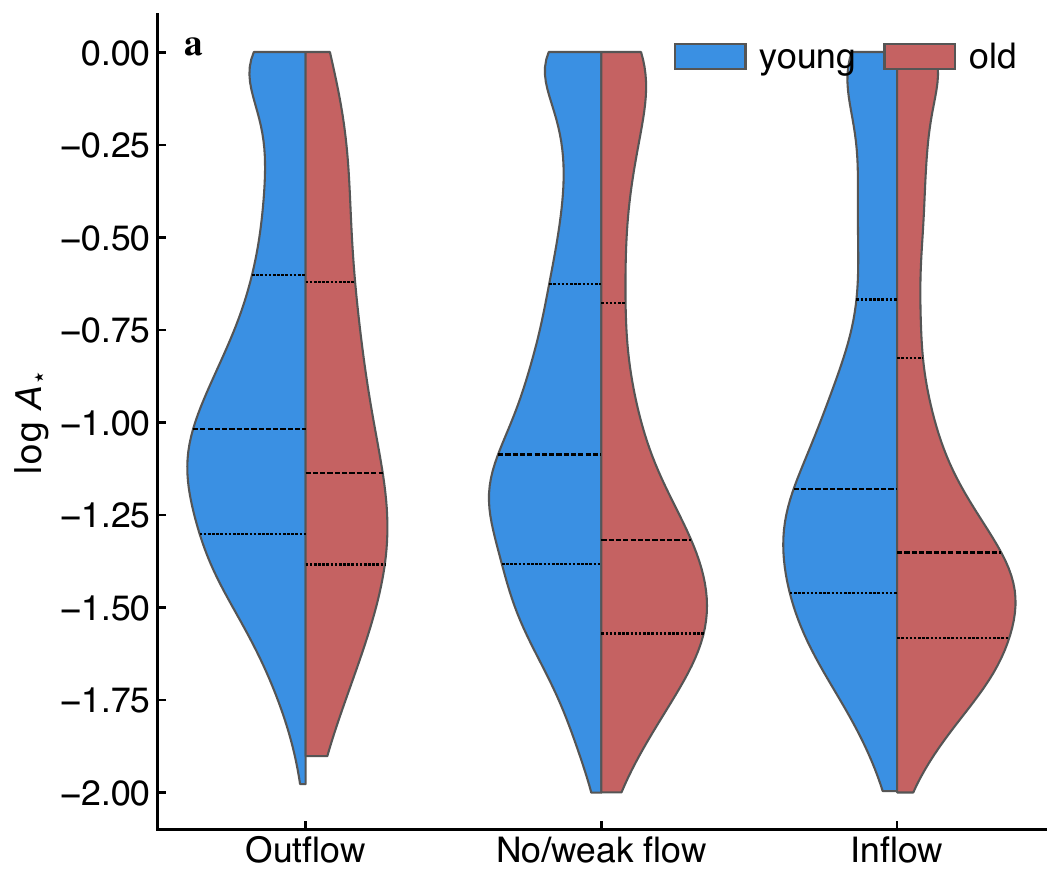}
\includegraphics[width=0.49\linewidth]{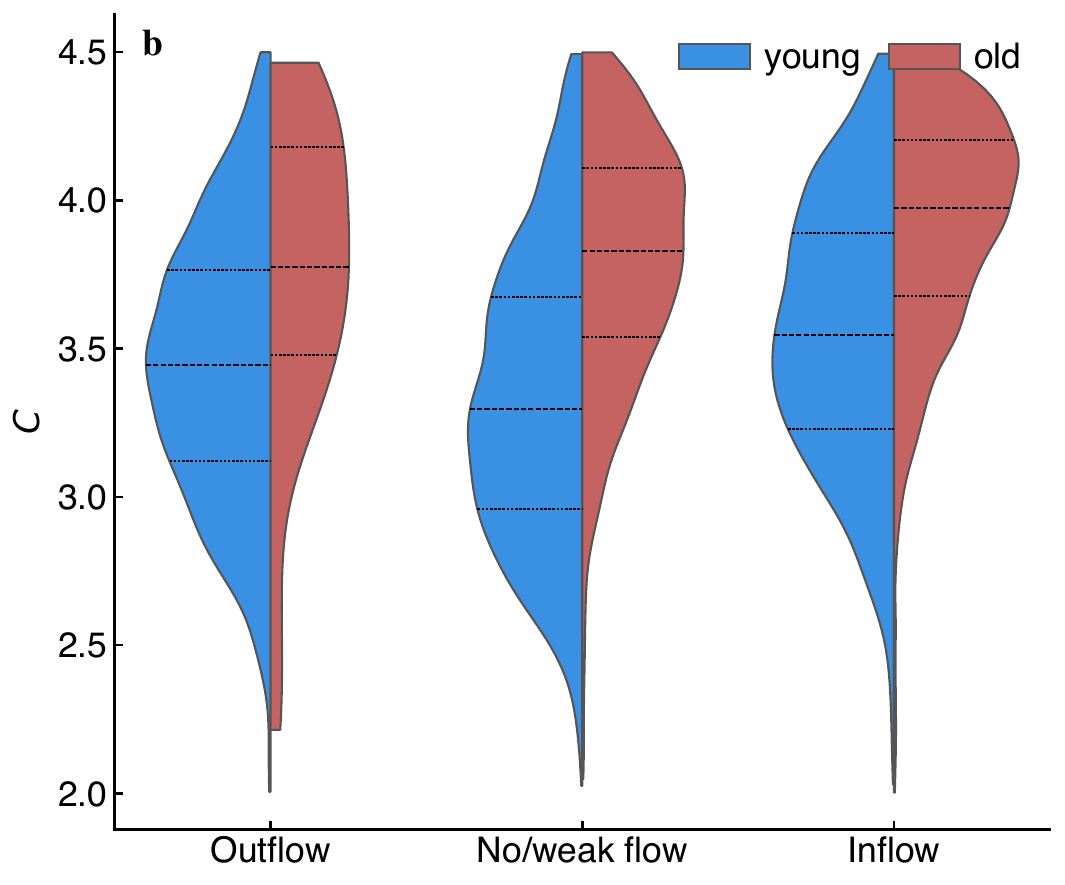}
\caption{\textbf{Distribution of asymmetry and concentration at fixed stellar population age.}
\textbf{a,} Distribution of asymmetry for galaxies divided by gas flow state (inflow, no/weak flow, and outflow) and stellar population age, traced by the $D_n(4000)$ index. 
\textbf{b,} Corresponding distribution of concentration index for the same populations. The sample is restricted to galaxies with $\log(M_\star/\mathrm{M_\odot}) > 10$, and asymmetry and concentration measurements are available for nearly all galaxies.}
\label{fig:CA_VP}
\end{figure}

\subsection*{Details on galaxy structures and gas-flow connections}

Bars and spiral structure are difficult to identify in edge-on galaxies. Because gas flows are anisotropic and their detectability depends on orientation, we restrict this analysis to young, face-on disk galaxies ($n < 2.5$ and inclination $i < 60^\circ$). Within this subset, outflow hosts show higher morphological feature fractions (bar fraction $45 \pm 3\%$; spiral fraction $65 \pm 3\%$) than inflow hosts (bar fraction $27 \pm 5\%$; spiral fraction $39 \pm 5\%$) and no/weak-flow systems (bar fraction $31 \pm 2\%$; spiral fraction $51 \pm 2\%$). 

The quoted uncertainties reflect Poisson--Binomial counting statistics only and do not include systematic uncertainties associated with machine-learning feature identification. The classifier achieves an accuracy of 83\% for bars and 93\% spirals, with comparable completeness and purity \citep{YeR+25}. Because systematic uncertainties likely dominate, statistical errors are not reported in the main section.

The fraction of mergers among young galaxies is $\sim$8--11\% across the three flow states based on machine classification. Although this is roughly half the fraction inferred from asymmetry ($A>0.3$), both metrics indicate that major mergers are not the dominant drivers of gas cycling and accretion. While major mergers are expected to funnel gas to galaxy centers and trigger strong outflows \citep[e.g.,][]{Hopkins+08}, this behavior is rarely observed, in contrast to the more frequent association with secular processes such as bars and spiral arms.

For completeness, we repeat Fig.~3d showing distributions of S\'{e}rsic indices of the three flow states using a non-parametric light concentration measure in z-band. In particular, the concentration index, $C$, quantifies the central light concentration of a galaxy and is defined as $C = 5 \log_{10}(r_{80}/r_{20})$, where $r_{20}$ and $r_{80}$ are the radii enclosing 20\% and 80\% of the galaxy's total flux, respectively \citep{Rodriguez-Gomez+19}. Higher $C$ values indicate more centrally concentrated light profiles.

Supplementary Fig.~\ref{fig:CA_VP} shows the asymmetry and concentration distributions for galaxies separated by flow state and $D_n(4000)$. Consistent with their older ages and lower star formation rates, inflow galaxies are the least asymmetric and most concentrated. Because these populations differ structurally, comparisons must be made at fixed age. Even among young galaxies, inflow systems remain less asymmetric and more concentrated than outflow galaxies. Young no/weak-flow galaxies are the least concentrated and exhibit asymmetries comparable to outflows. In contrast, old no/weak-flow galaxies are as symmetric as old inflow systems but are significantly less concentrated (KS test $p < 10^{-6}$).

\subsection*{Angular momentum transport and galaxy structure}

The MACER simulations establish a direct link between angular momentum redistribution and observable galaxy structure. Observed young, star-forming galaxies span a wide range of bulge prominence (traced observationally by S\'{e}rsic and concentration indices), reflecting differences in angular momentum content and transport efficiency. Non-axisymmetric structures such as bars and spiral arms both trace and facilitate angular momentum transfer, enabling sustained gas inflows and prolonged star formation without immediate quenching. In contrast, episodes of strong angular momentum loss --- driven by mergers, violent disk instabilities, or high-momentum inflows --- trigger centrally concentrated starbursts followed by rapid suppression of star formation. Subsequent dynamical heating and disk fading erase non-axisymmetric features, producing smooth, bulge-dominated morphologies.

\subsection*{Photoionization models and ISM metallicity}

To interpret the observed emission-line ratios in terms of gas metallicity, we use the MAPPINGS-V photoionization model grids \cite{Flury+25}, which include both AGN and BPASS stellar population inputs. The grids cover a wide range of metallicities ($Z$) and ionization parameters ($U$), enabling comparison with observations. Star-forming models employ BPASS stellar population SEDs, accounting for stellar age, rotation, and binary evolution, and include both single stellar populations and continuous star formation histories. AGN models adopt disk plus power-law SEDs parameterized by black hole mass and accretion rate. All models incorporate dust and radiation pressure, assuming an isobaric density profile to approximate realistic H\,\textsc{ii} regions and AGN narrow-line regions.

Supplementary Fig.~9 shows the predicted emission-line ratios [N\,\textsc{ii}]/[S\,\textsc{ii}] versus [N\,\textsc{ii}]/H$\alpha$ for AGN and star-forming MAPPINGS-V models \cite{Dopita+16, Flury+25}. Panels (a) and (b) display the model ratios colored by metallicity relative to solar ($\log\,Z/\mathrm{Z_\odot}$) and ionization parameter ($\log\,U$), respectively. Within the observed parameter space, [N\,\textsc{ii}]/[S\,\textsc{ii}] varies primarily with metallicity, while its dependence on $\log\,U$ is weak, confirming its robustness as a metallicity tracer. Panels (c) and (d) show contours of the observed ratios for galaxies with inflows, outflows, and weak/no detected flows, including (c) or excluding (d) AGN hosts. The ISM metallicities of galaxy centers hosting inflows or outflows are broadly consistent with solar within a factor of two, suggesting the absence of massive metal-poor inflows that would otherwise reduce central metallicities by an order of magnitude, consistent with expectations for massive galaxies in the EAGLE cosmological simulations \citep{Wright+21}.

We caution that converting the observed emission-line ratios into metallicity and ionization parameter is inherently model dependent. While [N\,\textsc{ii}]/[S\,\textsc{ii}] provides a relatively robust metallicity diagnostic within the parameter space probed here, the ionization parameter is only weakly constrained by [N\,\textsc{ii}]/[S\,\textsc{ii}] and [N\,\textsc{ii}]/H$\alpha$ alone. Additional diagnostics such as [O\,\textsc{iii}]/[O\,\textsc{ii}] improve the constraints but introduce further uncertainties related to dust corrections and line detectability in weakly star-forming systems. Exploratory analyses using BPASS H\,\textsc{ii}-region models and additional line ratios yield metallicity trends qualitatively consistent with those inferred using the Dopita et al. calibration, while suggesting similar median ionization parameters across the three flow classes ($\log U \sim -3.5$). Given the remaining systematic uncertainties and calibration dependence, we defer a full $(Z,U)$ inversion analysis to future work.

Accretion from the circumgalactic medium could, in principle, dilute central metallicities through mixing with lower-metallicity gas, although such dilution is not uniquely associated with external accretion. Our measurements, however, probe the metallicity of the central ISM through strong-line diagnostics rather than the metallicity of the inflowing gas itself, and therefore provide only indirect constraints on the origin of the accreted material. Within this framework, the absence of systematically lower ISM metallicities in inflow systems---together with the similarity in ISM metallicity between inflow and non-flow galaxies at fixed stellar age---argues against scenarios in which the detected inflows are dominated by pristine or strongly metal-poor gas. Some lower-metallicity accretion may still occur and mix efficiently with the existing ISM, but it is unlikely to dominate the gas mass budget. Instead, the near-solar ISM metallicities and lack of strong central dilution are more naturally explained if a substantial fraction of the accreting material is already chemically enriched, consistent with recycled gas or the cooling of previously processed halo material, as predicted in feedback-regulated models such as MACER.

\begin{figure}
\includegraphics[width=0.49\linewidth]{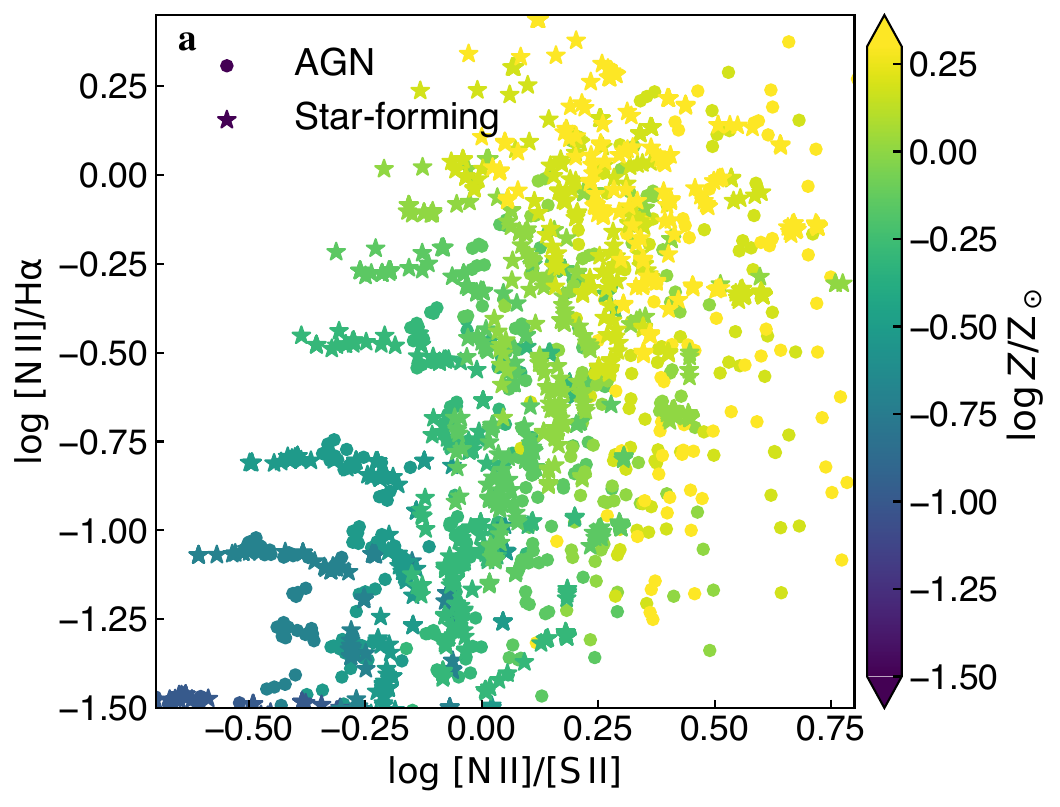}
\includegraphics[width=0.49\linewidth]{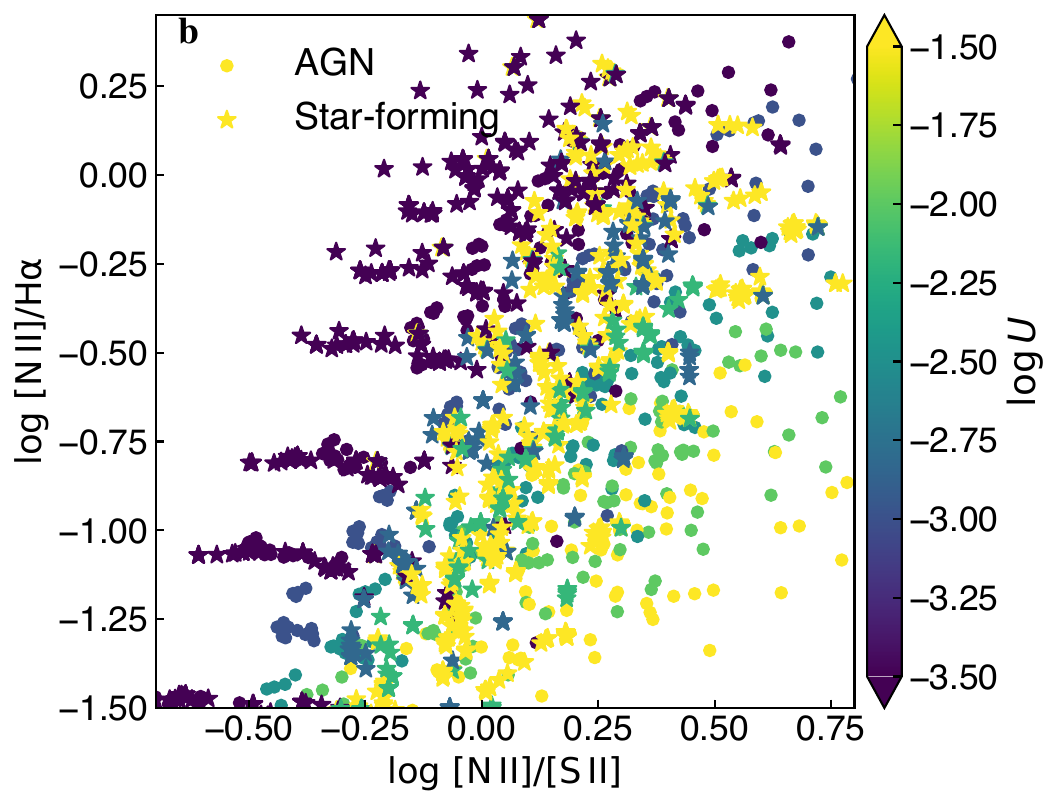}
\includegraphics[width=0.49\linewidth]{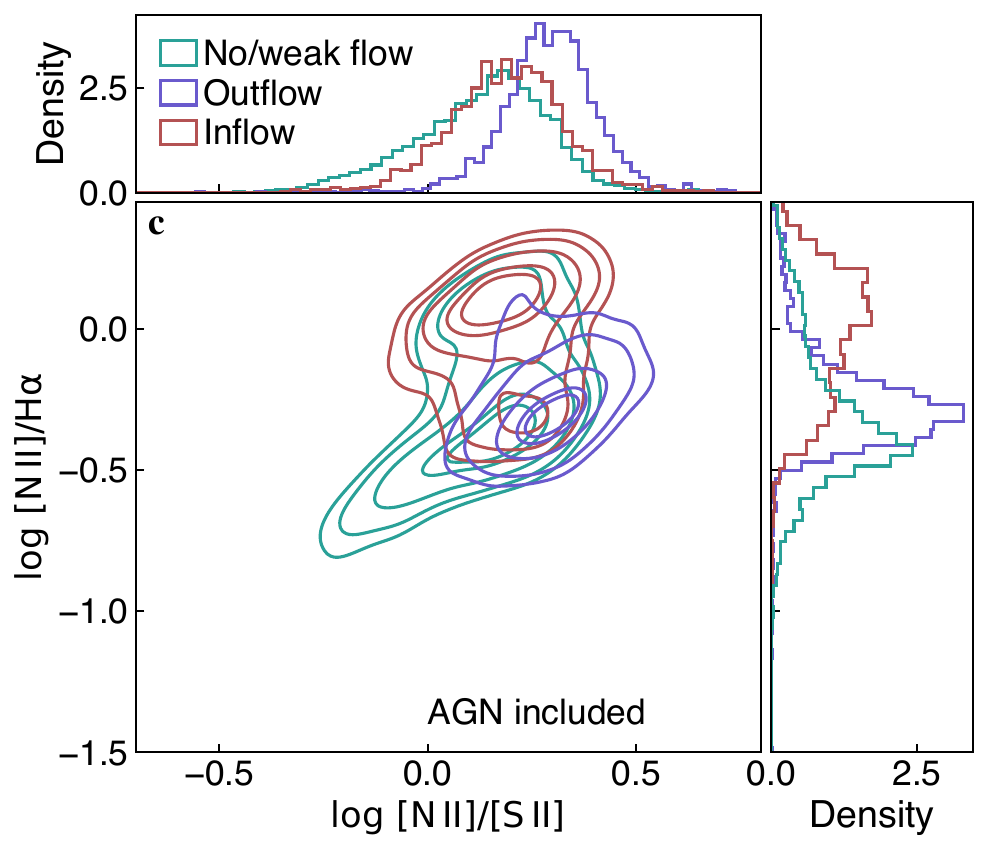}
\includegraphics[width=0.49\linewidth]{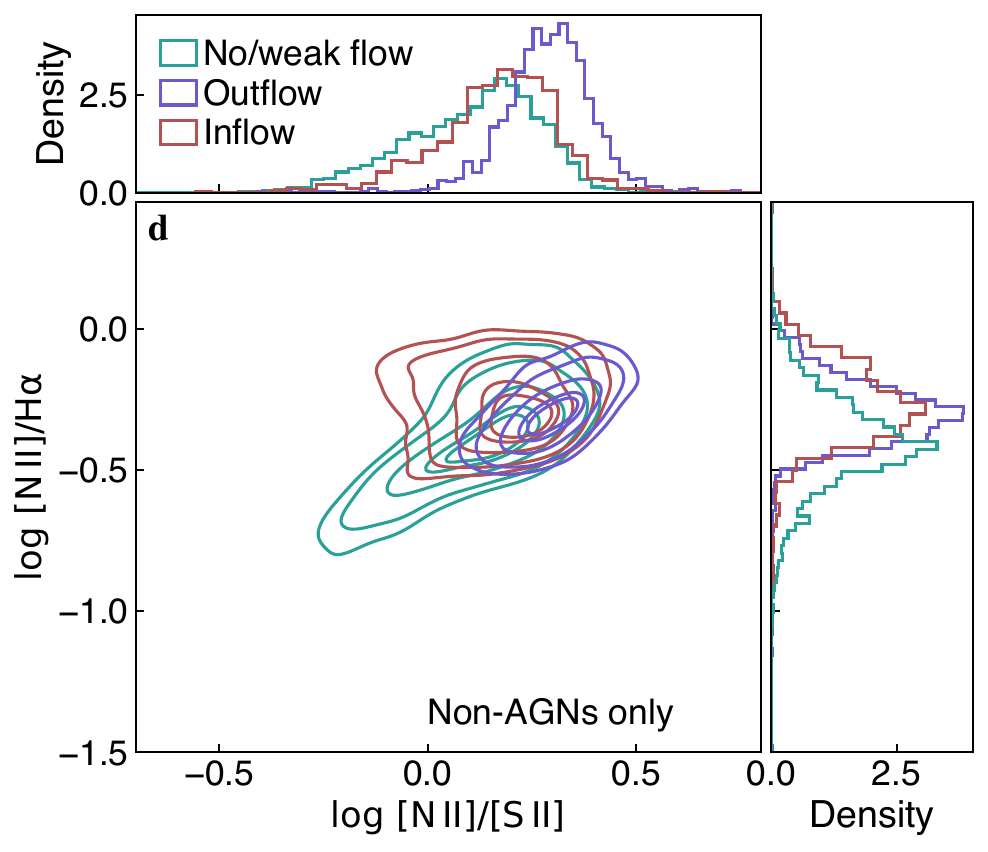}
\caption{\textbf{Nitrogen emission-line ratios as tracers of gas-phase metallicity.} 
    \textbf{a--b,} MAPPINGS-V photoionization model grids \cite{Flury+25} showing the predicted emission-line ratios 
    [N\,\textsc{ii}]$\lambda6583$/[S\,\textsc{ii}]$\lambda\lambda6717,6731$ versus [N\,\textsc{ii}]$\lambda6583$/H$\alpha$ 
    for AGN and star-forming regions. Colors indicate (\textbf{a}) gas-phase metallicity, and (\textbf{b}) ionization parameter. \textbf{c--d,} Observed contours of the same line ratios for galaxies with inflows (red), outflows (blue), and no/weak flows (teal), 
    including (\textbf{c}) or excluding (\textbf{d}) AGN hosts.}
\end{figure}

\medskip

\textbf{Synthesis and outlook}

\medskip

Individually, alternative scenarios can reproduce subsets of the observations. However, no single mechanism simultaneously explains the full combination of kinematic, chemical, demographic, and structural constraints. Our adopted gas-flow framework (Fig.~2)---dominated by galactic fountain recycling in star-forming systems and slow cooling of enriched halos in quiescent systems---provides the most self-consistent interpretation of the data. This framework naturally explains (i) the demographic continuity across flow classes, (ii) the absence of strongly metal-poor inflows, (iii) the correlations with stellar age and star formation activity, (iv) the role of geometry and dust in shaping flow detectability, and (v) the predominantly gravitationally bound nature of the observed outflows. Taken together, these results suggest that most cool inflows observed in massive nearby galaxies do not arise from pristine cosmological accretion, but instead originate primarily from recycled or previously enriched gas.

Future progress will require spatially resolved observations capable of tracing gas from ejection to re-accretion, multi-phase measurements linking the hot, warm, and cool gas components, and tighter constraints on star formation histories, low-level AGN activity, and the physics of halo cooling in quiescent systems. While the present results point to a gas cycle dominated primarily by internal recycling in nearby young galaxies, the relative importance of recycling and external accretion is expected to evolve with redshift as cosmic gas inflow rates, gas fractions, star formation activity, and black hole growth all increase.

\bibliography{references_flow}
\end{document}